%% file: main.tex
\documentclass[final,authoryear]{elsarticle}
\usepackage[utf8]{inputenc}
\usepackage[margin=1in]{geometry}
\usepackage[english]{babel}
\usepackage{booktabs}
\usepackage[labelfont=bf, skip=2pt, font=small]{caption}
\usepackage{graphicx}
\usepackage{hyperref}
\usepackage{bbm}
\usepackage{amsmath,amssymb}

\makeatletter
\def\ps@pprintTitle{%
  \let\@oddhead\@empty
  \let\@evenhead\@empty
  \let\@oddfoot\@empty
  \let\@evenfoot\@empty
}
\makeatother

\begin{document}

\input{frontmatter}

\input{introduction}

\input{methodology}

\input{stylised_facts}

\input{scenario_analysis}

\input{discussion}

\input{conclusion}

\input{acknowledgements}

\input{appendix}

\bibliographystyle{elsarticle-harv.bst}

\bibliography{references}

\end{document}

%% file: frontmatter.tex
\begin{frontmatter}

\title{Implications of zero-growth economics analysed with an agent-based model}

\author[aff1]{D. C. Terry-Doyle\corref{cor1}}
\ead{d.c.terry-doyle@sussex.ac.uk}

\author[aff1,aff2]{A. B. Barrett}
\ead{adam.barrett@sussex.ac.uk}

\cortext[cor1]{Corresponding author}

\address[aff1]{Department of Informatics, University of Sussex, Brighton BN1 9QJ, UK}
\address[aff2]{Sussex Centre for Consciousness Science, University of Sussex, Brighton BN1 9QJ, UK}

\begin{abstract}
    The breaching of planetary boundaries and the potentially catastrophic consequences of climate change are leading researchers to question the endless pursuit of economic growth. Several macroeconomic modelling studies have now examined whether a zero-growth trajectory in a capitalist system with interest-bearing debt can be economically stable, with mixed results. However, stability has not previously been explored at the microeconomic level, where it is important to know the consequences of zero-growth on e.g., distribution of firm sizes, market instability and risk of individual firm bankruptcy. Here we address this by developing an agent-based model incorporating Minskyan financial dynamics, the Post-Growth DYNamic Agent-based MINskyan (PG-DYNAMIN) model, and carrying out simultaneous macro- and microeconomic analyses. Accounting for the fact that growing capitalist economies are unstable and produce crises, we compare the relative stability of growth and zero-growth scenarios. This is achieved by tweaking an exogenous productivity parameter. We find zero-growth scenarios are viable yet exhibit distinct dynamics from growth scenarios. Under zero-growth, GDP was less volatile, there was reduced systemic risk in the credit network, lower unemployment rates, a higher wages share of GDP for workers,  lower corporate debt to GDP ratio, and a reduction in market instability. Additionally, there was a higher rate of inflation, lower profit share of GDP for firms, increased market concentration, more economic crises with higher severity, and increased default probabilities for firms during periods of crises.
\end{abstract}


\end{frontmatter}

%% file: introduction.tex
\section{Introduction} 
\label{Introduction}
For as long as we have been modelling economies, they have existed in states of growth, at least in the long-run; and discussion of an end to growth has been confined outside the economics mainstream for more than a century. However, growth rates across the developed world have, in recent years, slowed to what appears to be a “new normal” of 1\% \citep{malmaeus2017potential, kallis2025postgrowth}, and further increases in wealth from material and energy use are becoming increasingly bound by the Earth's biophysical limits. There is evidence to suggest that six out of nine `planetary boundaries' are currently transgressed \citep{rockstrom2009safe, steffen2015planetary, richardson2023earth}, and that the Earth system has fallen outside the `safe operating space for humanity'. It is therefore timely and imperative to develop models that seek insight into economic dynamics in a post-growth era. 

It remains a concern that the prosperity predicted by \cite{keynes1930granchildren} has not been reached. Inequality and poverty remain widespread in our economic systems \citep{piketty2014capital, chancel2021inequality}, and a stationary state of affluence and abundance remains a distant ideal. Further growth is still the remedy prescribed by the 2030 Agenda for Sustainable Development, adopted in 2015 by all UN member states, which explicitly states in the 8$^{th}$ Sustainable Development Goal\footnote{For Sustainable Development Goal 8, see here: \url{https://sdgs.un.org/goals/goal8}} an aim to ``promote sustained, inclusive and sustainable economic growth, full and productive employment and decent work for all". Achieving this goal while simultaneously tackling climate change, and the other planetary boundaries, relies on the concept of `decoupling', to promote economic growth while reducing the use of natural resources and emissions. However, there is a growing body of evidence that decoupling and green growth policies will not enable sufficiently rapid reductions in emissions \citep{haberl2020systematic} and there are calls to look beyond these to more quickly reduce our impact on the environment \citep{elder2019design}. 

The idea of decoupling is bound to an unrealistic circular flow vision of economies, which is inconsistent with the physics of thermodynamics. As outlined in Georgescu-Roegen's seminal work \citep{georgescuroegen1972entropy}, and later discussed by \cite{daly1996beyond}, fundamental to the economic process is the transformation of low-entropy input from the environment and the simultaneous return of high-entropy polluting output. Hence, it is impossible for the global economy to continuously grow, because low-entropy inputs, such as the finite flow of solar energy and the finite set of terrestrial materials, become depleted, and high-entropy outputs, such as material waste and emissions, are accumulated. Explicit IPAT\footnote{The IPAT model is a simple formulation of environmental impact (I) as a function of population ($P$), affluence ($A$), and technology ($T$): $I=PAT$.} modelling in \cite{ward2016decoupling} concurred, demonstrating that economic growth cannot be decoupled from growth in material and energy use. \cite{wiedmann2020scientists} discussed how resource use and pollutant emissions grow more rapidly than the efficiency of technology, which relates to Jevons' paradox, whereby gains in efficiency cause a rise in production and consumption \citep{alcott2005jevons}. Looking empirically, \cite{parrique2019decoupling} and \cite{hickel2020green} could find no real-world evidence to support decoupling and green growth theory. 

Given these findings, and the increasingly urgent need to stabilise the Earth system, research on post-growth economics is burgeoning \citep{lauer2025comparative,edwards2025towards}. For example, modelling by \cite{slamervsak2024post_growth} demonstrated that low growth rates would make it more feasible to reach mitigation goals consistent with a 1.5°C-2°C average temperature rise above pre-industrial levels. A common conclusion for \textit{advanced nations} is that: while sustained further growth may in any case be difficult to achieve \citep{malmaeus2017potential}, a policy shift away from seeking this out will be vital for tackling instability in the Anthropocene.

The important question arises here whether an end to growth would lead to greater economic instability in a capitalist system, particularly due to the presence of interest-bearing debt. \cite{barrett2018stability} analysed this on a macroeconomic model by tweaking a productivity growth input parameter, which was set at either 2\% or zero. The model, based on work by \cite{keen1995finance}, incorporated elements of Goodwin's growth cycle model \citep{goodwin1967growth_cycle} and Minsky’s Financial Instability Hypothesis (FIH). The FIH highlights the general instability of a capitalist market, and the role of debt dynamics in creating economic crises \citep{minsky1977financial,minsky1986stabilizing,minsky1992financial}. On Barrett's model, zero-growth scenarios are generally no less stable than growth scenarios. Several other modelling studies have found an economically desirable zero-growth fixed point, i.e.~equilibrium \citep{berg2015stock,rosenbaum2015zerogrowth,jackson2015growth_imperative,cahen2016ecological,jackson2020transition,alessandro2020green_growth}. One model \citep{binswanger2009growth} did not produce an economically desirable equilibrium. However, the study of ongoing dynamics, as exemplified in \cite{barrett2018stability}, is more pertinent to the study of stability than the positions of fixed points.

The aforementioned literature on zero-growth modelling only considers macroeconomic variables. As yet unexplored is the impact zero-growth will have on variables derived from microeconomic data such as bankruptcy rates, market instability, market concentration, and credit network risk. The modelling framework that best suits an incorporation of the micro is that of agent-based models (ABMs). ABMs are a type of computational model in which there are numerous `agents’ interacting with each other, such as firms, households, or, in other fields, particles or cells. This computational framework enables the modeller to create a system from the micro level upwards, that can have emergent properties\footnote{Emergent properties are the phenomenon in which an entity has attributes that its individual parts do not have on their own.}, and analyse outcomes in both micro and macro level dynamics. ABMs do not depend on a supply of historical data to enable a change in policy or regime, which is useful here given the lack of long-run historical data from a zero-growth state. 

With the increase in computing power available to researchers in recent decades, the use of ABMs has been growing, see \cite{axtell2025abm} for a review of the literature. Several existing ABMs have stable zero-growth paths, for example \cite{lengnick2013baseline}, \cite{assenza2015macro_abm}, \cite{klimek2015bail}, and \cite{caiani2016benchmark_model} but this has not before been the main focus of analysis \citep{lauer2025comparative}, and the relative stability of growth versus zero-growth scenarios has not previously been studied on an ABM. 

This paper introduces the Post-Growth DYNamic Agent-based MINskyan (PG-DYNAMIN) model, and employs it to present an unprecedentedly detailed view of the nature of a post-growth economy, covering both macroeconomic and microeconomic analyses. As in \cite{barrett2018stability}, growth and zero-growth scenarios are compared by tweaking a parameter that controls the average growth of productivity. In addition, maintaining the idea from Minsky, that debt behaviour is important for stability, scenarios with different debt appetite levels are also compared. PG-DYNAMIN draws from a range of ABMs, including the CATS (Complex Adaptive Trivial Systems) models \citep{gatti2003financial,gatti2005new,russo2007industrial,gaffeo2008emergent_macro,gatti2010financial,assenza2015macro_abm}, the SK (Schumpeter meets Keynes) models \citep{dosi2006evolutionary,dosi2009microfoundations,dosi2010schumpeter,dosi2013income, dosi2017evolutionary}, as well as models from \cite{deissenberg2008eurace}, \cite{lengnick2013baseline}, \cite{klimek2015bail}, \cite{caiani2016benchmark_model}, \cite{lamperti2018faraway}, and \cite{poledna2023economic_forecasting} and strives to capture the core features of a capitalist economy with interest-bearing debt. It contains four sectors: households, consumption firms (which sell goods to households), capital firms (which sell capital to consumption firms), and banks.

The structure of the paper is as follows. Section \ref{sec:methodology} provides an overview of the methodology and model. Section \ref{sec:stylised_facts} demonstrates that PG-DYNAMIN reproduces a broad range of stylised facts. Section \ref{sec:scenario_analysis} focuses on a stability analysis comparing growth and zero-growth scenarios at both the macro- and microeconomic levels. Section \ref{sec:discussion} presents a discussion regarding the results. Finally, section \ref{sec:conclusion} contains a summary and concluding remarks. 

%% file: methodology.tex
\section{Methodology \& Modelling} 
\label{sec:methodology}
The PG-DYNAMIN model contains four sectors: (i) a household sector comprising $\mathbf{N}_H$ households; (ii) a consumption firm sector with $\mathbf{N}_C$ consumption firms (C-firms) selling a homogenous consumption good (C-good); (iii) a capital firm sector with $\mathbf{N}_K$ capital firms (K-firms) selling a homogenous capital good (K-good); and (iv) a banking sector with $\mathbf{N}_B$ banks. As we are interested in the instability of the capitalist system, we chose to omit the government sector, which should act as a damping force against the cyclical dynamics of a purely capitalist economy, given the right choice of fiscal policy. With these sectors in place, we endeavoured to keep the model as simple as possible for a focused analysis of financial stability under growth and zero-growth scenarios. The model is thus less detailed than some of the larger models in the literature \citep{dosi2017evolutionary, lamperti2018faraway, botte2021transition, reissl2025dsk}, yet it is capable of producing a rich set of stylised facts from real-world economies. The focus is fully on economic dynamics, and thus incorporating the environment into the model, as in \cite{lamperti2018faraway,lamperti2020climate,lamperti2021three} and \cite{reissl2025dsk}, is left to future research. The  model is stock-flow consistent (SFC) \citep{godley2006monetary_economics}: each financial asset is related to an equivalent financial liability; and each payment flow goes explicitly from one agent in the model to another. Here we give an overview of the most important features and assumptions of the model. For a detailed description of all the equations and algorithms included in the model, see \ref{app:model}.

\begin{figure}[!htb]
    \centering
    \includegraphics[width=0.9\textwidth]{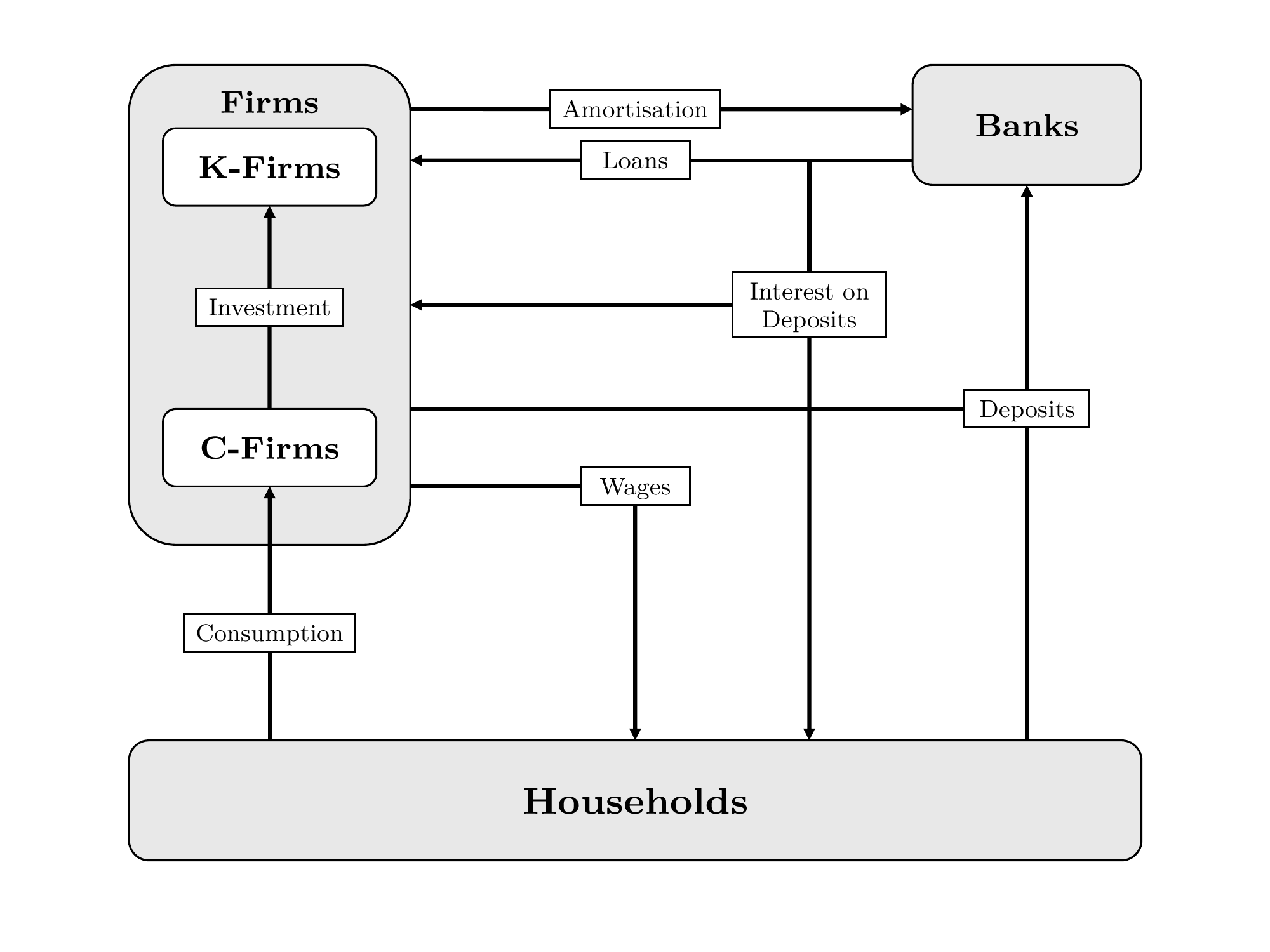}
    \caption{Model ontology, showing the flow of payments between each category of agent in the model. Arrows point from paying agents to receiving agents.}
    \label{figure_1}
\end{figure}

Fig.~\ref{figure_1} gives a visual representation of the model and how the agents interact with one another through financial flows. In summary, households work at either C-firms or K-firms and are paid a wage for their work each period; they put any savings at banks as deposits, and receive interest from the bank. Households also consume consumption goods from C-firms using their wage from employment and/or their deposits at the bank. C-firms invest in new capital from K-firms using their internal finance and loans from banks. Additionally, both C-firms and K-firms also seek external finance from banks if their wage bill exceeds internal finance capabilities. Firms use an amortisation schedule to pay down both the interest and principal of the loan. Furthermore, banks pay interest to firms and households for any deposits they hold at banks. Additionally, each period is assumed to correspond to a quarter of a year, where the sequence of events that occur during each time period is given below.

To initialise the PG-DYNAMIN model we create a stylised macroeconomic balanced growth version of the model, following methodology from \cite{caiani2016benchmark_model}. In balanced growth, it is assumed that the economy grows at real rate $g$ (average labour productivity) and there is a constant inflation rate $g_P$, thus, a nominal growth rate $g_N=g+g_P$. Therefore, all real stocks must grow at rate $g$ and all nominal stocks must grow at rate $g_N$ in balanced growth. This model can be then be solved to produce initial values for the stocks of households, firms, and banks, see \ref{app:balanced_growth} for details of the balanced growth model and \ref{app:initial_values} for the derived initial values.

\subsection{Sequence of events}
\label{sec:sequence_events}
Each period, the model moves discretely through a pre-determined sequence of events as follows:
\begin{enumerate}
  \item At the beginning of each time step $t$, new entrants enter their respective markets to replace firms that defaulted in the previous period.
  \item A decentralised market for labour opens where firms post vacancies and update their wage rate to attract new employees. Households then send out applications to firms that have open vacancies, and firms hire employees from their application pool.
  \item C-firms and K-firms engage in production, pay their employees for their work, and update the price of their goods.
  \item A decentralised market for consumption goods opens, and households visit and buy goods from C-firms for consumption. 
  \item A decentralised market for capital goods opens, and C-firms visit and buy goods from K-firms to use as investment in their capital stock for production in the next period.
  \item A decentralised credit market opens, and C-firms and K-firms demand loans from banks, which are supplied depending on the bank's risk tolerance.
  \item C-firms and K-firms update their accounts and exit the market if they have run out of deposits to cover their payments. Households employed by bankrupt firms become unemployed, and banks absorb any outstanding loans to bankrupt firms with their equity. 
  \item Banks become bankrupt if they have negative equity, and are bailed out by their depositors, both firms and households. 
\end{enumerate}

\subsection{Consumption firms}
Consumption firms' output is given by a Leontief production function:
\begin{equation}
    Y_i(t)=\min\left\{a_i(t)N_i(t),\frac{K_i(t)}{\nu}\right\},
\end{equation}
where $Y_i(t)$ is the output of C-firm $i$ in period $t$, $a_i(t)$ is labour productivity, $N_i(t)$ is labour, $K_i(t)$ is the capital stock (stock of accumulated K-goods), and $\nu$ is a fixed capital-to-output ratio, uniform across C-firms. 

The key variable underpinning growth is firms' labour productivity, which is assumed to be given by a discrete-time geometric Brownian motion (GBM):
\begin{equation}
\label{eq:productivity}
    a_i(t)=a_i(t-1)\exp\left\{g - \frac{1}{2} \sigma_a^2 + \sigma_a \varepsilon_i(t)\right\}.
\end{equation}
Here $g$ is the key parameter that specifies the mean growth rate of labour productivity. This is set to a fixed positive value ($2\%$ mean per annum) for growth scenarios, and to zero for zero-growth scenarios, as in \cite{barrett2018stability}. The variability of productivity growth is captured by the dispersion parameter $\sigma_a$, and $\varepsilon_i(t)\sim \mathcal{N}(0,1)$ is distributed according to a standard normal distribution.

C-firms set their desired output to expected demand using adaptive expectations (exponentially decaying weights on previous demand). They pay households wages each period, and calculate their profits as revenue minus net wage bill and interest payments.
The pricing mechanism is similar to the CATS framework \citep{russo2007industrial, gaffeo2008emergent_macro, assenza2015macro_abm},  C-firms either mark their prices up (down) by a random amount if they have excess demand (supply), with the addition that prices adjust towards the average price level in order to remain competitive. Similarly, C-firms mark their wages up (down) if they have excess labour demand (supply) and adjust towards the average wage. C-firms' investment depends on both internal and external financing capabilities. Similarly, to the model in \cite{barrett2018stability}, C-firms use a debt-to-output ratio to determine their desired level of external financing for investment, given by:
\begin{equation}
    d^d_i(t+1) = d_0 + d_1 \alpha_i(t) + d_2 \pi_i(t)
\end{equation}
where $d_0,d_1,d_2>0$ are parameters, $\alpha_i(t)=\ln(a_i(t))-\ln(a_i(t-1))$ is the log-difference in labour productivity (approximates percentage change), and $\pi_i(t)=\Pi_i(t)/P_i(t)Y_i(t)$ is the profit share of nominal output. Hence, $d_1$ and $d_2$ control how strongly changes in $\alpha_i(t)$ and $\pi_i(t)$ lead to changes in the desired debt-to-output ratio. The $d_1$ and $d_2$ parameters allow us to change the financial behaviour of C-firms in the model; increasing $d_1$ and $d_2$ will generally lead to higher desired debt-to-output ratios for positive $\alpha_i(t)$ and $\pi_i(t)$. Additionally, C-firms can take on extra debt to cover their wage bill if their internal funds are low.

Consumption firms place investment orders from capital firms each period, using a random matching process, similar to \cite{assenza2015macro_abm}. This can lead to imperfect market allocation of capital goods, with C-firms having excess demand and K-firms having excess supply. Furthermore, when C-firms run out of deposits they become bankrupt and it is assumed that there is a one-to-one replacement of bankrupt firms in the model. See \ref{app:cfirms} for a detailed description of C-firm behaviour, \ref{app:entry_exit} for C-firms' entry and exit dynamics, \ref{app:labour_market} for the labour market, \ref{app:consumption_market} for the consumption market, and \ref{app:capital_market} for the capital market.

\subsection{Capital firms}
Capital firms' output is their labour productivity, $a_j(t)$, multiplied by their labour employed, $N_j(t)$, hence, K-firm $j$'s production of K-goods is given by:
\begin{equation}
    Y_j(t)=a_j(t)N_j(t),
\end{equation}
where K-firm labour productivity is also given by a discrete-time GBM, Eq.~\ref{eq:productivity}, with mean gain in productivity per year given by the same parameter $g$ as for consumption firms. Capital firms use the same mechanism to increase (decrease) their prices and wages as consumption firms, depending on their excess demand (supply). Capital firms take on a more passive role in finance, only taking on debt to cover their wage bill if they have insufficient internal funds. Similarly to C-firms, there is a one-to-one replacement of bankrupt K-firms in the model. See \ref{app:kfirms} for a detailed description of K-firm behaviour, \ref{app:entry_exit} for K-firms' entry and exit dynamics, \ref{app:labour_market} for the labour market, and \ref{app:capital_market} for the capital market.

\subsection{Households}
Households are classified as either employed or unemployed, depending on whether they currently have a position at a firm. If households are employed they inelastically supply one unit of labour to their employer each period, for which they receive a wage. Otherwise, unemployed households enter the labour market each period and send job applications to firms. Households calculate their desired expenditure using a simple Keynesian consumption function; they desire to spend a fixed proportion of their income and savings. Markets are inherently imperfect, with households having bounded rationality when they search for C-firms to visit: households visit a small random selection of C-firms and consume as much as they can, they have a higher probability of visiting large firms and prioritise whichever firm has the cheapest goods. Any unconsumed income is kept as savings at the bank for future expenditure. See \ref{app:households} for a detailed description of household behaviour, \ref{app:labour_market} for the labour market, and \ref{app:consumption_market} for the consumption market.

\subsection{Banks}
Banks play a crucial role in the model, determining the number of loans issued based on their risk tolerance. In line with \cite{mcleay2014money} and \cite{werner2014money}, the money creation and destruction process is endogenous in the model. Loans from banks to firms create new deposits for C-firms to use for investment or for both C-firms and K-firms to cover their wage bill in case of cash flow shortages. Additionally, the repayment of debt by firms destroys money. 

Banks can vary the supply of credit (new loans) to firms based on their current risk tolerance, which is determined based on the difference between their desired and actual capital ratio.\footnote{Banks capital ratio is the quotient of their equity to loans.} Banks' desired capital ratio is the minimum of a regulatory required minimum and their expected loss to loans ratio. Expected loss is determined by estimated default probabilities for firms,\footnote{Banks estimate firm default probabilities by applying a logistic regression to firm leverage ratios, as in \cite{assenza2015macro_abm}.} and loans outstanding \citep{chatterjee2015credit_risk}. Banks default when their equity becomes negative, and following default, it is assumed banks are bailed-in by their depositors, namely households and firms. If depositors of defaulted banks are unable to sufficiently cover the losses incurred, then an unmodelled central-bank acts as a lender of last resort. See \ref{app:banks} for a full description of bank behaviour, \ref{app:entry_exit} for banks' entry and exit dynamics, and \ref{app:credit_market} for details of the credit market.

During times of stability when firms' profit rates and banks' capital ratios are high, credit is built up as C-firms demand more debt for investment and banks are willing to supply the credit. Credit (the change in debt) can then collapse during a downturn as banks restrict the supply of credit due to defaults causing capital ratios to drop below desired levels, and C-firms desiring less debt for investment as profitability reduces. Once banks have sufficiently deleveraged, they once again start to issue new loans, and the economy can recover. This increases firms' profitability and thus willingness to take on more debt relative to their size. Hence, the model produces an endogenous credit cycle in accordance with the financial instability hypothesis proposed by \cite{minsky1977financial, minsky1986stabilizing, minsky1992financial}. Credit relations fluctuate between those that create a stable economy to those that create an unstable economy.

\subsection{Stock-flow consistency}
It is demonstrated here that the model is stock-flow consistent (SFC) at the aggregate level, ensuring that all financial transactions are accounted for, such that each payment from one agent is directed to another agent in the model. Therefore, every financial stock is recorded as a liability for one agent and an asset for another agent. 

Table~\ref{tab:balance_sheets} provides a financial balance sheet representation of the model for each agent class, where values in the table represent the aggregate stock variables for each type of agent\footnote{Each variable with a subscript refers to that agent's share of the total, e.g. $M_H$ are household deposits, $M_C$ are C-firm deposits, and $M_K$ are K-firm deposits. The same variable without a subscript refers to the total of that variable, e.g. total deposits are the sum of all agents' deposits: $M=M_H+M_C+M_K$.}. It is evident from Table~\ref{tab:balance_sheets} that all the columns and rows dealing with financial assets must sum to zero, where tangible capital ($K$) is the only asset that appears once on the balance sheet and thus does not sum to zero. The sum of all agents' equities must be equal to tangible capital, in line with the reasoning set out by \cite{godley2006monetary_economics}.

\begin{table}[!htb]
    \small
    \centering
        \begin{tabular}{lccccccc}
        \toprule
            & \textbf{Households} & \textbf{C-Firms} & \textbf{K-Firms} & \textbf{Banks} & \textbf{Central Bank} & $\sum$ \\
        \midrule
        Capital     &  & $K$ &  &  &  & $K$ \\
        Deposits    & $M_H$ & $M_C$ & $M_K$ & $ -M$ &  & 0 \\
        Debt        &  & $-D_C$ & $-D_K$ & $D$ &  & 0 \\
        Reserves    &  &  &  & $R$ & $-R$ & 0 \\
        Advances    &  &  &  & $-A$ & $A$ & 0 \\
        Equity      & $-E_H$ & $-E_C$ & $-E_K$ & $-E_B$ & $-E_{CB}$ & $-K$ \\
        \midrule
        $\sum$ & 0 & 0 & 0 & 0 & 0 & 0 \\
        \bottomrule
        \end{tabular}
    \caption{Macro financial balance sheet matrix.}
    \label{tab:balance_sheets}
\end{table}

Table~\ref{tab:transaction_flows} presents a transaction flow matrix showing the aggregate flows of financial transactions among the agents in the model. Each column and row of the transaction flow matrix must sum to zero for the model to be SFC. The upper part of the transaction flow matrix in Table~\ref{tab:transaction_flows} reproduces the national income statistics presented in Table~\ref{tab:balance_sheets}. The lower part of Table~\ref{tab:transaction_flows} represents the inter-sectoral flow of funds in the model. The variables in the transaction flow matrix are defined as follows: $W$ is the wage bill, $C$ is household consumption, $I$ is C-firm investment, $\rho D$ is debt repayment, $IP$ is loan interest payments, $i^M M$ is deposit interest payments, $\Pi$ is profits, $\Delta M$ is the change in deposits, $L$ is new loans, $\Delta R$ is the change in reserves, $\Delta A$ is the change in central bank (CB) advances, and $B$ is firm bad debt (total non-performing loans). 

\begin{table}[!htb]
    \scriptsize
    \centering
        \begin{tabular}{lcccccccccc}
        \toprule
            & \textbf{Households} & \multicolumn{2}{c}{\textbf{C-Firms}} & \multicolumn{2}{c}{\textbf{K-Firms}} & \multicolumn{2}{c}{\textbf{Banks}} & \textbf{CB} & $\sum$ \\
        \cmidrule(l{10pt}r{5pt}){3-4}
        \cmidrule(l{10pt}r{5pt}){5-6}
        \cmidrule(l{10pt}r{5pt}){7-8}
            &  & Current & Capital & Current & Capital & Current & Capital & \\
        \midrule
        Wages              & $W$ & $-W_C$ &  & $-W_K$ &  &  &  &  & 0 \\
        Consumption        & $-C$ & $C$ &  &  &  &  &  &  & 0 \\
        Investment         &  &  & $-I$ & $I$ &  &  &  &  & 0 \\
        Loan repayments    &  &  & $-\rho D_C$ &  & $-\rho D_K$ &  & $\rho D$ &  & 0 \\
        Loan Interest      &  & $-IP_C$ &  & $-IP_K$ &  & $IP$ &  &  & 0 \\
        Deposit Interest   & $i^M M_H$ & $i^M M_C$ &  & $i^M M_K$ &  & $-i^MM$ &  &  & 0 \\
        Profits            &  & $-\Pi_C$ & $\Pi_C$ & $-\Pi_K$ & $\Pi_K$ & $-\Pi_B$ & $\Pi_B$ &  & 0 \\
        Inventories        &  &  &  & $\Delta V$ & $-\Delta V$ &  &  &  & 0 \\
        \midrule
        Change in Deposits & $-\Delta M_H$ &  & $-\Delta M_C$ &  & $-\Delta M_K$ &  & $\Delta M$ &  & 0 \\ 
        Change in Debt     &  &  & $L_C$ &  & $L_K$ &  & $-L$ &  & 0 \\
        Change in Reserves &  &  &  &  &  &  & $-\Delta R$ & $\Delta R$ & 0 \\
        Change in Advances &  &  &  &  &  &  & $\Delta A$ & $-\Delta A$ & 0 \\
        Loan Defaults      &  &  & $B_C$ &  & $B_K$ &  & $-B$ &  & 0 \\
        \midrule
        $\sum$             & 0 & 0 & 0 & 0 & 0 & 0 & 0 & 0 & 0 \\
        \bottomrule
        \end{tabular}
    \caption{Macro transaction flow matrix.}
    \label{tab:transaction_flows}
\end{table}

\subsection{Parameters and simulation scenarios}
Simulations were run for four different sets of parameter choices, in order to compare growth and zero-growth scenarios for two different levels of C-firm desired debt. Table~\ref{tab:scenario_parameters} shows the parameter choices that are distinct for each scenario. Growth scenarios have mean productivity growth set to 2\% ($g=0.02$), while zero growth scenarios set this to zero ($g=0$). In scenarios `Growth S1' and `Zero-Growth S1' C-firms have low desired debt, while in scenarios `Growth S2' and `Zero-Growth S2' C-firms have high desired debt. 

\begin{table}[!htb]
    \centering
        \begin{tabular}{lcccc}
        \toprule
        \textbf{Parameters} & \textbf{Growth S1} & \textbf{Growth S2} & \textbf{Zero-Growth S1} & \textbf{Zero-Growth S2}\\
        \midrule
        $g$     & 0.02  & 0.02  & 0     & 0     \\
        $d_1$   & 3     & 5     & 3     & 5     \\
        $d_2$   & 2     & 3     & 2     & 3     \\
        \bottomrule
        \end{tabular} 
    \caption{Parameter values used for each scenario}
    \label{tab:scenario_parameters}
\end{table}

Values for the parameters that were held constant across the different scenarios were chosen where possible based on estimated values from real data, or else by imposing that the model produce non-degenerate outcomes for the growth scenario, specifically, outcomes with no hyperinflation, collapse in GDP toward zero, sustained unemployment rate over $50\%$, or sustained bankruptcy rates over $50\%$. See \ref{app:parameters} for details of the parameters used.

Each scenario was run 100 times, in order to average stochastic effects. The same set of 100 random seeds ($s\in\mathcal{S}=\{1,2,...,100=S\}$) were used to generate the 100 runs for each of the four scenarios (to ensure differences between scenarios are not due to random variables taking different values). Each simulation was run for a total of 800 periods, with the first 400 periods removed to avoid any transient behaviour at the start of the simulation. Hence, the results analyse the last 400 periods of each simulation, and given the assumption that each time step represents a quarter of a year, this equates to a total simulation length of 100 years.

%% file: stylised_facts.tex
\section{Stylised Facts}
\label{sec:stylised_facts}
To confirm the usefulness of a model for performing a scenario analysis, it is first imperative to evidence the model's ability to produce realistic data. As such, we demonstrate that the baseline Growth S1 scenario (corresponding to the first column in Table~\ref{tab:scenario_parameters}) of the PG-DYNAMIN model reproduces a broad range of macro- and microeconomic stylised facts from real world economies and the ABM literature \citep{gaffeo2008emergent_macro, assenza2015macro_abm, caiani2016benchmark_model, lamperti2018faraway}. 

\begin{figure}[!htb]
    \centering
    \includegraphics[width=0.9\textwidth]{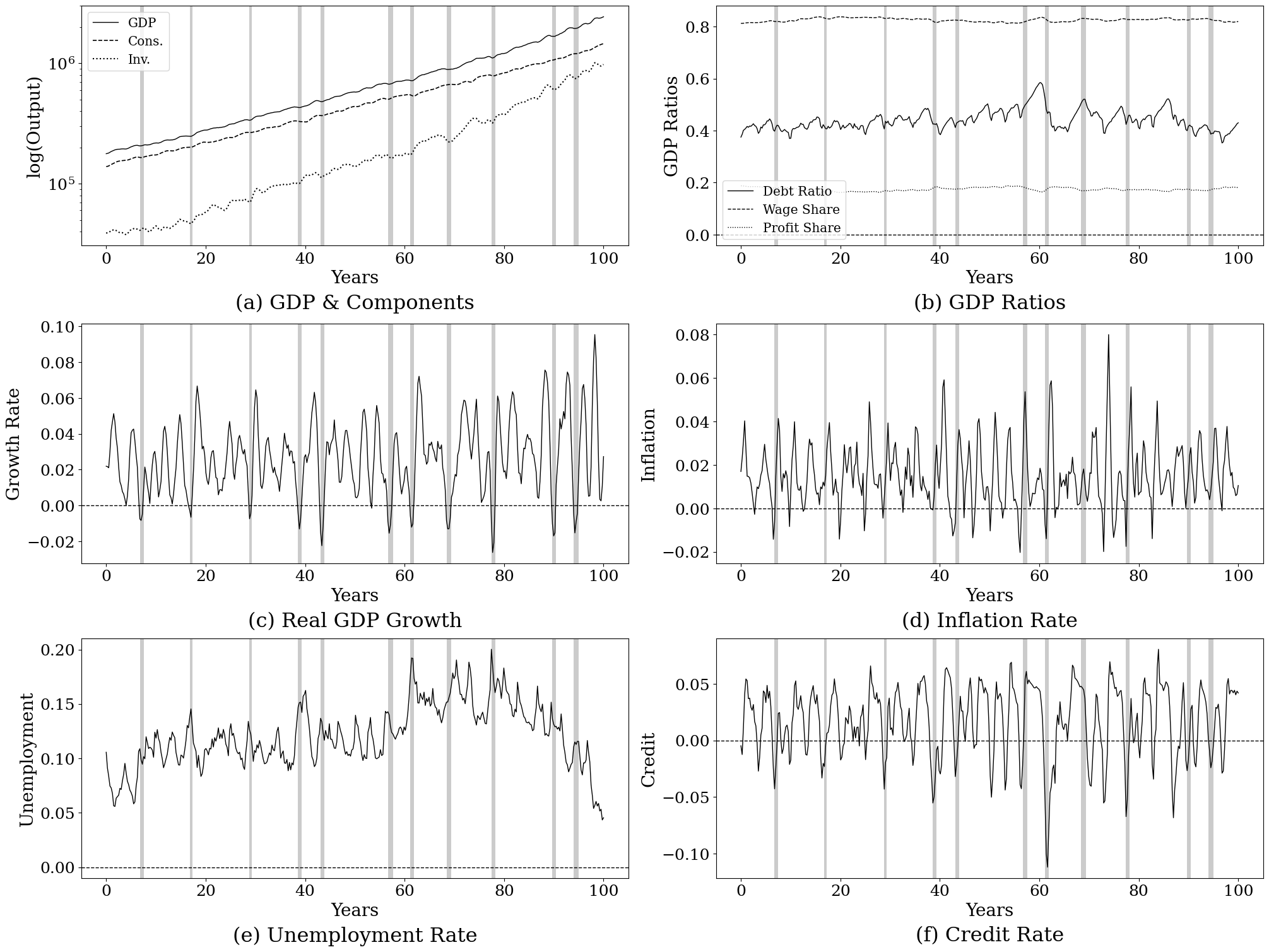}
    \caption{Macroeconomic time-series for a randomly-chosen run of the baseline Growth S1 scenario ($s=5$). The shaded areas highlight economic recessions (two or more quarters of negative real GDP growth). Panel (a) shows real GDP (solid), consumption (dashed), and investment (dotted). Panel (b) shows the debt to GDP ratio (solid), wages share of GDP (dashed), and profit share of GDP (dotted). Panel (c) shows the growth rate of real GDP. Panel (d) shows the CPI inflation rate. Panel (e) shows the unemployment rate. Finally, panel (f) shows the credit rate, defined as the annual change of debt as a percentage of GDP.}
    \label{figure_2}
\end{figure}

First, we show in Fig.~\ref{figure_2} time-series for key macroeconomic variables for a randomly chosen simulation.\footnote{We observed the other runs and always saw similar behaviour.} Several realistic stylised facts can be observed, for example: (i) recessions mostly occur during periods in which the debt to GDP ratio is falling and the credit rate is negative, suggestive of a Minskyan debt dynamic; (ii) recessions usually coincide with a period of deflation, implying a debt deflation dynamic; (iii) the unemployment rate typically spikes during a recession.

\begin{table}[!htb]
    \centering
    \scriptsize
        \begin{tabular}{lll}
        \toprule
        \textbf{Stylised facts} & \textbf{Empirical Evidence} & \textbf{Model evidence}\\
        \midrule
        \textbf{Macroeconomic} & & \\
        1. Minskyan debt cycles. & \cite{stockhammer2023debt_cycles} & \ref{app:stylised_fact_minskyan}\\
        2. Debt deflation. & \cite{fisher1933debt_deflation} & \ref{app:stylised_fact_debt_deflation}\\
        3. Fat-tailed real GDP growth rate distribution. & \cite{fagiolo2008output_distributions} & \ref{app:stylised_fact_gdp_growth}\\
        4. Recession duration distributed exponentially. & \cite{ausloos2004recession_durations} & \ref{app:stylised_fact_recessions}\\
        5. Autocorrelation of macro variables. & \cite{fiorito1994stylized} & \ref{app:stylised_fact_acf}\\
        6. Cross-correlation between macro variables. & \cite{stock1999business} & \ref{app:stylised_fact_ccf}\\
        7. Volatility hierarchy of macro growth rates. & \cite{stock1999business} & \ref{app:stylised_fact_volatility_hierarchy}\\
        8. Credit and unemployment relationship. & \cite{keen2014endogenous} & \ref{app:stylised_fact_relationships}\\
        9. Okun curve. & \cite{okun1962potential_gnp} & \ref{app:stylised_fact_relationships}\\
        10. Nominal wage-inflation Phillips curve. & \cite{phillips1958relation} & \ref{app:stylised_fact_relationships}\\
        11. Weak inflation Phillips curve. & \cite{stock2020slack} & \ref{app:stylised_fact_relationships}\\
        \textbf{Microeconomic} & & \\
        1. Fat-tailed distribution of firm growth rates. & \cite{bottazzi2003common, bottazzi2006firm_distributions} & \ref{app:stylised_fact_firm_growth}\\
        2. Right skewed and fat-tailed firm size distribution. & \cite{axtell2001zipf}, \cite{wit2005firm_size} & \ref{app:stylised_fact_firm_size} \\
        3. Right skewed and fat-tailed bank size distribution. & \cite{ennis2001bank_size} & \ref{app:stylised_fact_bank_size} \\
        4. Lumpiness of investment rates across firms. & \cite{doms1998capital} & \ref{app:stylised_fact_investment_lumpiness}\\
        5. Heterogeneous productivity across firms. & \cite{bartelsman2000understanding} & GBM Process\\
        6. Persistent productivity differences across firms. & \cite{bartelsman2000understanding} & GBM Process\\
        \bottomrule
        \end{tabular} 
    \caption{Stylised facts of real world economies replicated by the PG-DYNAMIN model for the baseline Growth S1 scenario.}
    \label{tab:stylised_facts}
\end{table}

Table~\ref{tab:stylised_facts} provides a list of stylised facts that the model has replicated. For details and evidence of these, see \ref{app:stylised_facts}. At the macroeconomic level, Minskyan debt cycles were present across the majority of simulations, whereby a build-up of debt precedes a recession \citep{minsky1977financial,minsky1986stabilizing,minsky1992financial,stockhammer2023debt_cycles}, see \ref{app:stylised_fact_minskyan}. Furthermore, a debt-deflation dynamic often occurred around recessions \citep{fisher1933debt_deflation}, see \ref{app:stylised_fact_debt_deflation}. Real GDP growth rates follow a fat-tailed distribution and were found to be statistically different to a normal distribution \citep{fagiolo2008output_distributions,castaldi2009patterns}, see \ref{app:stylised_fact_gdp_growth} and \ref{app:stat_tests}. The duration of recessions is best fit by an exponential distribution (see \ref{app:stylised_fact_recessions}) as observed empirically by \cite{ausloos2004recession_durations}. The autocorrelation (\ref{app:stylised_fact_acf}) and cross-correlation (\ref{app:stylised_fact_ccf}) for the cyclical components (decomposed using a Hodrick-Prescott filter \citep{hodrick1997postwar}) of real GDP, consumption, investment, debt, and unemployment closely match those of empirical US time-series \citep{stock1999business}. Additionally, the model replicates the volatility hierarchy of investment, real GDP, and consumption growth rates \citep{stock1999business}, see \ref{app:stylised_fact_volatility_hierarchy}. Finally, the model reproduces well-known relationships between the credit rate and unemployment \citep{keen2014endogenous}, Okun curve \citep{okun1962potential_gnp}, nominal wage-inflation Phillips curve \citep{phillips1958relation}, and a weak price-inflation Phillips curve \citep{stock2020slack}, see \ref{app:stylised_fact_relationships}. 

Microeconomic stylised facts included: the fat-tailed distribution of firm growth rates \citep{bottazzi2003common,bottazzi2006firm_distributions}, see \ref{app:stylised_fact_firm_growth}; right skewed and fat-tailed distribution of firm size \citep{gibrat1931inegalites,axtell2001zipf,wit2005firm_size} and bank size \citep{ennis2001bank_size} (see \ref{app:stylised_fact_firm_size} and \ref{app:stylised_fact_bank_size}, respectively); and the lumpiness of investment rates \citep{doms1998capital}, see \ref{app:stylised_fact_investment_lumpiness}. Additionally, the assumption of firm level productivity following a GBM implies that productivity is heterogeneous across firms, and that productivity differences show persistence \citep{bartelsman2000understanding}. 

%% file: scenario_analysis.tex
\section{Scenario Analysis} 
\label{sec:scenario_analysis}
In this section, an analysis of the stability and implications of all four growth and zero-growth scenarios is presented, with analyses involving both macroeconomic variables and dynamics of distributional microeconomic variables relating to firm and bank fortunes. 

\subsection{Leveraging emergence: stability of the macro}
\begin{figure}[!htb]
    \centering
    \includegraphics[width=0.9\textwidth]{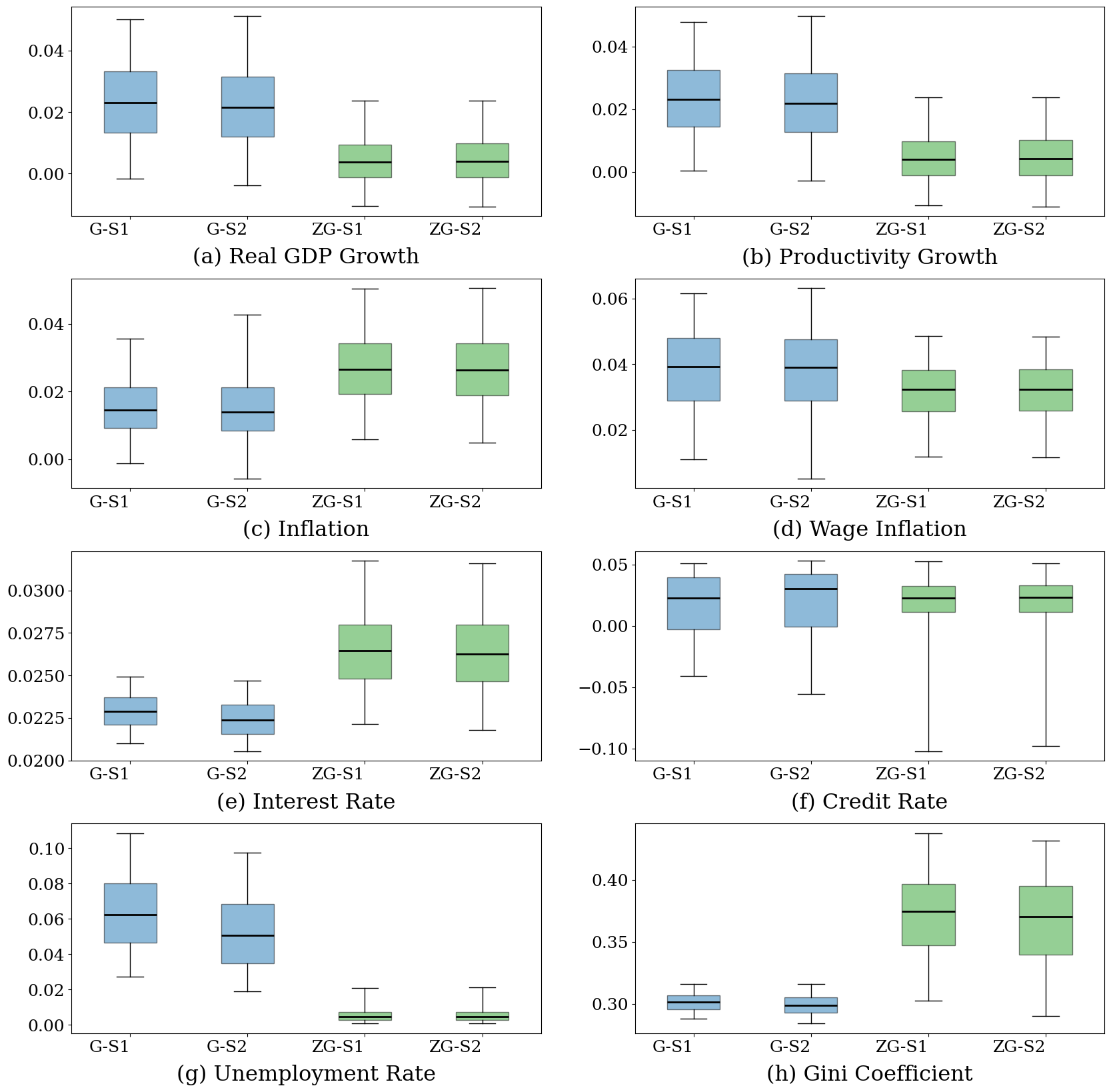}
    \caption{Box plots of key macroeconomic variables across all simulations. The horizontal line shows the median value, the box gives the inter-quartile range, and the whiskers indicate the middle 90\% inter-percentile range. The blue values correspond to the Growth S1 (G-S1) and Growth S2 (G-S2) scenarios, and the green values correspond to the Zero-Growth S1 (ZG-S1) and Zero-Growth S2 (ZG-S2) scenarios.}
    \label{figure_3}
\end{figure}

The results for the primary macroeconomic variables of the model for each scenario over all time periods and simulations is presented in Fig.~\ref{figure_3} in the form of box plots. It can first be noted that the median values for real GDP growth (log difference of real GDP) are close to their given parameter values. This suggests that the discrete-time GBM used to model firm labour productivity (Eq.~\ref{eq:productivity}) was successful in controlling the aggregate growth rate of the simulated economy. The volatility of real GDP growth, as measured by the interquartile range (IQR), is greater in both growth scenarios when compared to both zero-growth scenarios, with little impact from varying debt dynamics. Additionally, productivity growth (log difference of aggregate productivity) displays similar results, with growth scenarios having a higher IQR than zero-growth scenarios. There is more inflationary pressure in zero-growth scenarios, as evidenced by higher consumer price index (CPI) inflation rates (the CPI is defined using a Paasche price index of C-firm prices weighted by output, see \ref{app:cpi_inflation}). Median wage inflation is higher in both growth scenarios, with a higher IQR, suggesting that growth scenarios produce higher and more volatile wage inflation. The median nominal interest rate is higher in both zero-growth scenarios, likely due to the increase of inflation in the zero-growth scenarios, since banks are modelled as having a target interest rate that involves a mark-up from inflation. The median credit rate (change in debt as a percentage of nominal GDP) is similar between scenarios, with the growth S2 scenario having the highest value. The IQR of credit rate is larger in the growth scenarios, however, the lower tail of the credit rate distribution is larger in zero-growth scenarios. The zero-growth scenarios indicate better employment outcomes for households, with lower and more stable unemployment rates. Lastly, the inequality between households (workers) as measured by the Gini Coefficient (see \ref{app:gini}) is higher in the zero-growth scenarios.

While real GDP growth rates are overall less volatile in terms of IQR in zero-growth scenarios, it is important to also consider frequency and severity of crises in the different scenarios. We define a crisis period, following \cite{dosi2013income}, as a period in which real GDP growth is below $-3\%$.\footnote{An analysis of recessions, defined as two quarters of negative real GDP growth, is not suitable here, because this would be biased towards perceiving growth scenarios as more stable - the random fluctuations in real GDP growth around the average value for $g$ mean that the zero-growth scenarios would have more recessions even if they overall exhibit less volatility in real GDP.} Note that in the macroeconomic analysis of zero-growth scenarios in \cite{barrett2018stability}, a crisis was defined as a breakdown of the model, specifically by aggregate macro level investment going negative. PG-DYNAMIN (typical for ABMs) rarely produces complete breakdowns, and so the \cite{dosi2013income} definition of a crisis is more appropriate here. Using this crisis indicator, we are able to construct a probability of crisis (probability of at least one crisis occurring in a given year) and a crisis severity measure (cumulative percentage of GDP loss below the crisis threshold during a crisis period), see \ref{app:crisis_measures} for definitions. From Table~\ref{tab:crises} it is evident that zero-growth scenarios have a higher crisis probability. Additionally, increasing firm debt volatility increases the average crisis probability in growth scenarios, however, this decreases average crisis probability in zero-growth scenarios. Moreover, the severity of crises is higher in zero-growth scenarios. Increasing firm debt volatility worsens crisis severity in growth scenarios, and reduces crisis severity in zero-growth scenarios. In summary, although volatility (IQR) of real GDP growth rates is lower in zero-growth scenarios, the probability and severity of crises is higher.

\begin{table}[!htb]
    \centering
    \begin{tabular}{lcccc}
        \toprule
        & \multicolumn{2}{c}{\textbf{Crisis Probability}} 
        & \multicolumn{2}{c}{\textbf{Crisis Severity}} \\
        \cmidrule(lr){2-3} \cmidrule(lr){4-5}
        \textbf{Scenario} & \small Average & \small Std. Dev. 
        & \small Average & \small Std. Dev. \\
        \midrule
        Growth S1       & 0.15\%  & 0.41\%  & 0.69\%  & 1.00\% \\
        Growth S2       & 0.46\%  & 0.68\%  & 1.08\%  & 1.09\% \\
        Zero-Growth S1  & 0.74\%  & 1.19\%  & 2.75\%  & 3.01\% \\
        Zero-Growth S2  & 0.66\%  & 1.03\%  & 2.44\%  & 2.88\% \\
        \bottomrule
    \end{tabular}
    \caption{Crisis probability and severity, displaying averages and standard deviations (std. dev.) across simulations for each scenario.}
    \label{tab:crises}
\end{table}

\begin{figure}[!htb]
    \centering
    \includegraphics[width=0.9\textwidth]{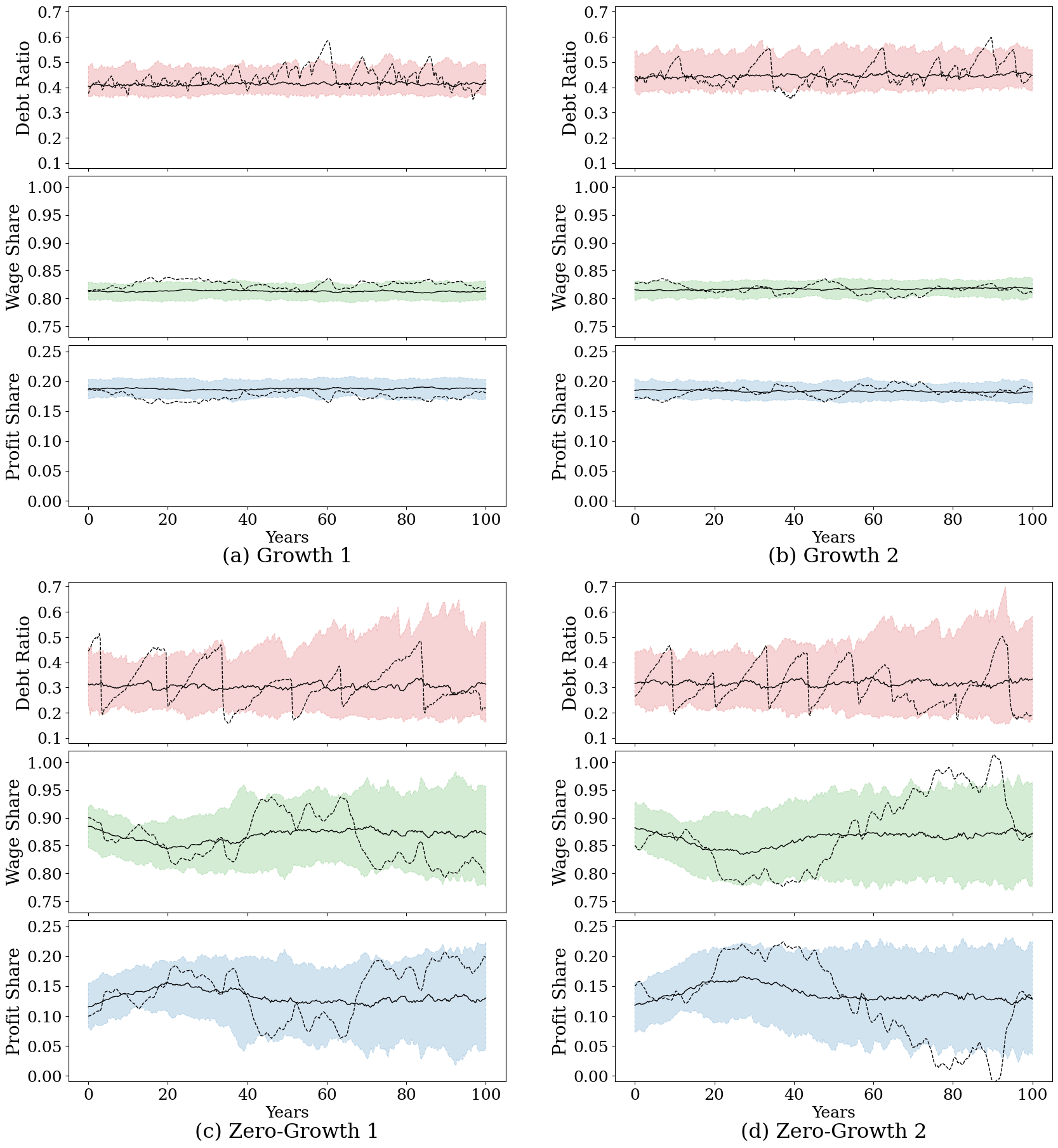}
    \caption{Time-series of nominal GDP ratios for each scenario. Panels respectively show debt ratio (red), wage share (green), and profit share (blue). Solid black lines are the median for each variable over simulations, dashed black lines show a single randomly selected simulation ($s=5$), and the shaded area is the middle 90\% inter-percentile range (IPR) across simulations.}
    \label{figure_4}
\end{figure}

Finally, we present an analysis of the debt to nominal GDP ratio, wages share of nominal GDP, and profit share of nominal GDP. Fig.~\ref{figure_4} shows that the median debt ratio is higher in both growth scenarios and increasing firm debt volatility seems to increase the debt ratio more in the growth scenarios when compared to the zero-growth scenarios. However, the volatility of the debt ratio is significantly increased in the zero-growth scenarios, suggesting that growth dynamics are an important determination of firms' financial dynamics in aggregate. It is further evident from Fig.~\ref{figure_4} that the median wages share is higher in both zero-growth scenarios, concurrent with \cite{barrett2018stability}. Thus, on average, workers receive a higher share of nominal GDP in zero-growth scenarios. This implies that, on average, firms/capitalists receive a smaller profit share of nominal GDP in zero-growth scenarios (also plotted in Fig.~\ref{figure_4}). However, there is again more volatility in both the wages share and profit share of GDP in zero-growth scenarios. Additionally, there seems to be little impact on either the wages share or profit share for differing firm debt volatility dynamics between scenarios S1 and S2 in both growth and zero-growth scenarios.

\subsection{Leveraging the micro: stability \& distribution}
This section analyses microeconomic data from the model, employing distributional variables relating to the stability and fortune of firms and banks in the different scenarios. We utilise the Hymer-Pashigian Instability Index (HPI) to measure the instability of market shares (see \ref{app:HPI}), and the normalised Herfindahl-Hirschman Index (HHI*) to measure market concentration (see \ref{app:HHI}). Furthermore, we compute the probability of default during both normal times and crises to compare firm and bank defaults between scenarios, as well as comparing ages between large and small firms and banks. Finally, using micro-data from the credit network, we employ a measure of the expected systemic loss of assets as a percentage of nominal GDP (see \ref{app:expected_systemic_loss}), in order capture the systemic risk present in the credit network.

\begin{figure}[!htb]
    \centering
    \includegraphics[width=0.9\textwidth]{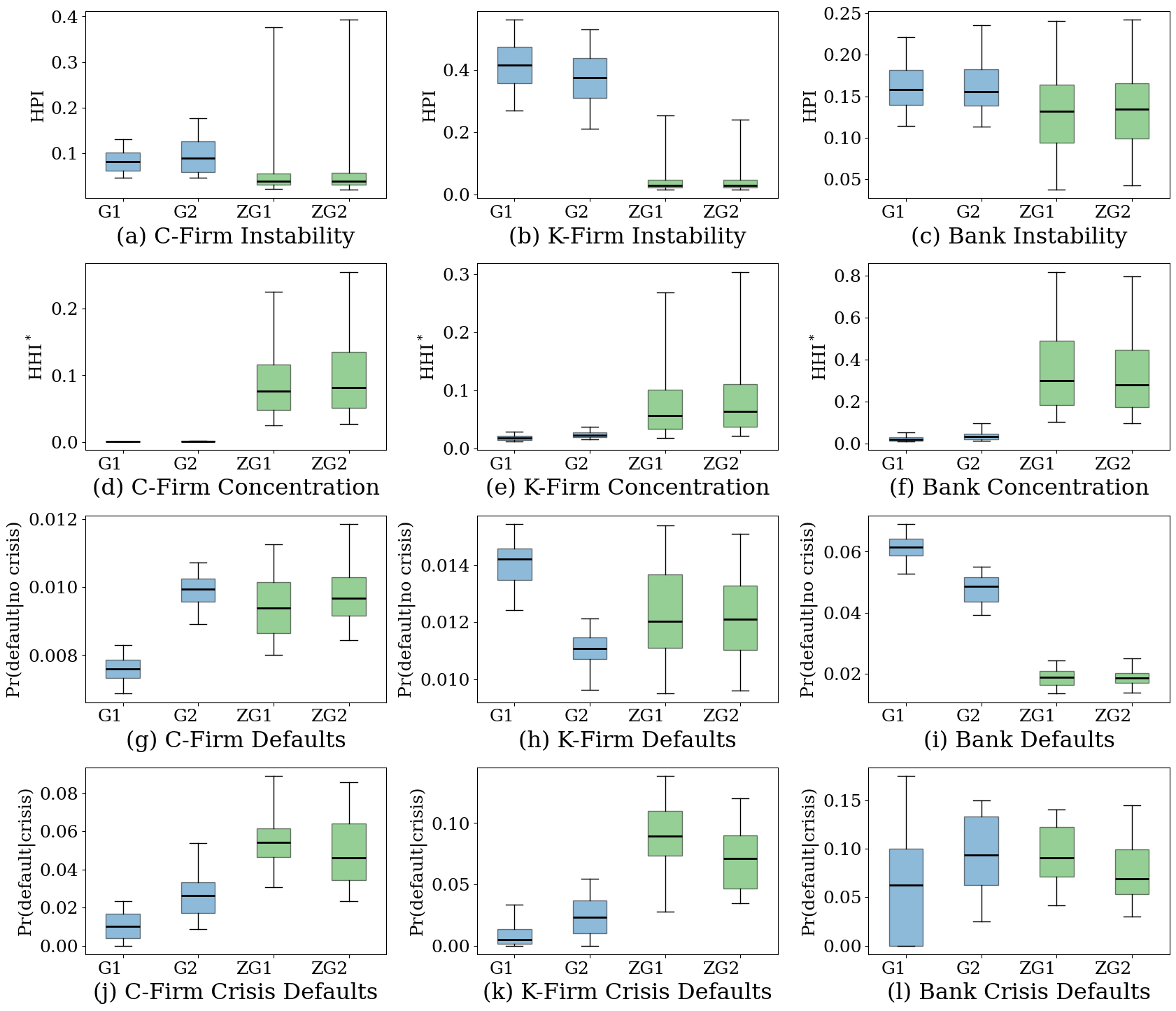}
    \caption{Box plots of key distributional variables across all simulations. The horizontal line shows the median value, the box gives the inter-quartile range, and the whiskers show the farthest value within the middle 90\% inter-percentile range. The blue values correspond to the Growth S1 (G1) and Growth S2 (G2) scenarios, and the green values correspond to the Zero-Growth S1 (ZG1) and Zero-Growth S2 (ZG2) scenarios. HPI is the Hymer-Pashigian Instability index, HHI* is the normalised Herfindahl-Hirschman Index, $\Pr(\text{default}|\text{no crisis})$ is the probability of default per annum during normal times, and $\Pr(\text{default}|\text{crisis})$ is the probability of a default per annum during a crisis.}
    \label{figure_5}
\end{figure}

Fig.~\ref{figure_5} shows box plots for market instability (HPI), market concentration (HHI*), and default probabilities over all simulations and time periods for each scenario. Markets are on average less stable (higher HPI) in both growth scenarios when compared to the zero-growth scenarios, most notably in the K-firm sector. However, market concentration (HHI*) is higher for both zero-growth scenarios in each market, suggestive of a few firms/banks dominating the market. For C-firms, the median probability of default during normal times is lowest in the growth S1 scenario and highest in the growth S2 scenario, with both zero-growth scenarios between these values, indicating that C-firm debt dynamics have a larger impact on the probability of default during normal times for growth scenarios. However, for both K-firms and banks, the probability of default during normal times is highest in the growth S1 scenario. For K-firms the lowest value is in the growth S2 scenario, likely due to the increased demand for investment goods from C-firms in the growth S2 scenario, and the median values for zero-growth scenarios are between both growth scenarios. Banks have a significantly reduced probability of default during normal times under zero-growth, with similar values for both zero-growth scenarios. Next, we explore the probability of default for C-firms, K-firms, and banks during a crisis period ($\Pr(\text{default}|\text{crisis})$). It is evident for C-firms and K-firms that the probability of default during a crisis increases more and is higher in both zero-growth scenarios, which is in line with the increased severity of crises in the zero-growth scenarios from the macroeconomic analysis (Table~\ref{tab:crises}). Banks' probability of default during a crisis also increases more from the normal times level in zero-growth scenarios. However, compared to C-firms and K-firms, banks' probability of default during a crisis is relatively similar between growth and zero-growth scenarios.

\begin{figure}[!htb]
    \centering
    \includegraphics[width=0.9\textwidth]{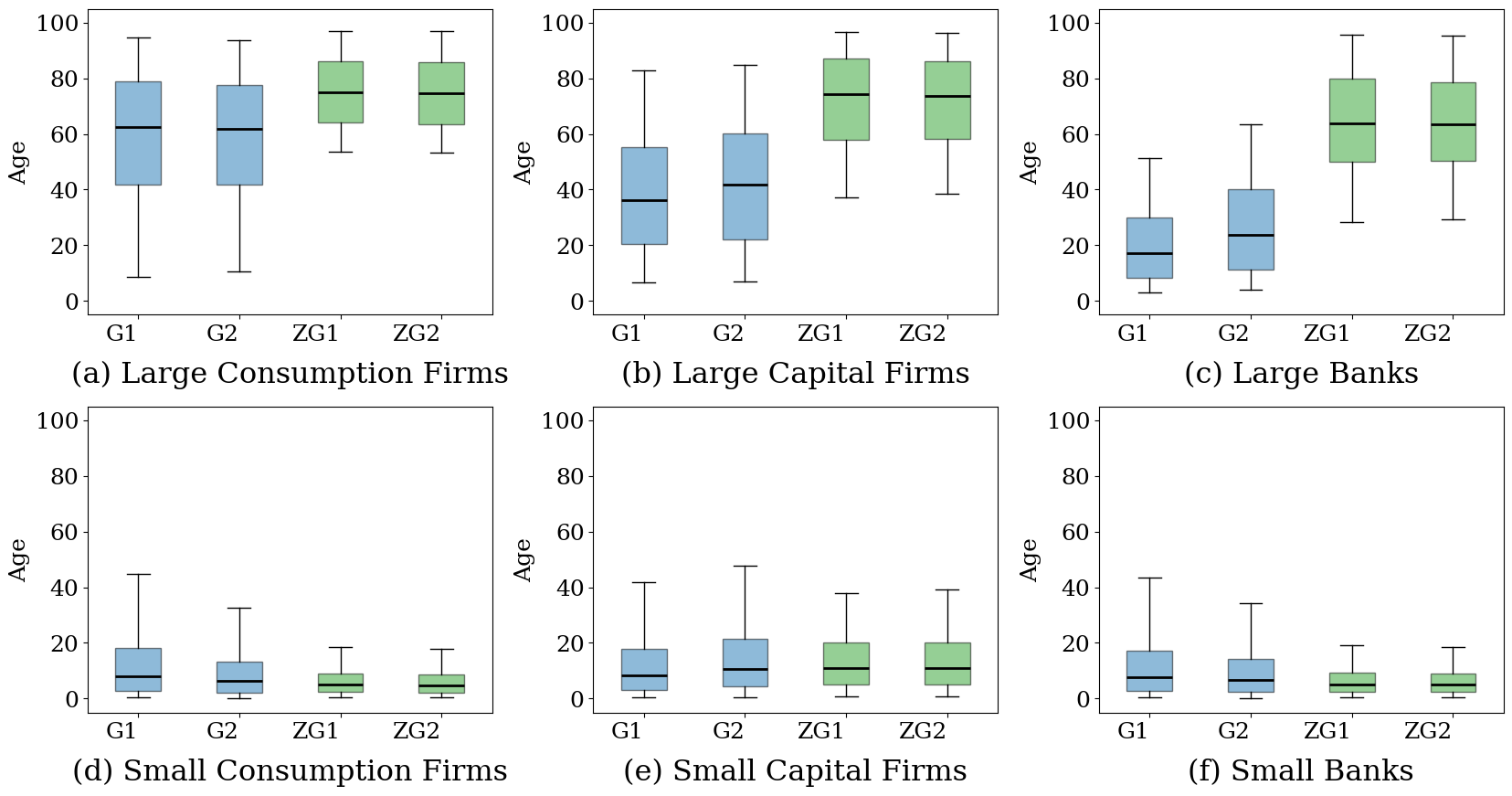}
    \caption{Box plots of C-firm, K-firm and banks' ages (years) by size. Large firms and banks are those in the top 1\% of market shares and small firms and banks are those in the bottom 50\% of market shares.}
    \label{figure_6}
\end{figure}

To further investigate the structure and demographics of firms and banks beyond default probabilities, we explore the distribution of firm and bank ages\footnote{Bank ages are counted from the time period in which the bank was previously bailed-in.} across scenarios. From Fig.~\ref{figure_6}, it is evident that large firms and banks (top 1\% of market share) in the zero-growth scenarios survive for longer on average compared to those in the growth scenarios; whereas small firms and banks (bottom 50\% of market share) have similar or slightly lower median ages in zero-growth scenarios. Additionally, there seems to be little impact for differing C-firms debt dynamics on firm and bank ages. Therefore, this confirms the results from the market concentration evidence (HHI*) in Fig.~\ref{figure_5}, suggesting that large, and also old, firms and banks tend to dominate markets in zero-growth scenarios. Moreover, this helps to explain the increased market instability in growth scenarios, because large firms do not live for as long throughout the simulations.

\begin{figure}[!htb]
    \centering
    \includegraphics[width=0.9\textwidth]{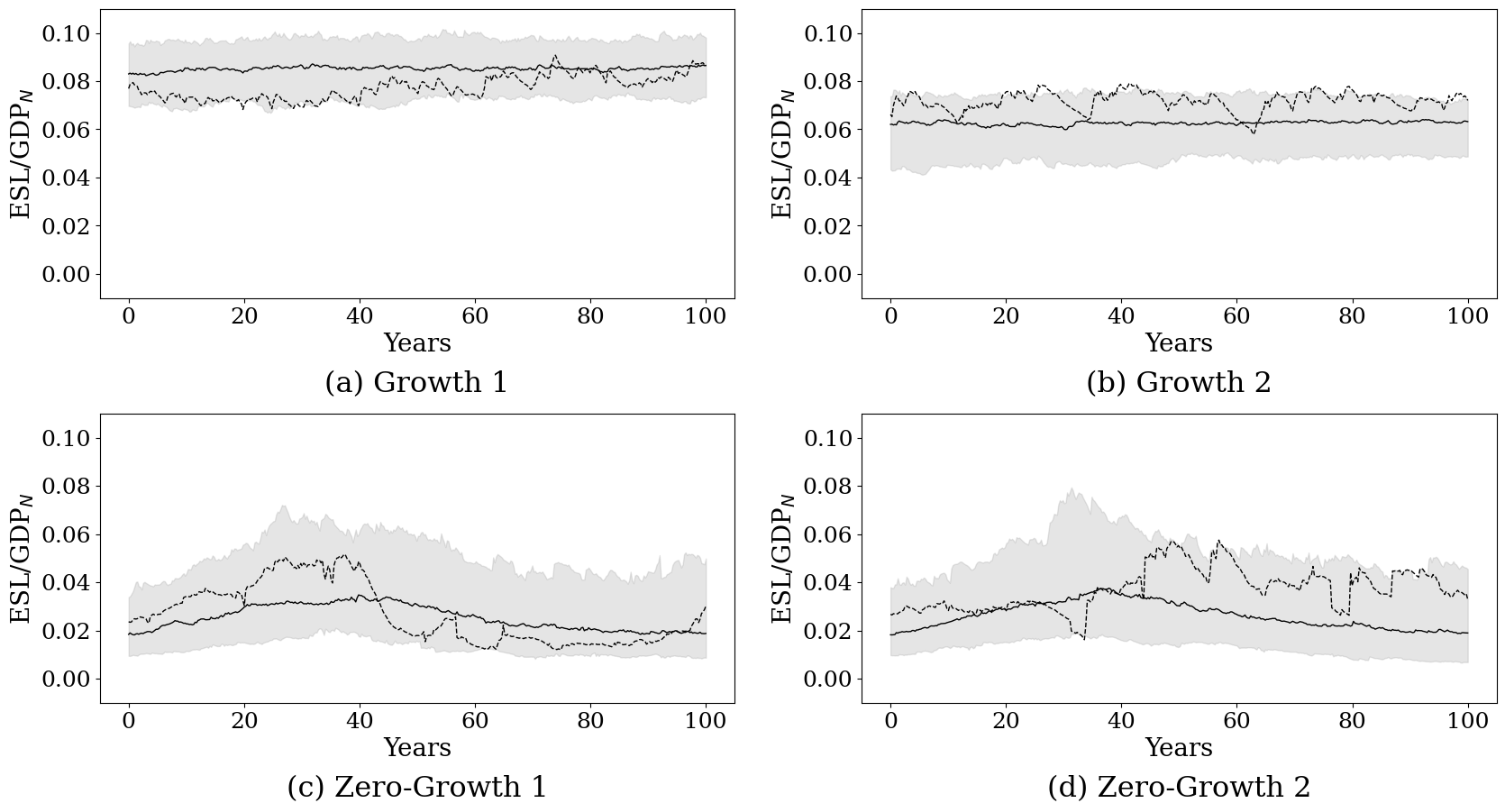}
    \caption{Time-series of expected systemic loss as a proportion of nominal gross domestic product (ESL/GDP$_N$) for each scenario. Solid lines are the median over simulations, dashed lines are for a randomly selected simulation ($s=5$), and the shaded area is the middle 90\% inter-percentile range (IDR) over simulations.}
    \label{figure_7}
\end{figure}

To capture the level of systemic risk in the bipartite credit network between banks and firms, we computed the approximate Expected Systemic Loss (ESL), using methodology from \cite{poledna2015systemic_risk}. This involves calculating, for each bank, the DebtRank index, following \cite{battiston2012debtrank} and \cite{aoyama2013debtrank}.\footnote{DebtRank captures the amount of economic value that could potentially be affected by a node in a credit network.} See \ref{app:expected_systemic_loss} for detailed definitions of these measures. Fig.~\ref{figure_7} shows time series of ESL divided by GDP$_N$. It can be noted that the systemic risk is relatively similar between high and low debt behaviour in the growth S1 and S2 scenarios, with slightly lower systemic risk in the growth S2 scenario. Additionally, zero-growth S1 and S2 scenarios have relatively similar systemic risk. Comparing growth versus zero-growth scenarios, it is evident that the median ESL/GDP$_N$ is higher in both growth scenarios. Thus, even though crises in zero-growth scenarios are more common and severe when measured through loss of GDP (see Table~\ref{tab:crises}), the ESL/GDP$_N$ informs us that the expected loss of the total stock of assets as a percentage of GDP is higher in the growth scenarios. Hence, the risk associated with the credit network is decreased in zero-growth scenarios. 

%% file: discussion.tex
\section{Discussion}
\label{sec:discussion}
This paper has introduced the Post-Growth DYNamic Agent-based MINskyan (PG-DYNAMIN) model, which has enabled, for the first time, a comparison of both macroeconomic and microeconomic dynamics on growth versus zero-growth scenarios. This work is timely, given recent reviews that pointed out significant uncertainty about the extent to which a post-growth economics is compatible with the current capitalist system \citep{edwards2025towards}, and that there has been a lack of post-growth studies employing ABMs \citep{lauer2025comparative}. This work continues the methodological approach applied to a purely macroeconomic model in \cite{barrett2018stability}, of comparing the relative stability between positive and zero mean growth scenarios on the same model. This approach is motivated by the fact that growth is never constant in any realistic economic scenario. Importantly, following \cite{minsky1986stabilizing, minsky1992financial}, this work considers interest-bearing private debt as a potential source of instability in both growing and non-growing economies. It was shown that the model reproduces well-known stylised facts from the real world, supporting its use when applied to a scenario analysis. The model was run for four scenarios: two growth scenarios with low and high debt volatility dynamics, and two zero-growth scenarios with low and high debt volatility dynamics. Results were then compared across scenarios to determine the relative stability of long-run zero-growth, and whether this is a viable possibility for a capitalist economy with interest-bearing debt. Whether a scenario exhibited growth or not was determined by an exogenous productivity growth parameter that was set to either $2\%$ or zero. Thus the zero-growth scenarios could be considered as either arising from policy or from natural causes, or from a combination of these. Overall, the scenario analysis found that zero-growth scenarios were just as economically viable as growth scenarios. 

The macroeconomic results from PG-DYNAMIN both support and diverge with those from the previous purely macroeconomic models that have considered the effects of degrowth, post-growth, and/or zero-growth. Similar to the results found in \cite{barrett2018stability} -- zero-growth scenarios have a lower debt share of GDP, higher wage share, and lower profit share than growth scenarios. Most of the post-growth literature is in agreement that post-growth scenarios should exhibit an increased labour share and a decreased profit share or profit rate \citep{cahen2016ecological, alessandro2020green_growth, jackson2020transition, kallis2025postgrowth}. It is noteworthy that there are now several explicit modelling analyses that contradict the theoretical long-run equilibrium analysis in \cite{piketty2014capital} that suggested there must be a trade-off between having a more sustainable lower growth rate and reducing inequality between workers and capitalists. At the microeconomic level, however, results do suggest greater inequality between different firms and households (workers) - we discuss this below. The lower and more stable unemployment rates concur with analyses in \cite{barrett2018stability}, \cite{alessandro2020green_growth} and \cite{kallis2025postgrowth}, but diverge from \cite{rosenbaum2015zerogrowth}. In terms of stability of unemployment, both \cite{barrett2018stability} and \cite{jackson2020transition} found this to be more volatile under zero-growth, which is consistent with the findings here. The lower and increased stability of unemployment rates in PG-DYNAMIN is likely due to the increased market stability and concentration under zero-growth, which reduces firms' cyclical labour adjustments. In contrast to \cite{barrett2018stability}, it was found that increasing consumption firms' desired debt ratio for investment did not significantly increase the instability of the model across the majority of indicators measured.

Unlike the model in \cite{barrett2018stability}, PG-DYNAMIN does not produce crises from which the economy does not recover. Thus, a distinct definition of a crisis was required here. The definition from \cite{dosi2013income} was taken, whereby a crisis is defined as a period in which real GDP growth is less than $-3\%$. With this, it was found that zero-growth scenarios had more economic crises than growth scenarios, and with greater severity, which is opposite to what was seen in the simulations in \cite{barrett2018stability}. Most of the time, however, real GDP growth showed less variance in zero-growth scenarios, as evidenced by a smaller inter-quartile range of real GDP growth. It was also found that economic crises in growth scenarios were more sensitive to firms' financial conditions than those under zero-growth scenarios. Increasing firm debt volatility behaviour tended to raise the probability and severity of crises on average in growth scenarios, whereas in zero-growth scenarios it slightly reduced both.

A key departure from \cite{barrett2018stability} was the inclusion of endogenously determined prices set by firms. This enabled a consideration of inflation, which has often been lacking in the post-growth literature \citep{lauer2025comparative}. A higher inflation rate was observed in the zero-growth scenarios than in the growth scenarios. To understand this result, one can consider the argument of \cite{soddy1926wealth}, that to be able to pay off interest on debt there is an imperative to either: (1) grow the physical goods within an economy; or (2) increase the price of goods in an economy. Another explanation comes from the fact that, when the wages share of output is approximately constant, inflation will be approximated by growth in wages minus growth in productivity.\footnote{Derived from the fact that taking the logarithmic time derivative of the wage share ($\omega=w/Pa$, where $w$ is wages, $P$ is a price index, and $a$ is labour productivity), then setting to zero (constant wage share) yields: $g_P=g_w-g_a$. Hence, inflation ($g_P$) is given by the difference between nominal wage growth rates ($g_w$) and productivity growth rates ($g_a$)} Given zero-growth scenarios have higher wage growth rates (see Fig.~\ref{figure_3}), as well as lower productivity growth rates by design, higher inflation arises via this accounting identity. (It is verified in Table~\ref{tab:inflation_stats} in \ref{app:discussion_results} that this inflation identity produces very similar expected inflation rates for each scenario when compared to the actual inflation rate.) Note that this identity is satisfied without wage-rise offset being explicitly modelled in the equations for price-setting. It has been evidenced that inflation necessarily redistributes income \citep{nitzan2009casp}, and typically towards capital owners. This is somewhat misaligned with our findings, where zero-growth scenarios have higher inflation but an increased workers' share. This may be due to model simplifications, such as exclusion of a complex industrial base, capitalisation, and a stock market. The other distribution to consider is that between small and large firms, and on that our results concur with \cite{nitzan2009casp}, that large firms are better off in an inflationary environment, at the expense of small firms. This also concords with Post-Keynesian reasoning on the recent causes of inflation \citep{weber2023sellers_inflation}. The increased risk of inflation under zero-growth suggests that active government fiscal policy might be necessary to counteract the redistributive forces of inflation.

An important advantage of employing an ABM is the ability to analyse the micro level in addition to the macro and obtain more nuance in conclusions. For example, while zero-growth scenarios have more economic crises, market instability and expected systemic loss are generally lower. We now summarise and discuss the microeconomic findings. Increased market stability under zero-growth is likely a result of increased market concentration, such that large firms tend to dominate markets. Default probabilities are generally lower during normal times for zero-growth scenarios, however, during periods of crisis default probabilities increase more and are higher for both C-firms and K-firms. Additionally, because zero-growth poses a higher risk of large firm dominance, the government may have to uphold competition policy and support for small firms during a crisis to ensure that small firms can survive and remain competitive. The macroeconomic results show a decrease in inequality between the wages and profit share of GDP under zero-growth, whereas, the microeconomic results show increased inequality of deposits between households (workers), evidenced by a higher Gini Index. This is likely caused by the reduced and more stable unemployment rates, allowing households to accumulate more deposits from their work and interest payments on deposits. Moreover, a novel contribution made here is the application of expected systemic loss (ESL) as a percentage of nominal GDP (ESL/GDP$_N$) to measure the systemic risk in the credit network across scenarios - methods from finance and network theory \citep{battiston2012debtrank, aoyama2013debtrank, poledna2015systemic_risk} were employed. It was found that systemic risk in the credit network between banks and firms (ESL/GDP$_N$) was generally lower in the zero-growth scenarios, suggesting that zero-growth may reduce the amount of assets lost from a default cascade throughout the credit network. This is in contradiction to the macroeconomic results, where real GDP growth rates where more likely to be in crisis in zero-growth scenarios, which captures the flow rate risk. However, the ESL/GDP$_N$ measure highlights the potential loss of the stock of assets, thus capturing financial and real loss outside of real GDP measures.

PG-DYNAMIN is closed by design, and there is still an open question of how a single country under zero-growth will compare when embedded into an international context with global trade \citep{barrett2018stability}. Additionally, a country's place on the currency hierarchy \citep{cohen2018geography,herr2022currency,orsi2025currency_hierarchy} will likely affect the ability with which it can diverge to a degrowth or post-growth regime as an international outlier. Reduced profit rates may prompt capital flight from such a region, in which case, the government's ability to respond fiscally will likely impact the possibility of sustaining a post-growth scenario. Addressing the question of whether post-growth is conditioned on an economy being relatively closed and local rather than open and international is an important topic for future research. 

We have looked at the effects of post-growth on a purely capitalist system using a relatively simplified ABM when compared to the real world and wider literature. Particularly, our analysis has focused on the impact that post-growth will have on the financial stability of a purely capitalist system. The addition of a government sector using countercyclical fiscal policies and a central bank that can leverage monetary policy would greatly increase the scope of analysis for post-growth policies, which we leave to future research. Moreover, the inclusion of the environment, with economy-ecology feedback loops, such as those proposed in integrated assessment modelling (IAM) and ecological macroeconomics (EM) literature is omitted in this model. There has been limited research linking ABMs to IAMs \citep{lamperti2018faraway, lamperti2020climate, lamperti2021three, reissl2025dsk}, specifically, the effects of zero-growth and post-growth policies on the financial and ecological stability of such models at the disaggregated level. Hence, future research will endeavour to extend our capitalist Minskyan ABM with a government sector and embed the economy in the ecology, with a finite flow of solar energy and limited non-renewable resources. Moreover, as discussed above, the question as to whether post-growth is feasible in an open economy embedded in an international trade system predicated on growth remains ambiguous; further work in line with \cite{dosi2019endogenous} and \cite{dosi2021public}, applying multi-country ABMs or multi-country EM models, to the analysis of post-growth, is needed \citep{edwards2025towards}. Additionally, one might incorporate a more realistic treatment of workers' hierarchical position within firms, as modelled e.g.~by \cite{simon1957compensation} and \cite{lydall1959distribution}, to concur with empirical data on power-law-distributed incomes that scale with firm size \citep{fix2018hierarchy}, as implemented into ABM by \cite{ciarli2010effect} and \cite{botte2021transition}. We postulate that post-growth will increase income and wealth inequality in our model due to higher firm concentration leading to increased hierarchical size and CEO pay. The addition of endogenous technological change as in \cite{dosi2010schumpeter} and \cite{botte2021transition} is also worth consideration. One might then compare economic outcomes when diverse mechanisms of innovation are tweaked or encounter ecological limits \citep{lamperti2018faraway}. Furthermore, recent work has begun to tie ABMs more closely to data \citep{delli_gatti2020rising, chen2022regression, poledna2023economic_forecasting, hommes2025canvas}, utilising machine learning and regression-based techniques to calibrate the free parameters of a model, and agent-level real-world data to initialise models from starting points that correspond to actual economies.

Although we have argued that continuous economic growth is likely unsustainable given the fundamental nature of the economy's embedding in the biosphere, we nevertheless have presented a model that is devoid of such constraints for the analytical purpose to compare the stability of a capitalist economy under growth and zero-growth scenarios. The general tendency for the slowdown of growth rates in developed capitalist countries \citep{malmaeus2017potential, kallis2025postgrowth} gives motivation for this research, even without consideration of environmental limits and climate change. Hence, findings and conclusions of this paper remain of interest whether approached from the consideration of ecological limits or from the necessity to understand how zero-growth will impact capitalism due to natural slowdowns of growth rates. 

%% file: conclusion.tex
\section{Concluding Remarks} 
\label{sec:conclusion}
A comparison of growth and zero-growth economic scenarios has been analysed for the first time on a multi-agent macroeconomic ABM. This paper has shown that a viable zero productivity growth scenario, with interest-bearing debt, was possible on the model. Moreover, it has been shown that the model, given a 2\% labour productivity growth rate, captures the cyclical dynamics present in real-world data. It was found that real GDP growth rates were more stable in the zero productivity growth scenarios. However, there were more economic crises, signifying increased tail risk under zero-growth scenarios. Some of the benefits of zero-growth were lower and more stable unemployment rates, a higher wages share, suggesting better outcomes for workers, markets were relatively more stable, and there was lower systemic risk in the credit network. However, some of the consequences of zero-growth include a higher rate of inflation, supported by theoretical reasoning, an increase in market concentration, and increased default probabilities during a time of crisis. Overall, the findings suggest that the end of growth may increase the prevalence and severity of crises in a capitalist system. At the same time however, we found reduced systemic credit risk in terms of the total stock of assets. Furthermore, this does not imply a complete economic breakdown, but rather a reconfiguration of economic dynamics with distinct distributional and structural consequences. This paper demonstrates that the end of growth is not completely instability-inducing on the economy, and could provide a new path for humanity to abate the strain of anthropogenic perturbations on the Earth's biophysical limits.

%% file: acknowledgements.tex
\section*{Acknowledgements}
We are grateful to the University of Sussex, which funded Dylan Terry-Doyle under the Junior Research Associate scheme for the initial version of this research, and to the Engineering and Physical Sciences Research Council (EPSRC) Doctoral Landscape Award for funding Dylan Terry-Doyle's PhD in Modelling Post-Growth Economics under the supervision of Dr Adam Barrett to continue this research. Additional thanks to Giordano Mion, as Dylan Terry-Doyle's undergraduate dissertation advisor, for giving valuable feedback on the model, and to Michael Taylor for his comments and feedback on the paper. 

%% file: appendix.tex
\appendix

\section{PG-DYNAMIN Model}
\label{app:model}
In the following appendix, we present a full model description including the equations and algorithms used to simulate the PG-DYNAMIN model. If the reader is interested in the code used to simulate the model, see \url{https://github.com/DylanTerryDoyle/Py-DYNAMIN}.

\subsection{Households}
\label{app:households}
\subsubsection{Income \& Expenditure}
The $h$th household receives both a wage from their employer, firm $\iota$, if employed and interest from their bank on deposits as income:
\begin{equation}
\label{eq:household_income}
    Y_{h}(t) = \mathbbm{1}_{h}(t) w_{\iota}(t) + r_M M_{h}(t),
\end{equation}
where $w_{\iota}(t)$ denotes the wage rate of firm $\iota$ defined in their labour contract, $r_M$ is bank $b$'s interest rate on deposits (uniform across all banks), $M_{h}(t)$ are household $h$'s deposits at bank $b$, and $\mathbbm{1}_{h}(t)$ is a dummy variable determining the employment status of household $h$:
\begin{equation}
    \mathbbm{1}_{h}(t)=
    \begin{cases}
        1 & \text{if household $h$ is employed}\\
        0 & \text{if household $h$ is unemployed.}\\
    \end{cases}
\end{equation}
Household $h$'s desired consumption expenditure, $E^d_{h}(t)$, is assumed, for simplicity, to be given by a simple Keynesian budget constraint in which household $h$ desires to spend a proportion of their income and deposits each period:
\begin{equation}
\label{eq:household_expenditure}
    E^d_{h}(t) =c_Y Y_{h}(t) + c_M M_{h}(t),
\end{equation}
where $c_Y\in(0,1)$ is the marginal propensity to consume out of income and $c_M\in(0,1)$ is the marginal propensity to consume out of deposits, with $c_Y>c_M$.

\subsubsection{Accounting}
Household $h$ keeps any savings at bank $b$ as deposits which earn an interest as in Eq.~\ref{eq:household_income}. Household $h$'s deposits are then updated by the difference between their income and expenditure:
\begin{equation}
\label{eq:household_deposits}
M_{h}(t+1) = M_{h}(t) + Y_{h}(t) - E_{h}(t),
\end{equation}
where $Y_h(t)$ is household $h$'s income and $E_{h}(t)$ is household $h$'s actual expenditure on C-goods, defined in the market for consumption goods, see \ref{app:consumption_market}. 

\subsection{Consumption Firms}
\label{app:cfirms}
\subsubsection{Production}
The output of the $i$th C-firm, $Y_{i}(t)$, is assumed to be given by a Leontief production function with constant returns to scale for both labour, $N_{i}(t)$, and capital, $K_{i}(t)$:
\begin{equation}
\label{eq:cfirm_output}
    Y_{i}(t) = \min\left\{a_{i}(t) N_{i}(t), \frac{K_{i}(t)}{\nu}\right\},
\end{equation}
where $a_{i}(t)$ is firm $i$'s labour productivity and $\nu$ is a constant capital-to-output ratio. Labour productivity, which is the essential variable of growth in the model, determines the efficiency of firm $i$'s production process. It is assumed that labour productivity evolves according to the stochastic differential equation (SDE) for geometric Brownian motion (GBM):
\begin{equation}
\label{eq:cfirm_productivity_sde}
    \text{d}a_{i}(t)=a_{i}(t)(g \text{d}t + \sigma_a \text{d}W_{i}(t)),
\end{equation}
where $g$ is the constant average growth rate of labour productivity, $\sigma_a$ is the standard deviation of labour productivity, and $W_{i}(t)$ is a Brownian motion, orthogonal between firms. This equation for labour productivity was chosen because it captures exponential growth with random fluctuations from the deterministic growth path set by the $g$ parameter. This will allow the growth rate of the model to be set exogenously. Furthermore, the expected value of the SDE in Eq.~\ref{eq:cfirm_productivity_sde} is given by $\mathbbm{E}[a_{i}(t)]=a_{i}(0)\exp\{gt\}$, which corresponds to the deterministic model used in \cite{barrett2018stability}. The SDE for labour productivity can be simulated on a discrete time grid using the exact solution for GBM:
\begin{equation}
\label{eq:cfirm_productivity}
    a_{i}(t)=a_{i}(t-1)\exp\left\{g - \frac{1}{2} \sigma_a^2 + \sigma_a \varepsilon_{i}(t)\right\},
\end{equation}
where $\varepsilon_{i}(t) \sim \mathcal{N}(0,1)$ is a random variable drawn from a standard normal distribution. Hence, when $g>0$, labour productivity grows exponentially. Furthermore, due to the functional form in Eq.~\ref{eq:cfirm_productivity}, for a given time period $t$, firm labour productivity will be distributed according to a log-normal distribution, such that $\ln(a_{i}(t))\sim \mathcal{N}\left(\ln(a_{i}(0)) + (g-\sigma^2_a/2)t,\sigma^2_a t\right)$.

\subsubsection{Desired Output \& Inventories}
C-firm $i$'s desired output, $Y^d_{i}(t+1)$, is assumed to be equal to their expected demand, $Z^e_{i}(t+1)$:
\begin{equation}
\label{eq:cfirm_desired_output}
    Y^d_{i}(t+1) = Z^e_{i}(t+1),
\end{equation}
where expected demand is adaptively updated according to firm $i$'s actual demand:
\begin{equation}
\label{eq:cfirm_expected_demand}
    Z^e_{i}(t+1) = Z^e_{i}(t) + \gamma_Z (Z_{i}(t)-Z^e_{i}(t)),
\end{equation}
where $\gamma_Z\in(0,1)$ determines the speed of adjustment to actual demand, and $Z_{i}(t)$ is firm $i$'s actual demand, which is determined on the market for consumption goods, see \ref{app:consumption_market}. From Eq.~\ref{eq:cfirm_expected_demand}, it is evident after repeated substitution that the expected demand is equal to the sum of all past demands with geometrically decaying weights.

It is assumed that C-goods are perishable and do not last longer than one period. Therefore, at the beginning of period $t$, C-firms set their inventories equal to output $V_{i}(t)=Y_{i}(t)$. Then, households consume from C-firm $i$'s inventories until they are depleted or C-firm $i$ is left with involuntary inventories. Additionally, any leftover inventories are disposed of by firm $i$ at no extra cost. Thus, firm $i$'s level of inventories after household consumption is given by:
\begin{equation}
    V_{i}(t) = Y_{i}(t) - Q_{i}(t),
\end{equation}
where $Q_{i}(t)$ is the actual quantity of sold C-goods, determined on the consumption good market, see \ref{app:consumption_market}.

\subsubsection{Prices}
C-firm $i$ sets their price according to both internal and external factors. In particular, C-firm $i$ will stochastically increase (decrease) their price if involuntary inventories are zero (positive), as this signals there is excess demand (supply) for C-firm $i$'s goods. Furthermore, C-firm $i$ will also adjust their price towards the average price level of other C-firms to stay competitive. Hence, this pricing mechanism is similar to the CATS framework \citep{russo2007industrial, gaffeo2008emergent_macro, assenza2015macro_abm}, with the addition of prices tending to adjust towards the average price level as in \cite{kalecki1954dynamics}. Hence, the pricing mechanism of C-firm $i$ is given by:
\begin{equation}
\label{eq:prices}
    P_{i}(t)=
    \begin{cases}
        P_{i}(t-1)(1 + \sigma_P|\varepsilon_{i}(t)|) + \gamma_P(\bar{P}^C(t-1) - P_{i}(t-1)) & \text{if  $V_{i}(t-1) = 0$} \\
        P_{i}(t-1)(1 - \sigma_P|\varepsilon_{i}(t)|) + \gamma_P(\bar{P}^C(t-1) - P_{i}(t-1)) & \text{if $V_{i}(t-1) > 0$}\\
    \end{cases}
\end{equation}
where $\sigma_P$ is the standard deviation of C-firm $i$'s price growth, $\varepsilon_{i}(t) \sim \mathcal{N}(0,1)$, therefore $|\varepsilon_{i}(t)|$ is distributed according to a folded normal distribution, $\gamma_P\in(0,1)$ is a speed of adjustment parameter, $V_{i}(t-1)$ are C-firm $i$'s inventories in the previous period, a proxy for excess demand ($V_{i}(t-1)=0$) or supply ($V_{i}(t-1)>0$), and $\bar{P}^C(t-1)$ is the weighted average of consumption goods prices:
\begin{equation}
    \bar{P}^C(t) = \frac{\sum^{\mathbf{N}_C}_{k=1}P_{k}(t)Y_{k}(t)}{\sum^{\mathbf{N}_C}_{k=1}Y_{k}(t)}.
\end{equation}

\subsubsection{Investment \& capital}
C-firms finance new investment using both internal and external finance. C-firm $i$ determines external investment financing using their desired debt-to-output ratio for investment finance, given by:

\begin{equation}
\label{eq:cfirm_desired_debt_ratio}
    d^d_{i}(t+1) = d_0 + d_1 \alpha_{i}(t) + d_2 \pi_{i}(t)
\end{equation}
where $d_0,d_1,d_2>0$ are parameters, $\alpha_{i}(t) = \ln(a_{i}(t)) - \ln(a_{i}(t-1))$ is the log difference of C-firm $i$'s labour productivity, and $\pi_{i}(t) = \Pi_{i}(t) / (P_{i}(t) Y_{i}(t))$ is firm $i$'s profit share. This is the essential equation determining the level of debt in the model, which has been taken from \cite{barrett2018stability}. $d_0$ controls the level of the desired investment debt ratio, given zero labour productivity growth or profit share, and $d_1$ and $d_2$ determine the strength with which the desired debt ratio is influenced by C-firm $i$'s productivity growth and profit share, respectively. Therefore, when C-firm $i$ is successful, and so has higher $\alpha_{i}(t)$ and $\pi_{i}(t)$, this will lead to a higher desired debt-to-output ratio for investment purposes. Additionally, C-firm $i$'s nominal desired debt level for investment can be derived from Eq.~\ref{eq:cfirm_desired_debt_ratio} as: 
\begin{equation}
    D^d_{i}(t+1) = d^d_{i}(t+1) P_{i}(t) Y_{i}(t).
\end{equation}
Hence, C-firm $i$'s desired loan for investment financing is given by:
\begin{equation}
    IL^d_{i}(t+1) = \max\big\{D^d_{i}(t+1) - D_{i}(t), 0\big\},
\end{equation}
where $D_{i}(t)$ is C-firm $i$'s current debt. C-firm $i$'s desired investment expenditure is then given by:
\begin{equation}
\label{eq:desired_investment_expenditure}
    IE^d_{i}(t+1) = \max\big\{IL^d_{i}(t+1) + \Pi_{i}(t) + M_{i}(t) - \zeta W_{i}(t), 0\big\},
\end{equation}
where $M_{i}(t)$ is C-firm $i$'s deposits at bank $b$ and $W_{i}(t)$ is C-firm $i$'s current wage bill. C-firms are assumed to keep enough internal finance to cover a multiple, $\zeta$, of their future wage bill, which they assume will be equal to the current wage bill.

After C-firm $i$ has purchased their desired amount of K-goods on the capital market, C-firm $i$ updates their capital stock by:
\begin{equation}
\label{eq:capital}
    K_{i}(t+1) = K_{i}(t)(1 - \delta) + I_{i}(t+1),
\end{equation}
where $\delta$ is the depreciation rate of capital, and $I_{i}(t+1)$ is C-firm $i$'s new investment orders. C-firm $i$ also updates their total capital expenditure, which is similarly given by:
\begin{equation}
    KE_{i}(t+1) = KE_{i}(t)(1 - \delta) + IE_{i}(t+1),
\end{equation}
where $IE_{i}(t+1)$ is C-firm $i$'s actual investment expenditure. $I_{i}(t+1)$ and $IE_{i}(t+1)$ are both determined on the capital good market, see \ref{app:capital_market}.

\subsubsection{Labour \& wages}
C-firm $i$'s desired labour in the next period is determined using Eq.~\ref{eq:cfirm_output} and their desired capital utilisation:
\begin{equation}
    N^d_{i}(t+1) = \upsilon^d_{i}(t+1) \frac{K_{i}(t+1)}{\nu a^e_{i}(t+1)},
\end{equation}
where $\upsilon^d_{i}(t+1)$ is C-firm $i$'s desired capital utilisation, similar in form to that specified in \cite{caiani2016benchmark_model}:
\begin{equation}
\label{eq:desired_utilisation}
    \upsilon^d_{i}(t+1) = \min\bigg\{\frac{\nu Y^d_{i}(t+1)}{K_{i}(t+1)}, 1\bigg\},
\end{equation}
where, $Y^d_{i}(t+1)$ is the desired output of firm $i$, defined in Eq.~\ref{eq:cfirm_desired_output}. The expected labour productivity in the next period, $a^e_{i}(t+1)$, is derived from Eq.~\ref{eq:productivity} as the expectation of a discrete-time GBM:
\begin{equation}
    a^e_{i}(t+1) = a_{i}(t)\exp\{g\}.
\end{equation}
C-firm $i$ then decides how many employees they wish to hire or fire in the next period using:
\begin{equation}
\label{eq:employment_decision}
    \eta_{i}(t+1) = N^d_{i}(t+1) - N_{i}(t).
\end{equation}

The wage that C-firm $i$ offers their employees depends on C-firm $i$'s employment decisions and the average wage in the previous period. When C-firm $i$ desires to hire $\eta_{i}(t)\geq 0$ (or fire $\eta_{i}(t) < 0$) labour then C-firm $i$ stochastically increases (decreases) their nominal wage. Furthermore, C-firm $i$ will also adjust their wage towards the average wage to stay competitive. Hence, C-firm $i$'s wage is updated as: 
\begin{equation}
\displaystyle w_{i}(t)= \begin{cases}
  w_{i}(t-1)(1 + \sigma_w|\varepsilon_{i}(t)|) + \gamma_w(\bar{w}(t-1) - w_{i}(t-1)) & \text{if $\eta_{i}(t) \geq 0$} \\
  w_{i}(t-1)(1 - \sigma_w|\varepsilon_{i}(t)|) + \gamma_w(\bar{w}(t-1) - w_{i}(t-1)) & \text{if $\eta_{i}(t) < 0$},\\
 \end{cases}
\end{equation}
where $\sigma_w$ is the standard deviation of C-firm $i$'s wage growth, $\varepsilon_{i}(t) \sim \mathcal{N}(0,1)$, hence $|\varepsilon_{i}(t)|$ is distributed according to a folded normal distribution, $\gamma_w\in(0,1)$ is a speed of adjustment parameter, and $\bar{w}(t-1)$ is the average wage in the previous period. The actual labour hired by C-firm $i$, $N_{i}(t)$, is determined on the labour market, see \ref{app:labour_market}.

\subsubsection{Debt}
Due to imperfect market conditions, the cost of external finance is greater for C-firm $i$ than using internal funds, hence, in line with the pecking order theory of finance \citep{myers1984capital_structure} and the methodology of other ABMs \citep{gaffeo2008emergent_macro,assenza2015macro_abm,caiani2016benchmark_model}, C-firm $i$ will seek external financing when their internal flow of funds (profits and deposits) have been exhausted. However, it is rare that a firm will ever completely exhausts their internal finances, therefore, C-firm $i$ keeps a buffer of internal funds for precautionary reasons. This buffer is assumed to be proportional to their wage bill. Thus, C-firm $i$'s actual desired loan is given by:
\begin{equation}
\label{eq:desired_loan}
    L^d_{i}(t+1) = \max\big\{IE_{i}(t+1) + \zeta W_{i}(t) - \Pi_{i}(t) - M_{i}(t), 0\big\}.
\end{equation}
where $IE_{i}(t+1)$ is firms $i$'s actual investment expenditure, $W_{i}(t)=w_{i}(t)N_{i}(t)$ is firm $i$'s wage bill, $\zeta$ is the amount of precautionary internal finance saved by C-firm $i$ in relation to their wage bill, $\Pi_{i}(t)$ is firm $i$'s profits, and $M_{i}(t)$ is firm $i$'s deposits at bank $b$.

C-firm $i$ is assumed to request a single loan $\ell$ each period from a given bank $b$, see \ref{app:credit_market} for the credit market, denoted $L_{ib,\ell}(t)$, with an interest rate $r^L_b(t)$. C-firm $i$ then calculates the amortisation cost to bank $b$ on this loan using the following formula:
\begin{equation}
\label{eq:amortisation}
    A_{ib,\ell}(t) = L_{ib,\ell}(t) \frac{r^L_b(t)(1 + r^L_b(t))^n}{(1 + r^L_{b}(t))^n - 1},
\end{equation}
where $n$ is the number of repayment periods. In each period, C-firm $i$ repays a proportion of the principal of loan $\ell$ to bank $b$, $\rho L_{ib,\ell}(t)$, where $\phi=1/n$ is the repayment rate and $L_{ib,\ell}(t)$ is the value of the loan that was taken out in period $t$. Hence, loan repayments remain constant until the loan has been repaid in full, a linear amortisation schedule. Assuming that a loan was taken out in period $\tau$, where $t\geq \tau \geq t-n$, the current outstanding value of the loan $\ell$ (amount left to be repaid), is given by:
\begin{equation}
\label{eq:loan_value}
    L_{ib,\ell}(t)=(t-\tau)\rho L_{ib,\ell}(\tau),
\end{equation}
where $\tau$ is the period in which the loan was taken out from bank $b$, and $L_{ib,\ell}(\tau)$ is the initial value of the loan $\ell$, the loan principal. Therefore, assuming no defaults, it will take C-firm $i$ $n$ periods to repay the loan in full. C-firm $i$ pays interest to bank $b$ on the loan $\ell$, as the difference between the initially calculated amortisation cost, $A_{ib,\ell}(t)$ and principal payment, $\rho L_{ib,\ell}(t)$. Thus, for a loan taken out in period $\tau$, the interest cost in period $t$, for $\tau+n\geq t\geq \tau$, is given by: 
\begin{equation}
\label{eq:loan_interest}
    IP_{ib,\ell}(t)=A_{ib,\ell}(\tau) - \rho L_{ib,\ell}(\tau).
\end{equation}
Therefore, interest payment remains constant until the loan has been repaid in full in period $\tau+n$. Hence, total interest payments made by C-firm $i$ in period $t$ is given by the sum of interest payments to each bank $b$ that C-firm $i$ still has an outstanding loan with in time $t$:
\begin{equation}
\label{eq:interest_payments}
    IP_{i}(t) = \sum_{b\in\mathcal{B}_{i}(t)} \sum_{\ell\in\mathcal{L}_{ib}(t)} IP_{ib,\ell}(t),
\end{equation}
where $\mathcal{B}_{i}(t)$ is the set of banks that firm $i$ still has an outstanding loan with in period $t$ and $\mathcal{L}_{ib}(t)$ is the set of loans $\ell$ that firm $i$ still has outstanding with bank $b$. Additionally, C-firm $i$'s total debt in period $t$ is given by the sum of all C-firm $i$'s outstanding loans with each bank $b$:
\begin{equation}
\label{eq:debt}
    D_{i}(t) = \sum_{b\in\mathcal{B}_{i}(t)} \sum_{\ell\in\mathcal{L}_{ib}(t)} L_{ib,\ell}(t).
\end{equation}
The actual loan, $L_{ib,\ell}(t)$, given to firm $i$ from bank $b$ in period $t$ is determined on the credit market, see \ref{app:credit_market}, and bank loan supply, Eq.~\ref{eq:loan_supply}. 

\subsubsection{Accounting}
C-firm $i$'s profits are given by their revenue and interest on deposits minus the wage bill and total interest payments:
\begin{equation}
\label{eq:cfirm_profits}
    \Pi_{i}(t) = P_{i}(t)Q_{i}(t) + r_M M_{i}(t) - W_{i}(t) - IP_{i}(t),
\end{equation}
where $P_{i}(t)$ denotes C-firm $i$'s price, $Q_{i}(t)$ is the quantity of C-goods sold, $r_M$ is the interest rate on deposits, $M_{i}(t)$ are deposits at the bank, and $IP_{i}(t)$ are total interest payments on all outstanding loans.

The balance sheet identity of C-firm $i$ implies that their assets must be equal to their liabilities and equity: 
\begin{equation}
\label{eq:cfirm_balance_sheet}
    KE_{i}(t) + M_{i}(t) = D_{i}(t) + E_{i}(t),
\end{equation}
where $KE_{i}(t)$ is C-firm $i$'s total capital expenditure, $D_{i}(t)$ is firm $i$'s debt, and $E_{i}(t)$ is firm $i$'s equity. The equity of C-firm $i$ is updated based on their profits:
\begin{equation}
\label{eq:cfirm_equity}
    E_{i}(t+1) = E_{i}(t) + \Pi_{i}(t).
\end{equation}
Hence, given the above, the deposits of C-firm $i$ are derived as:
\begin{equation}
\label{eq:cfirm_deposits}
    M_{i}(t+1) = M_{i}(t) + \Pi_{i}(t) + L_{ib,\ell}(t+1) - \rho D_{i}(t) - IE_{i}(t+1),
\end{equation}
where $L_{ib,\ell}(t+1)$ is the actual $\ell$th loan C-firm $i$ takes out from bank $b$, $\rho D_{i}(t)$ is the cost of all principal repayments, and $IE_{i}(t)$ is C-firm $i$'s actual investment expenditure. If firm $i$'s deposits $M_{i}(t) \leq 0$, then firm $i$ becomes bankrupt, see \ref{app:entry_exit} for details on entry and exit dynamics.

\subsection{Capital Firms}
\label{app:kfirms}
\subsubsection{Production}
The $j$th K-firm is assumed only to use labour, $N_{j}(t)$, as a factor of production, with constant returns to scale. Hence, firm $j$'s production function is given by:
\begin{equation}
\label{eq:kfirm_output}
    Y_{j}(t) = a_{j}(t) N_{j}(t),
\end{equation}
where $a_{j}(t)$ is firm $j$'s labour productivity. As with C-firms, the $j$th K-firm updates their own production efficiency, which evolves according to a GBM:
\begin{equation}
\label{eq:capital_productivity}
    a_{j}(t)=a_{j}(t-1)\exp\left\{g - \frac{1}{2} \sigma_a^2 + \sigma_a \varepsilon_{j}(t)\right\},
\end{equation}
where $g$ is the average growth rate of labour productivity, $\sigma_a$ is the standard deviation of labour productivity, and $\varepsilon_{j}(t) \sim \mathcal{N}(0,1)$ is a standard normal random variable. 

\subsubsection{Desired output \& inventories}
Capital goods are assumed to be durable; hence, firm $j$ can keep inventories from one period to the next. Therefore, firm $j$'s desired output, $Y^d_{j}(t+1)$, is determined by their expected demand, $Z^e_{j}(t+1)$, and their level of inventories, $V_{j}(t)$:
\begin{equation}
\label{eq:kfirm_desired_output}
    Y^d_{j}(t+1)=Z^e_{j}(t+1)(1 + \xi) - V_{j}(t),
\end{equation}
where $\xi\in(0,1)$ is firm $j$'s desired excess capacity to meet future variations in demand that were not forecasted, $V_{j}(t)(1 - \delta)$ are firm $j$'s inventories from the previous period that have depreciated by $\delta$, and $Z^e_{j}(t+1)$ is firm $j$'s expected demand, adaptively updated according to their actual demand:
\begin{equation}
    Z^e_{j}(t+1)=Z^e_{j}(t)+\gamma_Z(Z_{j}(t)-Z^e_{j}(t)),
\end{equation}
where $\gamma_Z$ determines the speed of adjustment to actual demand, and  $Z_{j}(t)$ is firm $j$'s actual demand, which is determined on the market for capital goods, see \ref{app:capital_market}. 

Due to the assumption that K-goods are durable and depreciate at a rate $\delta$ each period, firm $j$ initially sets their inventories to $V_{j}(t)=V_{j}(t-1)(1-\delta)+Y_{j}(t)$ in period $t$. Then, after C-firms have consumed from K-firms, K-firm $j$'s leftover inventories are given by:
\begin{equation}
\label{eq:kfirm_inventories}
    V_{j}(t)=V_{j}(t-1) + Y_{j}(t) - Q_{j}(t),
\end{equation}
where $Q_{j}(t)$ is firm $j$'s actual quantity of sold K-goods, determined on the capital good market, see \ref{app:capital_market}. Additionally, because firm $j$'s desired excess capacity is positive, this implies that firm $j$ has desired inventories equal to their desired excess capacity:
\begin{equation}
\label{eq:kfirm_desired_inventories}
    V^d_{j}(t+1)=\xi Y_{j}(t).
\end{equation}

\subsubsection{Prices}
Similarly to C-firms, the $j$th K-firm sets their price according to both internal and external factors. In particular, firm $j$ will adjust their price according to their inventories and the average capital price, given by:
\begin{equation}
    P_{j}(t)=
    \begin{cases}
        P_{j}(t-1)(1 + \sigma_P|\varepsilon_{j}(t)|) + \gamma_P(\bar{P}^K(t-1) - P_{j}(t-1)) & \text{if  $V_{j}(t-1)\leq V^d_{j}(t)$} \\
        P_{j}(t-1)(1 - \sigma_P|\varepsilon_{j}(t)|) + \gamma_P(\bar{P}^K(t-1) - P_{j}(t-1)) & \text{if $V_{j}(t-1)>V^d_{j}(t)$}\\
    \end{cases}
\end{equation}
where $\sigma_P$ is the standard deviation of firm $j$'s price growth, $\varepsilon_{j}(t)\sim\mathcal{N}(0,1)$, so $|\varepsilon_{j}(t)|$ is distributed according to a folded normal distribution, $\gamma_P\in(0,1)$ is an adjustment parameter, $V_{j}(t-1)$ are firm $j$'s inventories in the previous period, $V^d_{j}(t)$ are firm $j$'s desired inventories, and $\bar{P}^K(t)$ is the weighted average of capital good prices:
\begin{equation}
    \bar{P}^K(t) = \frac{\sum^{\mathbf{N}_K}_{k=1}P_{k}(t)Y_{k}(t)}{\sum^{\mathbf{N}_K}_{k=1}Y_{k}(t)}.
\end{equation}

\subsubsection{Labour \& wages}
Firm $j$ uses their desired output and expected productivity in the next period to determine their desired amount of labour, derived from Eq.~\ref{eq:kfirm_output} as:
\begin{equation}
    N^d_{j}(t+1) = \frac{Y^d_{j}(t+1)}{a^e_{j}(t+1)},
\end{equation}
where $Y^d_{j}(t+1)$ is firm $j$'s desired output, defined in Eq.~\ref{eq:kfirm_desired_output}. Similarly to C-firms, K-firm $j$'s expected productivity is derived as the expectation of a discrete-time GBM from Eq.~\ref{eq:capital_productivity}:
\begin{equation}
    a^e_{j}(t+1) = a_{j}(t)\exp\{g\}.
\end{equation}
Firm $j$ then decides how many employees they wish to hire or fire in the next period:
\begin{equation}
    \eta_{j}(t+1) = N^d_{j}(t+1) - N_{j}(t).
\end{equation}

As with C-firms, K-firm $j$ updates their wage rate according to their labour demand and the average wage, given by:
\begin{equation}
    \displaystyle w_{j}(t)= \begin{cases}
        w_{j}(t-1)(1 + \sigma_w|\varepsilon_{j}(t)|) + \gamma_w(\bar{w}(t-1) - w_{j}(t-1)) & \text{if $\eta_{j}(t) \geq 0$} \\
        w_{j}(t-1)(1 - \sigma_w|\varepsilon_{j}(t)|) + \gamma_w(\bar{w}(t-1) - w_{j}(t-1)) & \text{if $\eta_{j}(t) < 0$},\\
    \end{cases}
\end{equation}
where $\sigma_w$ is the standard deviation of firm $i$'s wage growth, $\varepsilon_{j}(t) \sim \mathcal{N}(0,1)$, hence $|\varepsilon_{j}(t)|$ is distributed according to a folded normal distribution, $\gamma_w\in(0,1)$ is a speed of adjustment parameter, and $\bar{w}(t-1)$ is the average wage in the previous period. K-firm $j$'s actual labour, $N_{j}(t)$ is determined on the labour market, see \ref{app:labour_market}.

\subsubsection{Debt}
K-firms also experience higher costs for external finance; hence, they adhere to the pecking order theory of finance in which they initially use internal finance and then seek external funding in imperfect credit markets. As with C-firms, K-firms will never completely exhaust their internal finances. Therefore, firm $j$ keeps a buffer of internal funds for precautionary reasons. This buffer is assumed to be proportional to their wage bill. Thus, firm $j$'s desired new loan is given by:
\begin{equation}
    L^d_{j}(t+1) = \max\{W_{j}(t) - \Pi_{j}(t) - M_{j}(t),0\}.
\end{equation}
where $W_{j}(t)=w_{j}(t)N_{j}(t)$ is firm $j$'s wage bill, $\Pi_{j}(t)$ is firm $j$'s profits, and $M_{j}(t)$ is firm $j$'s deposits. K-firms use the same method as C-firms when calculating the amortisation cost of a new loan (Eq.~\ref{eq:amortisation}), total interest payments (Eq.~\ref{eq:interest_payments}), and total debt (Eq.~\ref{eq:debt}). Additionally, firm $j$'s actual loan $\ell$ given by the bank $b$, $L_{j b,\ell}(t)$, is determined on the credit market, see \ref{app:credit_market} and bank loan supply, Eq.~\ref{eq:loan_supply}.
Firm $j$ updates its debt with the new loan:
\begin{equation}
    D_j(t+1)=D_j(t)(1 - \phi)+L_j(t+1),
\end{equation}
where $L_j(t+1)$ is the actual new loan taken out from bank $b$. 

\subsubsection{Accounting}
Firm $j$'s profits are given by their revenue and interest on deposits minus the wage bill and total interest payments:
\begin{equation}
    \Pi_{j}(t) = P_{j}(t)Q_{j}(t) + r_M
M_{j}(t) - W_{j}(t) - IP_{j}(t),
\end{equation}
where $P_{j}(t)$ denotes firm $j$'s price, $Q_{j}(t)$ is the quantity of C-goods sold, $r_M$ is the interest rate on deposits, $M_{j}(t)$ are deposits at the bank, $W_{j}(t)$ is the wage bill, and $IP_{j}(t)$ are total interest payments on all outstanding loans.

The balance sheet identity of firm $j$ implies that their assets must be equal to their liabilities and equity. However, capital firms do not have any physical assets only monetary assets, hence firm $j$'s balance sheet is reduced to:
\begin{equation}
\label{eq:kfirm_balance_sheet}
    M_{j}(t) = D_{j}(t) + E_{j}(t),
\end{equation}
where $M_{j}(t)$ is firm $j$'s deposits, $D_{j}(t)$ is firm $j$'s debt and $E_{j}(t)$ is firm $j$'s equity. The equity of firm $j$ is updated based on their profits:
\begin{equation}
\label{eq:kfirm_equity}
    E_{j}(t+1) = E_{j}(t) + \Pi_{j}(t).
\end{equation}
Hence, given the above, the deposits of firm $j$ are derived as:
\begin{equation}
    M_{j}(t+1) = M_{j}(t) + \Pi_{j}(t) + L_{jb,\ell}(t+1) - \rho D_{i}(t),
\end{equation}
where $L_{jb,\ell}(t+1)$ is the actual $\ell$th loan firm $j$ takes out from bank $b$ and $\rho D_{i}(t)$ is the cost of all principal repayments. If firm $j$'s deposits $M_{j}(t) \leq 0$, then firm $j$ becomes bankrupt, see \ref{app:entry_exit} for details on entry and exit dynamics.

\subsection{Banks}
\label{app:banks}
\subsubsection{Accounting}
\label{Banks}
The balance sheet identity of bank $b$ is given by:
\begin{equation}
\label{eq:bank_balance_sheet}
    R_{b}(t) + L_{b}(t) = A_{b}(t) + M_{b}(t) + E_{b}(t),
\end{equation}
where $R_{b}(t)$ are the reserves of bank $b$, assumed to be deposited at an unmodelled central bank, $L_{b}(t)$ is bank $b$'s total stock of loans extended to firms, $A_{b}(t)$ are central bank advances, $M_{b}(t)$ is the total amount of household and firm deposits held by bank $b$, and $E_{b}(t)$ is bank $b$'s equity. 

Bank $b$'s reserves are simply derived from the above accounting identity:
\begin{equation}
\label{eq:bank_reserves}
    R_{b}(t) = A_{b}(t) + M_{b}(t) + E_{b}(t) - L_{b}(t).
\end{equation}
For simplicity, it is assumed that the central bank offers an interest free advance to banks when they have negative reserves, therefore, bank $b$'s advances from the central bank are given by:
\begin{equation}
    A_{b}(t+1) = \max\{-R_{b}(t),0\}.
\end{equation}
Bank $b$'s loans are given by the sum of all outstanding loans extended to both C-firms and K-firms:
\begin{equation}
\label{eq:bank_loans}
    L_{b}(t) = \sum_{\iota\in\mathcal{F}^L_{b}(t)}\sum_{\ell\in\mathcal{L}_b(t)} L_{\iota b}(t),
\end{equation}
where $\mathcal{F}^L_{b}(t)$ is the set of C-firms and K-firms with outstanding loans to bank $b$ in period $t$, which is determined on the credit market, see \ref{app:credit_market}. Bank $b$'s deposits are given by the sum of all deposits held by C-firms, K-firms and households at bank $b$:
\begin{equation}
\label{eq:bank_deposits}
    M_{b}(t) = \sum_{\iota\in\mathcal{F}^M_{b}(t)}M_{\iota}(t) + \sum_{h\in\mathcal{H}^M_{b}(t)}M_{h}(t),
\end{equation}
where $\mathcal{F}^M_{b}(t)$ is the set of C-firms and K-firms with deposits, $M_{\iota}(t)$, at bank $b$ in period $t$, and $\mathcal{H}^M_{b}(t)$ is the set of households with deposits, $M_{h}(t)$, at bank $b$ in period $t$. The equity of bank $b$ is updated by their profits, $\Pi_{b}(t)$, and bank $b$ is assumed to absorb all the losses from bad loans $B_{b}(t)$, which is the sum of all outstanding loans extended to insolvent firms. Moreover, when bank $b$'s equity becomes negative, they are assumed to be bailed in by their depositors, so that their equity is given by their desired capital ratio times their assets, see \ref{app:entry_exit} for details. Hence, bank $b$'s equity is given by:
\begin{equation}
\label{eq:bank_equity}
    \displaystyle E_{b}(t+1) = \begin{cases}
        E_{b}(t)+\Pi_{b}(t)-B_{b}(t) & \text{if $E_{b}(t) > 0$} \\
        CR^d_{b}(t+1)(L_{b}(t)+R_{b}(t)) & \text{if $E_{b}(t) \leq 0$},\\
    \end{cases}
\end{equation}
where $CR^d_{b}(t+1)$ is bank $b$'s desired capital ratio, defined in Eq.~\ref{eq:bank_desired_credit_ratio}. The profits of bank $b$ are given by the difference between the sum of interest received from loans and interest paid on deposits:
\begin{equation}
\label{eq:bank_profits}
    \Pi_{b}(t) = \sum_{\iota\in\mathcal{F}^L_{b}(t)}\sum_{\ell\in\mathcal{L}_{ib}(t)}IP_{\iota b,\ell}(t) - r_MM_{b}(t),
\end{equation}
where $\mathcal{F}^L_{b}(t)$ is the set of C-firms and K-firms that have outstanding loans at bank $b$ in period $t$, $\mathcal{L}_{ib,\ell}(t)$ is the set of loans $\ell$ that firm $\iota$ has outstanding with bank $b$ in period $t$, and $IP_{ib,\ell}(t)$ are firm $\iota$'s interest payments on loan $\ell$ to bank $b$ in period $t$, Eq.~\ref{eq:loan_interest}, and $r_M$ is bank $b$'s interest rate on deposits, which is assumed to be constant and uniform across all banks.

\subsubsection{Loans}
Bank $b$ is assumed to lend to firm $\iota$ based on bank $b$'s risk tolerance, which is measured by bank $b$'s desired capital ratio. It is assumed that if bank $b$'s ratio of expected bad loans to total loans increases, then bank $b$ becomes more risk averse, and vice versa. Therefore, bank $b$'s desired capital ratio is given by a linear function of their expected bad loans ratio:
\begin{equation}
\label{eq:bank_desired_credit_ratio}
    CR^d_{b}(t+1)=\max\{\kappa, \beta^e_{b}(t+1)\},
\end{equation}
derived from the fact that banks regulatory capital requirement is given by $\kappa$ and they also desire to hold additional capital to absorb expected losses. $\beta^e_{b}(t+1)=EL_{b}(t+1)/L_{b}(t)$ is bank $b$'s expected loss (EL) to loans ratio. Expected loss is given by the well known credit risk formula $EL_b(t+1)=\sum_{\iota\in\mathcal{L}_{\iota b}(t)} PD_\iota(t+1)\times LGD_{\iota b}(t) \times EAD_{\iota b}(t)$ \citep{chatterjee2015credit_risk}, where $\mathcal{L}_{b}(t)$ is the set of all firms in period $t$ with outstanding loans at bank $b$, $PD_\iota(t)$ is the probability of default of firm $\iota$, $LGD_\iota(t)=1$ is the loss given default, percentage of exposure bank $b$ expects to lose if firm $\iota$ defaults (assumed to be equal to 100\%), $EAD_{\iota b}(t)$ is the exposure at default of bank $b$ if firm $\iota$ defaults (sum of firm $\iota$'s outstanding loans). Therefore, the expected loss of bank $b$ is given by:
\begin{equation}
    EL_{b}(t+1)=\sum_{\iota\in\mathcal{F}^L_b(t)}PD_\iota(t+1)\sum_{\ell\in\mathcal{L}_{\iota b}(t)} L_{\iota b,\ell}(t).
\end{equation}
where $PD_\iota(t+1)$ is the probability of default of firm $\iota$ in the next period and the sum of loans ($L_{\iota b,\ell}(t)$) is bank $b$'s current EAD to firm $\iota$. Similar to \cite{assenza2015macro_abm}, firm $\iota$'s probability of default is estimated each period by bank $b$ using a logistic regression model of firm $\iota$'s expected leverage ratio, $\lambda^e_{\iota}(t+1)$:
\begin{equation}
    PD_\iota(t+1)=\frac{1}{1 + \exp\big\{-\big(\hat{\theta}_0 + \hat{\theta}_1\lambda_{\iota}(t)\big)\big\}},
\end{equation}
where $\hat{\theta}_0$ and $\hat{\theta}_1$ are estimated using a time series of firm defaults (0 no default, 1 default) and expected leverage ratios, where firm $\iota$'s expected leverage ratio is defined as:
\begin{equation}
\label{eq:expected_leverage}
    \lambda^e_{\iota}(t+1)=\frac{D^e_{i}(t+1)}{M_{\iota}(t)+\Pi_{\iota}(t)+D^e_{\iota}(t+1)},
\end{equation}
where $D^e_\iota(t+1)$ is firm $\iota$'s expected debt level, $D^e_{\iota}(t+1)=D_{\iota}(t)(1 - \rho) + L^d_{\iota}(t+1)$, $M_\iota(t)$ is firm $\iota$'s deposits, and $\Pi_\iota(t)$ is firm $\iota$'s profits. Hence, the expected leverage of firm $\iota$'s is bounded between 0 and 1, making for a more stable measure. Moreover, bank $b$ separates firms into C-firms and K-firms and estimates the bankruptcy probability of each firm type separately.

Bank $b$ supplies firm $\iota$'s desired loan in full if bank $b$'s desired capital ratio is lower than their actual capital ratio; bank $b$ does not supply the loan if their desired capital ratio is higher than their actual capital ratio, which ensures bank $b$ does not take on too much risk. Therefore, the loan that firm $\iota$ receives from bank $b$ is given by:
\begin{equation}
\label{eq:loan_supply}
    L_{\iota b,\ell}(t+1)=
    \begin{cases}
        L^d_{\iota}(t+1) & \text{if $CR^d_{b}(t+1) < CR_{b}(t)$} \\
        0               & \text{if $CR^d_{b}(t+1) \geq CR_{b}(t)$}, \\
    \end{cases}
\end{equation}
where $L^d_{\iota}(t+1)$ is firm $\iota$'s desired loan, Eq.~\ref{eq:desired_loan}, and bank $b$'s actual capital ratio is given by:
\begin{equation}
    CR_{b}(t) = \frac{E_{b}(t)}{L_{b}(t)},
\end{equation}
where $E_{b}(t)$ is bank $b$'s equity and $L_{b}(t)$ are bank $b$'s total outstanding loans to firms.

\subsubsection{Interest rate}
Banks update their interest rates based on a combination of two factors: where their capital ratio currently stands compared to their desired capital ratio, and where the current interest rate stands compared to a desired markup over inflation. Hence, bank $b$'s interest rate on loans is updated as:
\begin{equation}
\label{eq:bank_interest_rate}
    r^L_{b}(t+1)=
    \begin{cases}
        r^L_{b}(t)(1+\sigma_r|\varepsilon_{b}(t)|) + \gamma_r(r^{L,d}_b(t) - r^L_{b}(t)) & \text{if $CR^d_{b}(t+1) \geq CR_{b}(t)$} \\
        r^L_{b}(t)(1-\sigma_r|\varepsilon_{b}(t)|) + \gamma_r(r^{L,d}_b(t) - r^L_{b}(t)) & \text{if $CR^d_{b}(t+1) < CR_{b}(t)$}, \\
    \end{cases}
\end{equation}
where $\sigma_r$ is the standard deviation of bank $b$'s loan interest rate growth, $\varepsilon_{b}(t) \sim \mathcal{N}(0,1)$, hence $|\varepsilon_{b}(t)|$ is distributed according to a folded normal distribution, $\gamma_r\in(0,1)$ is a speed of adjustment parameter, $C^d_{b}(t+1)$ is bank $b$'s desired capital ratio, and $CR_{b}(t)$ is bank $b$'s actual capital ratio, and bank $b$'s desired target interest rate on loans, $r^{L,d}_b(t)$, is given by a markup over inflation:
\begin{equation}
\label{eq:bank_desire_interest_rate}
    r^{L,d}_b(t)=\max\{f(t)+r_N,r_N\}
\end{equation}
where $f(t)$ is the rate of inflation and $r_N$ is a desired real rate of interest on loans, uniform across banks. Additionally, it is assumed that bank $b$ will never set an interest rate below $r_N$ to ensure that interest rates stay positive, $r^L_b(t)>0$. 

\subsection{Entry \& Exit Dynamics}
\label{app:entry_exit}
The number of all agents apart from firms is fixed throughout the simulation. The bankruptcy condition of firm $\iota$ is assumed to be when their deposits are less than or equal to zero, $M_{\iota}(t)\leq 0$. There is a one-to-one replacement of firms; thus, if firm $\iota$ exits the market, a new firm will enter. New entrants are a random copy of incumbent firms with no debt, a single employee (chosen from the pool of unemployed households), and average market values for their price $P_{\iota}(t)=\bar{P}^{F}(t)$ ($F=\{C, K\}$ for either C-market or K-market respectively) and wage $w_{\iota}(t)=\bar{w}(t)$. It is assumed that bankrupt C-firms' capital is disposed of, and new C-firm entrants are endowed with the capital stock of the firm they imitate. 

Furthermore, banks become bankrupt when their equity is less than or equal to zero, $E_{b}(t)\leq 0$, however, unlike firms, banks are bailed in by their depositors (firms and households) and the new equity of the bankrupt bank in the next period is $E_{b}(t+1) = CR^d_{b}(t+1)(L_{b}(t)+R_{b}(t))$, which is the banks desired amount of equity to their assets. Each depositor's deposits are reduced proportionally by the size of their deposit account at the bankrupt bank. 

\subsection{The Market for Labour}
\label{app:labour_market}
The labour market is characterised by a search and matching mechanism similar to the one presented in \cite{russo2007industrial} and \cite{gaffeo2008emergent_macro}. Initially, firm $\iota$ determines their employment decision for the next period, to either hire or fire employees depending on their employment variable $\eta_{\iota}(t)$, Eq.~\ref{eq:employment_decision}. If $\eta_{\iota}(t)<0$, then firm $\iota$ wants to fire $|\eta_{\iota}(t)|$ employees, and when $\eta_{i}(t)>0$, firm $\iota$ will post vacancies equal to $\eta_{i}(t)$. Firm $\iota$ offers potential employees a single-period labour contract with a guaranteed wage of $w_{\iota}(t)$ for the current period. It is assumed that the labour contract is periodically renewed and updated with the current wage offered by the firm until the firm decides to terminate the employee's employment.

The labour market then opens, and firms post their vacancies with the corresponding wage rate. If a household $h$ is unemployed, they will randomly visit $n_F$ firms each period. Household $h$ incurs no travel costs when visiting the initial $n_F$ firms; however, after this number is reached, travel costs become prohibitively high. The probability that household $h$ visits firm $\iota$ is equal to firm $\iota$'s labour market share, $ms^N_{\iota}(t)=N_{\iota}(t)/(\sum^{\mathbf{N}_C}_{c=1}N_{c}(t) + \sum^{\mathbf{N}_K}_{k=1}N_{k}(t))$. Household $h$ then sorts the firms they visited by descending wage rate and will send an application to all firms on their list which have open vacancies, $\eta_{\iota}(t)>0$. 

Firms then hire or fire employees based on the value of $\eta_{\iota}(t)$. When $\eta_{\iota}(t)<0$, it is assumed that firm $\iota$ randomly chooses $\min\{|\eta_{\iota}(t)|, N_{\iota}(t)-1\}$ workers to fire from their set of employees, hence firm $\iota$ always keeps at least 1 employee. When $\eta_{\iota}(t)>0$, it is assumed that firm $\iota$ randomly hires $\min\{\eta_{\iota}(t),||\mathcal{A}_\iota(t)||\}$ new workers from their set of applications, denoted $\mathcal{A}_\iota(t)$, where $||\mathcal{A}_\iota(t)||$ is length of the set $\mathcal{A}_\iota(t)$, the number of households in firm $\iota$'s set of applications in period $t$. Thus, the number of households firm $\iota$ can hire is constrained by the number of applications they receive each period.

\subsection{The Market for Consumption Goods}
\label{app:consumption_market}
The consumption market uses a search and matching mechanism to match households' consumption demands with C-firms' output, similar to that proposed by \cite{russo2007industrial} and \cite{gaffeo2008emergent_macro}. At the beginning of period $t$, C-firm $i$ sets its inventories equal to output and discards any inventories from the previous period because C-goods are assumed to be perishable, hence $V_{i}(t)=Y_{i}(t)$ at the start of period $t$. Households randomly visit $n_C$ C-firms each period, the maximum amount that assumed travel costs permit. The probability that household $h$ will visit C-firm $i$ is equal to C-firm $i$'s consumption market share, $ms_{i}(t)=Y_{i}(t)/\sum^{\mathbf{N}_C}_{k=1}Y_{k}(t)$. Household $h$ then sorts the C-firms they visited by ascending price and demands to consume C-goods worth $C^d_{h}(t) = E^d_{h}(t)/P_{i}(t)$ from the first C-firm on their list. If C-firm $i$'s inventories run out before household $h$ exceeds their desired expenditure, household $h$ purchases the remaining C-goods from C-firm $i$ and moves to the next C-firm on their list. Households continue to demand C-goods from their list of C-firms to visit until they have reached their desired expenditure or exhausted their list of C-firms. Hence, household $h$'s actual expenditure is given by:
\begin{equation}
E_{h}(t) = \min\left\{E^d_{h}(t), \max\left\{\sum_{i\in\mathcal{C}_{h}(t)}P_{i}(t)V_{i}(t)-\sum_{h'\in\mathcal{H}'_{i}(t)} E^d_{h'}(t), 0\right\}\right\},
\end{equation}
where $E^d_{h}(t)$ is household $h$'s desired expenditure, $\mathcal{C}_{h}(t)$ is the set of C-firms visited by household $h$ in period $t$, $P_{i}(t)$ is visited C-firm $i$'s price, $Y_{i}(t)$ is C-firm $i$'s output, and $\mathcal{H}'_{i}(t)$ is C-firm $i$'s set of households that visited before household $h$ in period $t$. Therefore, the actual demand for C-goods faced by C-firm $i$ is given by:
\begin{equation}
Z_{i}(t) = \max\left\{\sum_{h\in\mathcal{H}_{i}(t)}\frac{E^d_{h}(t)}{P_{i}(t)}-\sum_{i'\in\mathcal{C}'_{h}(t)}Y_{i'}(t),0\right\},
\end{equation}
where $\mathcal{H}_{i}(t)$ is the set of households that visited C-firm $i$ in period $t$ and $\mathcal{C}'_{h}(t)$ is the set of C-firms $i'$ that household $h$ visited before C-firm $i$. C-firm $i$ then calculates their actual quantity of sold C-goods as:
\begin{equation}
Q_{i}(t) = \min\big\{Z_{i}(t), Y_{i}(t)\big\}.
\end{equation}
Then C-firm $i$'s involuntary inventories after the consumption market can be derived as:
\begin{equation}
    V_{i}(t)=Y_{i}(t)-Q_{i}(t).
\end{equation}

\subsection{The Market for Capital Goods}
\label{app:capital_market}
The capital market employs a search and matching mechanism to match C-firm investment demand with K-firm output, similar to the market for consumption goods as in \cite{assenza2015macro_abm}. At the start of the capital market, K-firm $j$ sets their inventories given previous inventories, less depreciation, and new production, $V_{j}(t)=V_{j}(t-1)(1-\delta)+Y_{j}(t)$. If the $i$th C-firm desires to invest in new capital, $I^d_{i}(t+1)>0$, then C-firm $i$ will randomly visit $n_K$ K-firms each period, the maximum amount that assumed travel costs will permit. The probability that C-firm $i$ will visit K-firm $j$ is equal to K-firm $j$'s capital market share, $ms_{j}(t)=Y_{j}(t)/\sum^{\mathbf{N}_K}_{k=1}Y_{k}(t)$. C-firm $i$ then sorts the K-firms they visited by ascending price and demands orders for K-goods worth $I^d_{i}(t+1)=EI^d_{i}(t+1)/P_{j}(t)$, where $E^d_{i}(t+1)$ is C-firm $i$'s desired investment expenditure and $P_{j}(t)$ is K-firm $j$'s price, from the first K-firm on their list for use in the next period's production. It is assumed that capital goods require additional time for delivery. If K-firm $j$'s inventories run out before C-firm $i$ exceeds their desired investment expenditure, then C-firm $i$ purchases the remaining K-goods from K-firm $j$ and moves to the next K-firm on their list. C-firms continue to demand K-goods from their list of K-firms until they have reached their desired investment or exhausted their list of K-firms. Hence, C-firm $i$'s actual investment expenditure is given by:
\begin{equation}
IE_{i}(t+1) = \min\left\{IE^d_{i}(t+1), \max\left\{\sum_{j\in\mathcal{K}_{i}(t)}P_{j}(t)V_{j}(t) - \sum_{i'\in\mathcal{C}'_{j}(t)}IE^d_{i'}(t+1), 0\right\}\right\},
\end{equation}
where $\mathcal{K}_{i}(t)$ is the set of K-firms visited by C-firm $i$ in period $t$, $IE^d_{i}(t+1)$ is C-firm $i$'s desired investment expenditure (Eq.~\ref{eq:desired_investment_expenditure}), and $\mathcal{C}'_{j}(t)$ is the set of C-firms $i'$ that visited K-firm $j$ before C-firm $i$ in period $t$. C-firm $i$'s investment is then derived as:
\begin{equation}
I_{i}(t+1) = \sum_{j\in\mathcal{K}_{i}(t)}\min\left\{\frac{IE^d_{i}(t+1)}{P_{j}(t)}, \max\left\{V_{j}(t) - \sum_{i'\in\mathcal{C}'_{j}(t)}\frac{IE^d_{i'}(t+1)}{P_{j}(t)}, 0\right\}\right\}.
\end{equation}
Furthermore, the actual demand for K-goods faced by K-firm $j$ is given by:
\begin{equation}
Z_{j}(t) = \max\left\{\sum_{i\in\mathcal{C}_{j}(t)}\frac{IE^d_{i}(t+1)}{P_{j}(t)}-\sum_{j'\in\mathcal{K}'_{i}(t)}Y_j(t)+V_{j'}(t-1),0\right\},
\end{equation}
where $\mathcal{C}_{j}(t)$ is the set of C-firms that visited K-firm $j$ in period $t$ and $\mathcal{K}'_{i}(t)$ is the set of K-firms $j'$ that C-firm $i$ visited before K-firm $j$. K-firm $j$ then calculates their actual quantity of sold K-goods as:
\begin{equation}
Q_{j}(t) = \min\big\{Z_{j}(t), Y_{j}(t)\big\}.
\end{equation}
Therefore, due to the durability of K-goods with a depreciation rate $\delta$, the inventories of K-firm $j$ after the capital goods market has closed is given by:
\begin{equation}
    V_{j}(t)=V_{j}(t-1)+Y_{i}(t)-Q_{i}(t).
\end{equation}

\subsection{The Market for Credit}
\label{app:credit_market}
C-firms and K-firms randomly visit a single bank $b$ each period. The probability that firm $\iota$ visits bank $b$ is equal to bank $b$'s credit market share, $ms_{b}(t)=L_{b}(t)/\sum^{\mathbf{N}_B}_{k=1}L_{k}(t)$. Firm $\iota$ then demands loans worth $L^d_{\iota}(t+1)$ from the selected bank $b$. Bank $b$ supplies the loan depending on their risk tolerance, Eq.~\ref{eq:loan_supply}. Hence, firms' actual loans are given by the loan supply of the bank they demand a loan from. When firm $\iota$ takes out a loan from bank $b$, bank $b$ simultaneously credits firm $\iota$'s deposit account, while also increasing bank $b$'s stock of loans. Hence, this action simultaneously increases total bank assets and liabilities, thus creating new money. When firm $\iota$ repays a proportion of a loan, this destroys money, because banks record this transaction as both a reduction in deposits and loans, which simultaneously decreases both assets and liabilities by the size of the loan repayment.

\section{Balanced Growth Model}
\label{app:balanced_growth}
In this section, we develop a stylised and simplified macro balanced growth equilibrium version of the PG-DYNAMIN model, following the methodology in \cite{caiani2016benchmark_model}. On such a model, it is assumed that the economy grows at a constant real growth rate $g$ (the average labour productivity growth rate), and inflation is constant at rate $g_P$. Hence, the nominal growth rate of the economy is $g_N=g+g_P$. The conditions for balanced growth imply that all real stocks must grow at rate $g$, and all nominal stocks must grow at rate $g_N$. It is also assumed that the nominal rate of interest on loans is given by $r_L=g_P+r_N$ in balanced growth. Additionally, it is assumed that there are no frictions in balanced growth, therefore, the market for labour clears and is in equilibrium (full employment), the market for consumption goods is in equilibrium (C-firm production is equal to household real consumption), the market for capital goods is in equilibrium (K-firm production is equal to C-firm investment), and the market for credit is in equilibrium (loan demand by firms is always supplied by the banks). 

Let sector shares be given by:
\begin{equation}
    s_C=\frac{Y_C(t)}{Y(t)} \quad \text{\&}\quad s_K=\frac{Y_K(t)}{Y(t)},
\end{equation}
where $Y_C(t)$ is total C-firm production, $Y_K(t)$ is total K-firm production, and $Y(t)=Y_C(t)+Y_K(t)$ is real GDP. C-firm production is given by $Y_C(t)=a(t)N_C$ and similarly K-firm production is given by $Y_K=a(t)N_K$, where $N_C$ is the number of employees in the consumption good sector and $N_K$ is the number of employees in the capital good sector. Labour productivity is equal across sectors in balanced growth, so here $a(t)=a_C(t)=a_K(t)=a_0\exp\{gt\}$. Thus, the sector shares are derived from the number of employees in each sector:
\begin{equation}
    s_C=\frac{N_C}{N_C+N_K} \quad \text{\&} \quad s_K=1-s_C=\frac{N_K}{N_C+N_K}.
\end{equation}

The wage share is constant in balanced growth, given by:
\begin{equation}
    \omega=\frac{w(t)}{P(t)a(t)},
\end{equation}
where $w(t)$ is the wage, equalised across sector due to perfect labour mobility in balanced growth, and $P(t)$ is the price, similarly equalised across sectors in balanced growth. Hence, given a constant rate inflation $g_P$, such that $P(t)=P_0\exp\{g_Pt\}$, the growth rate of nominal wages $g_w$, such that $w(t)=w_0\exp\{g_wt\}$, must be $g_w=g_N=g+g_P$ for the wage share to remain constant. Therefore, the wage share is given by:
\begin{equation}
\label{eq:initial_wages}
    \omega=\frac{w_0}{P_0a_0}.
\end{equation}
Furthermore, the profit share is given by:
\begin{equation}
    \pi=\frac{\Pi(t)}{P(t)Y(t)}=\pi_C+\pi_K+\pi_B,
\end{equation}
where $P(t)Y(t)$ is nominal GDP, $\pi_C$ is the C-firm sector profit share, $\pi_K$ is the K-firm sector profit share, and $\pi_B$ is the bank profit share. The debt to nominal GDP ratio is given by:
\begin{equation}
    d=\frac{D(t)}{P(t)Y(t)}=d_C+d_K,
\end{equation}
where $d_C$ is the C-firm debt ratio and $d_K$ is the K-firm debt ratio. Under balanced growth, K-firms do not take on any debt, hence, $d_K=0$ and $d=d_C$. Moreover, the deposit to nominal GDP ratio is given by:
\begin{equation}
    m=\frac{M(t)}{P(t)Y(t)}=\frac{M_C(t)+M_K(t)+M_H(t)}{P(t)Y(t)}=m_C+m_K+m_H,
\end{equation}
where $M(t)=M_C(t)+M_K(t)+M_H(t)$ are total deposits, sum of C-firm deposits ($M_C(t)$), K-firm deposits ($M_K(t)$), and household deposits ($M_H(t)$), $m_C$ is the C-firm deposit ratio, $m_K$ is the K-firm deposit ratio, and $m_H$ is the household deposit ratio. Furthermore, the equity to nominal GDP ratio is given by:
\begin{equation}
    e=\frac{E(t)}{P(t)Y(t)}=\frac{E_C(t)+E_K(t)+E_H(t)+E_B(t)}{P(t)Y(t)}=e_K+e_C+e_H+e_B,
\end{equation}
where total equity $E(t)=E_C(t)+E_K(t)+E_H(t)+E_B(t)$ is the sum
of C-firm equity ($E_C(t)$), K-firm equity ($E_K(t)$), household equity ($E_H(t)$), and bank equity ($E_B(t)$), $e_C$ is the C-firm equity ratio, $e_K$ is the K-firm equity ratio, $e_H$ is the household equity ratio, and $e_B$ is the bank equity ratio. 

Consumption market equilibrium implies that C-firm production is equal to real household consumption:
\begin{equation}
    Y_C(t)=C(t),
\end{equation}
where $C(t)$ is real household consumption. Additionally, capital market equilibrium implies that K-firm production is equal to C-firm investment:
\begin{equation}
    Y_K(t)=I(t),
\end{equation}
where $I(t)$ is C-firm investment. Moreover, credit market equilibrium implies that C-firm debt is equal to total bank loans:
\begin{equation}
    D(t)=L_B(t),
\end{equation}
where $L_B(t)$ are total bank loans.

\subsection{Household sector in balanced growth}
In the balanced growth model, household real consumption is equal to C-firm output. Hence, $C(t)=Y_C(t)$, and this also implies that household nominal consumption is given $P(t)C(t)=P(t)Y_C(t)$. Furthermore, from the households income equation in Eq.~\ref{eq:household_income}, household consumption to real GDP ratio is given by:
\begin{equation}
    c_H=s_C=c_Y(\omega+r_Mm_H)+c_Mm_H.
\end{equation}
Therefore, solving for the wage share $\omega$ gives:
\begin{equation}
    \omega=\frac{s_C}{c_Y}-m_H\left(\frac{c_M}{c_Y}+r_M\right).
\end{equation}
Then, in balanced growth, households deposits are given by:
\begin{equation}
    \Delta M_H(t)=Y_H(t) - P(t)C(t)=g_NM_H(t),
\end{equation}
where $Y_H(t)=w(t)N+r_MM_H(t)$ is household nominal income, and $P(t)C(t)$ is household nominal consumption. Therefore, the deposit to nominal GDP ratio can be found by substituting household nominal income and consumption into the above equation and solving for the deposit ratio, which yields:
\begin{equation}
    m_H=\frac{\omega(1-c_Y)}{c_M+g_N-r_M(1-c_Y)}.
\end{equation}
Therefore, the system of equations that defines the household sector is given by:
\begin{equation}
    \begin{aligned}
        \omega&=\frac{s_C}{c_Y}-m_H\left(\frac{c_M}{c_Y}+r_M\right)\\
        m_H&=\frac{\omega(1-c_Y)}{c_M+g_N-r_M(1-c_Y)}.
    \end{aligned}
\end{equation}
After solving this system of equations, the wage share is given by:
\begin{equation}
\label{eq:wage_share}
    \omega=\frac{s_C(c_M+g_N-r_M(1-c_Y))}{c_M+c_Yg_N-2c_Yr_M(1-c_Y)},
\end{equation}
and the household deposit ratio is given by:
\begin{equation}
\label{eq:household_deposit_ratio}
    m_H=\frac{s_C(1-c_Y)}{c_M+c_Yg_N-2c_Yr_M(1-c_Y)}.
\end{equation}

\subsection{Capital firm sector in balanced growth}
In the K-firm sector under balanced growth $Y_K=Q_K+\delta V_K(t)=Z_K(t)=Z^d_K(t)=Y^d_K(t)$. Hence, using K-firms inventories, Eq.~\ref{eq:kfirm_inventories} and desired inventories, Eq.~\ref{eq:kfirm_desired_inventories}, K-firm quantity of sold goods is given by $Q_K(t)=Y_K(t)(1-\delta\xi)$. Therefore, using the K-firms profit equation , the K-firm profit share is given by:
\begin{equation}
    \pi_K=s_K(1-\delta\xi)+r_Mm_K-\omega s_K.
\end{equation}
In balanced growth K-firm equity grows at the nominal growth rate $g_N$, thus, from Eq.~\ref{eq:kfirm_equity}:
\begin{equation}
    \Delta E_K(t)=\Pi_K(t)=g_NE_K(t).
\end{equation}
Therefore, the K-firm equity ratio is given by:
\begin{equation}
\label{eq:kfirm_equity_ratio}
    e_K=\frac{\pi_K}{g_N}.
\end{equation}
From the accounting identity in Eq.~\ref{eq:kfirm_balance_sheet}, and given that $D_K(t)=0$, the K-firm deposit ratio is given by:
\begin{equation}
\label{eq:kfirm_deposit_ratio}
    m_K=e_K.
\end{equation}
Hence, the K-firm sector is fully defined by the profit equation, substituting $m_K=e_K=\pi_K/g_N$ gives:
\begin{equation}
\label{eq:kfirm_profit_share}
    \pi_K=\frac{g_N\left(s_K(1-\delta\xi)-\omega s_K\right)}{g_N-r_M},
\end{equation}
given the wage share $\omega$ defined in by household sector. 

\subsection{Consumption firm sector in balanced}
In the C-firm sector under balanced growth $Y_C(t)=Q_C(t)=Z_C(t)=Z^d(t)=Y^d(t)=C$, that is output is equal to quantity sold, demand, desired demand, desired production, and household consumption, and desired debt is equal to actual debt $D^d(t)=D$. From the C-firm profit equation, Eq.~\ref{eq:cfirm_profits}, solving for the wage share yields:
\begin{equation}
    \pi_C=s_C+r_Mm_C-\omega s_C - r_Md.
\end{equation}
From the desired debt ratio equation, Eq.~\ref{eq:cfirm_desired_debt_ratio}, the debt ratio in balanced growth is give by:
\begin{equation}
\label{eq:debt_ratio}
    d=d_0+d_1g+d_2\frac{\pi_C}{s_C},
\end{equation}
where $\pi_C/s_C=\Pi_C(t)/P(t)Y_C(t)$ is the profit share in relation to C-firm output. 

In balanced growth, C-firm equity from Eq.~\ref{eq:cfirm_equity} grows at the nominal growth rate $g_N$, hence:
\begin{equation}
    \Delta E_C(t)=\Pi_C(t)=g_NE_C(t).
\end{equation}
Therefore, the equity ratio is given by:
\begin{equation}
\label{eq:cfirm_equity_ratio}
    e_C=\frac{\pi_C}{g_N}.
\end{equation}

From the accounting identity defined in Eq.~\ref{eq:cfirm_balance_sheet}, C-firms deposit ratio is given by:
\begin{equation}
\label{eq:cfirm_deposit_ratio}
    m_C=d+e_C-vs_C.
\end{equation}

Therefore, the system of equations that defines the consumption sector under balanced growth is given by:
\begin{equation}
    \begin{aligned}
        \pi_C&=s_C+r_Mm_C-\omega s_C - r_Md\\
        d&=d_0 + d_1g + d_2\frac{\pi_C}{s_C}\\
        m_C&=d+\frac{\pi_C}{g_N}-vs_C.
    \end{aligned}
\end{equation}
Then, after substituting $d$ into $m_C$, and $d$ and $m_C$ into $\pi_C$, the C-firms profit share as a function of the models parameters can be expressed as:
\begin{equation}
\label{eq:cfirm_profit_share}
    \pi_C = \frac{g_Ns_C(s_C(1 - \omega - r_M\nu) - (r_L - r_M)(d_0 + d_1g))}{s_C(g_N - r_M) + g_Nd_2(r_L - r_M)},
\end{equation}
from which $d$, $m_C$, and $e_C$ can easily be derived. 

\subsection{Bank sector in balanced growth}
In balanced growth, the bank profit share is derived from Eq.~\ref{eq:bank_profits}:
\begin{equation}
\label{eq:bank_profit_share}
    \pi_B=r_Ld-r_Mm,
\end{equation}
where $r_L=g_P+r_N$ is the nominal rate of interest, and $m=m_C+m_K+m_H$ is the total deposit share of nominal GDP. Furthermore, using Eq.~\ref{eq:bank_equity} and Eq.~\ref{eq:bank_desired_credit_ratio}, the bank equity ratio is given by:
\begin{equation}
\label{eq:bank_equity_ratio}
    e_B=\frac{\pi_B}{g_N}.
\end{equation}

\section{Parameters \& Initial Values}
\label{app:parameters_and_initial_values}

\subsection{Parameters}
\label{app:parameters}

Several parameters were chosen based on estimated values from real data, such as the average growth rate of labour productivity, where values were taken from the OECD, as in \cite{barrett2018stability}, see data from: \url{https://data.oecd.org/lprdty/labour-productivity-and-utilisation.htm\#indicator-chart}. 2\% was a typical value for our $g$ parameter, during the economically stable period from 1981-2006. Bank desired capital ratio (minimum capital ratio) was taken to reflect that recommended in the Basel III regulatory framework, which states that the minimum tier 1 capital ratio must be above 6\%, see \url{https://www.bis.org/fsi/fsisummaries/defcap_b3.htm}. Capital acceleration and the depreciation of capital were taken from \cite{jackson2015growth_imperative}. Other free parameter values were chosen such that the model did not display degenerate dynamics, specifically no hyperinflation, collapse of GDP to zero, sustained unemployment rate over 50\%, or sustained bankruptcy rates over 50\%.

\begin{table}[!htb]
    \centering
    \footnotesize
    \begin{tabular}{llll}
        \toprule
        \textbf{Symbol} & \textbf{Description} & \textbf{Value} & \textbf{Calibration}\\
        \midrule
        $T$ & Number of years & 100 & Free\\
        $\Delta t$ & Time step delta & $1/4$ & Free\\
        $\mathbf{N}_H$ & Number of households & 5000 & Free\\
        $\mathbf{N}_C$ & Number of C-firms & 400 & Free\\
        $\mathbf{N}_K$ & Number of K-firms & 100 & Free\\
        $\mathbf{N}_B$ & Number of banks & 20 & Free\\
        $n_C$ & Number of C-firms visited by households & 2 & Free\\
        $n_K$ & Number of K-firms visited by C-firms & 2 & Free\\
        $n_F$ & Number of job applications & 4 & Free\\ 
        $n_B$ & Number of banks visited by firms & 2 & Free\\
        $g$ & Average growth rate of labour productivity & $0.02\times\Delta t$ & OECD\\
        $\sigma_\alpha$ & Standard deviation of productivity growth & $0.03\times\sqrt{\Delta t}$ & Free\\
        $\sigma_P$ & Standard deviation of prices & $0.03\times\sqrt{\Delta t}$ & Free\\
        $\sigma_w$ & Standard deviation of wages & $0.03\times\sqrt{\Delta t}$ & Free\\
        $\sigma_r$ & Standard deviation of interest rates & $0.03\times\sqrt{\Delta t}$ & Free\\
        $\gamma_Z$ & Demand speed of adjustment & $0.1\times\Delta t$ & Free\\
        $\gamma_P$ & Price speed of adjustment & $0.1\times\Delta t$ & Free\\
        $\gamma_w$ & Wage speed of adjustment & $0.1\times\Delta t$ & Free\\
        $\gamma_r$ & Interest rate speed of adjustment & $0.1\times\Delta t$ & Free\\
        $c_Y$ & Household marginal propensity to consume income & $0.8$ & Free\\
        $c_M$ & Household marginal propensity to consume deposits & $0.1$ & Free\\
        $\nu$ & C-firm capital accelerator & 3 & \cite{jackson2015growth_imperative}\\
        $d_0$ & C-firm desired debt ratio intercept & 0.5 & \cite{barrett2018stability}\\
        $d_1$ & C-firm desired debt ratio productivity growth & 3 & \cite{barrett2018stability}\\
        $d_2$ & C-firm desired debt ratio profit share & 2 & \cite{barrett2018stability}\\
        $\delta$ & Depreciation rate of capital & $0.07\times\Delta t$ & \cite{jackson2015growth_imperative}\\
        $\xi$ & K-firm desired excess capacity & 0.1 & Free\\
        $n$ & Bank loan repayment periods & $10/\Delta t$ & Free\\
        $\kappa$ & Bank regulatory capital ratio & 0.06 & Basel III\\
        $r_M$ & Bank nominal interest rate on deposits & $0.001$ & Free\\
        $r_N$ & Bank natural real interest rate on loans & $0.02$ & Free\\
        $a_0$ & Initial labour productivity for C-firms and K-firms & $1$ & Free\\
        $P_0$ & Initial price for C-firms and K-firms & $1$ & Free\\
        \bottomrule
    \end{tabular}
\caption{Model parameters for baseline scenario.}
\label{tab:parameters}
\end{table}

\subsection{Initial values}
\label{app:initial_values}

To initialise the PG-DYNAMIN model we use the equilibrium values obtained from the macroeconomic balanced growth model, and where initial variables are needed for multiple agents, they are all initialised to the same value.

\subsubsection{Macroeconomic initial values}

Assuming full employment, initial real GDP is given by:
\begin{equation}
\label{eq:initial_real_gdp}
    Y(0)=a_0\mathbf{N}_H,
\end{equation}
where $a_0$ and $\mathbf{N}_H$ are both defined in Table \ref{tab:parameters}. Moreover, using the initial prices of C-firms and K-firms from Table \ref{tab:parameters}, $P_0$, initial nominal GDP is given by:
\begin{equation}
\label{eq:initial_nominal_gdp}
    Y_N(0)=P_0Y(0).
\end{equation}

\subsubsection{Household initial values}
Household initial deposits can be found by using initial nominal GDP (Eq.~\ref{eq:initial_nominal_gdp}) and the household deposit to nominal GDP ratio $m_H$ from the balanced growth model (Eq.~\ref{eq:household_deposit_ratio}). Household initial wages are derived from Eq.~\ref{eq:initial_wages} assuming full employment, and household initial income is then found using Eq.~\ref{eq:household_income}. Therefore, household $h$'s initial values are:
\begin{equation}
    \begin{aligned}
        M_h(0)&=\frac{m_HY_N(0)}{\mathbf{N}_H}\\
        w_h(0)&=\omega a_0 P_0\\
        Y_h(0)&=c_Y\left(w_h(0) + r_MM_h(0)\right).
    \end{aligned}
    \qquad \forall h \in\{1,\dots,\mathbf{N}_H\}
\end{equation}

\subsubsection{Capital Firm Initial Values}
The initial value for K-firm wages is derived from the wage share equation Eq.~\ref{eq:wage_share} and its relation to initial wages (Eq.~\ref{eq:initial_wages}). The initial value for K-firm equity and deposits is found using initial nominal GDP (Eq.~\ref{eq:initial_nominal_gdp}) and the balanced growth model K-firm profit share (Eq.~\ref{eq:kfirm_profit_share}), then making the substitutions into Eq.~\ref{eq:kfirm_equity_ratio} for equity and Eq.~\ref{eq:kfirm_deposit_ratio} for deposits. Initial output ($Y_j(0)$) is given by the full employment condition, where initial expected demand ($Z^e_j(0)$) and desired output ($Y^d_j(0)$) are equal to initial output. K-firm initial employees are derived from Eq.~\ref{eq:kfirm_output}. Therefore, the initial values for K-firm $j$ are given by:
\begin{equation}
    \begin{aligned}
        Y_j(0)&=\frac{s_KY(0)}{\mathbf{N}_K}\\
        Z^e_j(0)&=Y_j(0)\\
        Y^d_j(0)&=Y_j(0)\\
        N_j(0)&=\frac{Y_j(0)}{a_0}\\
        N^d_j(0)&=N_j(0)\\
        w_j(0)&=\omega a_0 P_0\\
        E_j(0)&=\frac{\pi_KY_N(0)}{g_N\mathbf{N}_K}\\
        M_j(0)&=E_j(0).
    \end{aligned}
    \qquad \forall j \in \{1,\dots,\mathbf{N}_K\}
\end{equation}

\subsubsection{Consumption Firm Initial Values}
Similarly to K-firms, C-firms initial wages are derived from the balanced growth model using the wage share (Eq.~\ref{eq:wage_share}) and its relation to initial wages (Eq.~\ref{eq:initial_wages}). The initial value for C-firm debt, equity, and deposits are found using initial nominal GDP (Eq.~\ref{eq:initial_nominal_gdp}) and the balanced growth values for C-firms profits share (Eq.~\ref{eq:cfirm_profit_share}), C-firms debt ratio (Eq.~\ref{eq:debt_ratio}), C-firms equity ratio (Eq.~\ref{eq:cfirm_equity_ratio}), and C-firms deposit ratio (Eq.~\ref{eq:cfirm_deposit_ratio}). Initial output ($Y_i(0)$) is given by the full employment condition, where initial expected demand ($Z^e_i(0)$) and desired output ($Y^d_i(0)$) are equal to initial output. The initial capital stock ($K_i(0)$) is derived from Eq.~\ref{eq:cfirm_output} assuming full capacity utilisation. The initial values for all C-firms is then equally distributed across agents, therefore, the initial values for C-firm $i$ are given by:
\begin{equation}
    \begin{aligned}
        Y_i(0)&=\frac{s_CY(0)}{\mathbf{N}_C}\\
        Z^e_i(0)&=Y_i(0)\\
        Y^d_i(0)&=Y_i(0)\\
        K_i(0)&=\nu Y_i(0)\\
        N_i(0)&=\frac{Y_i(0)}{a_0}\\
        N^d_i(0)&=N_i(0)\\
        w_i(0)&=\omega a_0 P_0\\
        D_i(0)&=\frac{dY_N(0)}{\mathbf{N}_C}\\
        E_i(0)&=\frac{\pi_CY_N(0)}{g_N\mathbf{N}_C}\\
        M_i(0)&=\left(d+\frac{\pi_C}{g_N}-vs_C\right)\frac{Y_N(0)}{\mathbf{N}_C}.
    \end{aligned}
    \qquad \forall i \in \{1,\dots,\mathbf{N}_C\}
\end{equation}

\subsubsection{Bank Initial Values}
To initialise each bank $b$, households and firms are randomly allocated a bank where they hold their deposits for the entire simulation. Additionally, firms are randomly allocated a bank where they initially take out their total initial debt as a loan. Banks' initial interest rate is set consistent with Eq.~\ref{eq:bank_desire_interest_rate}, such that initial nominal interest rate on loans is given by the balanced growth inflation $g_P$ and the real natural rate $r_N$. Additionally, total loans held by bank $b$ is easily derived as the sum of debts of C-firms that are allocated to bank $b$, and similarly deposits are given by the sum of firm and household deposits that were allocated to bank $b$. Equity can be derived from Eq.~\ref{eq:bank_equity_ratio}, and finally bank reserves from Eq.~\ref{eq:bank_reserves}. Therefore, bank $b$'s initial values are:
\begin{equation}
    \begin{aligned}
        r^L_b(0)&=g_P+r_N\\
        L_b(0)&=\sum_{i\in\mathcal{F}^L_b(0)}D_i(0)\\
        M_b(0)&=\sum_{\iota\in\mathcal{F}^M_b}M_\iota(0)+\sum_{h\in\mathcal{H}_b}M_h(0)\\
        \Pi_b(0)&=r^L_b(0)L_b(0) - r_MM_b(0)\\
        E_b(0)&=\frac{\Pi_b(0)}{g_N}\\
        R_b(0)&=M_b(0) + E_b(0) - L_b(0),
    \end{aligned}
    \qquad \forall b \in \{1,\dots,\mathbf{N}_B\}
\end{equation}
where $\mathcal{F}^M_b$ and $\mathcal{H}_b$ are the set of firms (C-firms and K-firms) and households, respectively, initially allocated to bank $b$ to hold their deposits throughout the simulation, and $\mathcal{F}^L_b(0)$ is the set of C-firms initially allocated at bank $b$ to supply the loan according to C-firm $i$'s initial debt $D_i(0)$.

\section{Stylised Facts}
\label{app:stylised_facts}

Here a further analysis of the stylised facts replicated by the baseline Growth S1 scenario of the PG-DYNAMIN model is presented. We show analyses for both the randomly selected simulation ($s=5$) and across all 100 simulations. 

\subsection{Macroeconomic Stylised Facts}

First we demonstrate that the baseline PG-DYNAMIN model can reproduce a range of macroeconomic stylised facts that have been observed empirically on real data.

\subsubsection{Minskyan debt cycles}
\label{app:stylised_fact_minskyan}

\begin{figure}[!htb]
    \centering
    \includegraphics[width=0.9\textwidth]{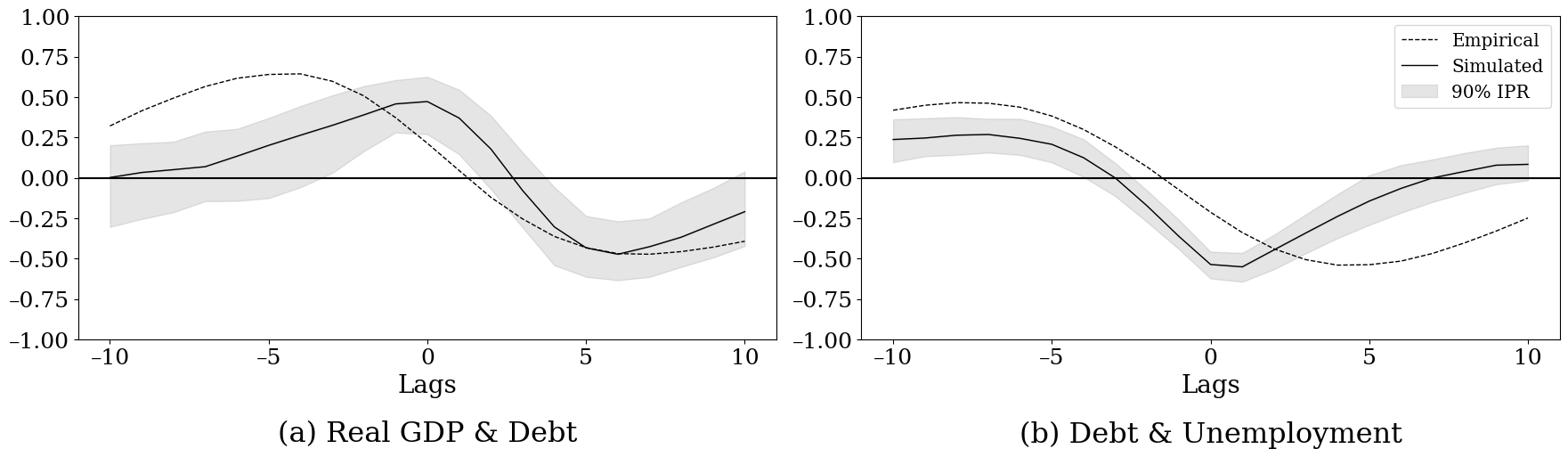}
    \caption{Cross correlation of the baseline Growth S1 scenario and empirical time-series. Panel (a) shows the cross correlation of real GDP in time $t$ with corporate debt in time $t\pm \text{lag}$. Panel (b) shows the cross correlation of corporate debt in time $t$ with the unemployment rate in time $t\pm\text{lag}$. Lags are a quarter of a year, the dashed line is the empirical cross-correlation, the solid line is the median cross-correlation across all simulation, and the shaded area is the 90\% inter-percentile range(IPR) across all simulations.}
    \label{figure_app_1}
\end{figure}

Minskyan debt cycles have recently been empirically validated by \cite{stockhammer2023debt_cycles}. Therefore, we thought it is essential when conducting a stability comparison between growth and zero-growth on the same model, that the baseline growth scenario is able to produce these at the macroeconomic level as an emergent property from the interaction of the microeconomic agents. Fig. \ref{figure_app_1} shows, at different lags, the cross-correlation between real GDP and corporate debt in Panel (a), and corporate debt and the unemployment rate in Panel (b), for the cyclical components of Hodrick-Prescott (HP) filtered time-series\footnote{We use a value of $\lambda=1600$ for the HP filter}. In can be noted that debt is pro-cyclical to real GDP and counter-cyclical to unemployment. Hence, a build up of debt precedes macroeconomic fluctuations in the business cycle.

\begin{figure}[!htb]
    \centering
    \includegraphics[width=0.9\textwidth]{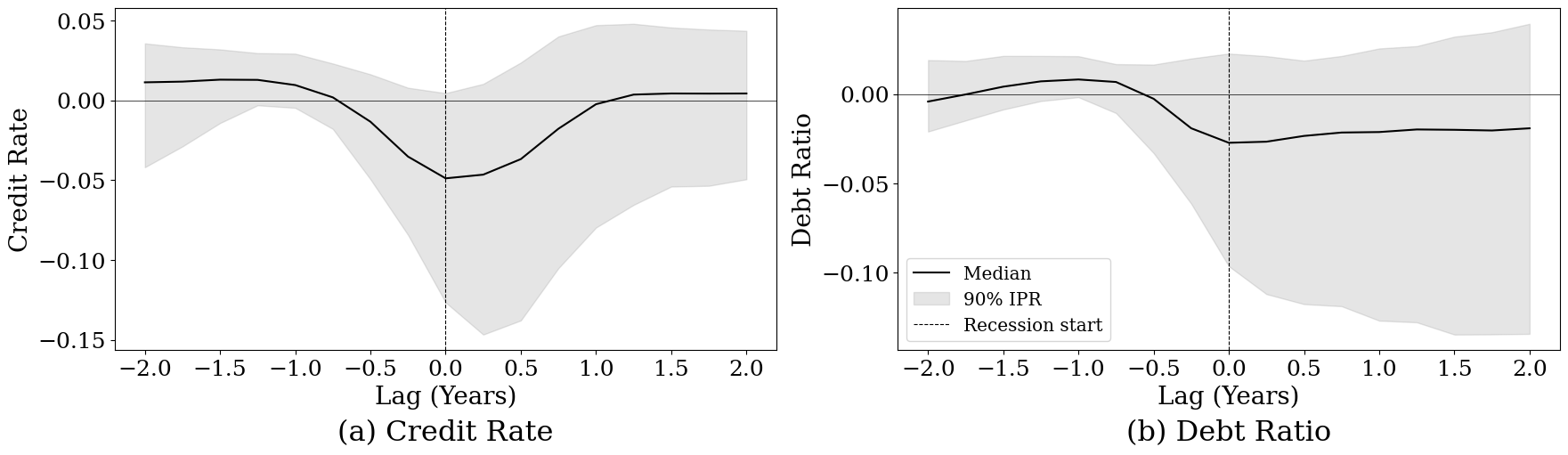}
    \caption{Behaviour of the debt ratio and credit rate around recessions. Black line is the median, the shaded area is the 90\% inter-percentile range (IPR), and the vertical dotted line marks the recession start.}
    \label{figure_app_2}
\end{figure}

Furthermore, we show in Fig.~\ref{figure_app_2} the growth rate of the debt to real GDP ratio and the credit rate (change in debt over nominal GDP) around the start of recessions in the model. Minsky's insights imply that these should be positive before a recession and then turn negative during a recession, as seen in Fig.~\ref{figure_app_2} this dynamic occurs on average. Additionally, we test the significance of a peak before a recession in both the debt ratio and credit rate growth rates and find a significant value above the 1\% significance level. Furthermore, for 71.35\% of recessions the debt ratio peaks before the recession, and for 53.92\% of recessions the credit rate peaks before the recession. Therefore, given the above evidence, it can be concluded that the model reproduces Minskyan debt cycles. 

\subsubsection{Debt deflation}
\label{app:stylised_fact_debt_deflation}
\begin{figure}[!htb]
    \centering
    \includegraphics[width=0.6\textwidth]{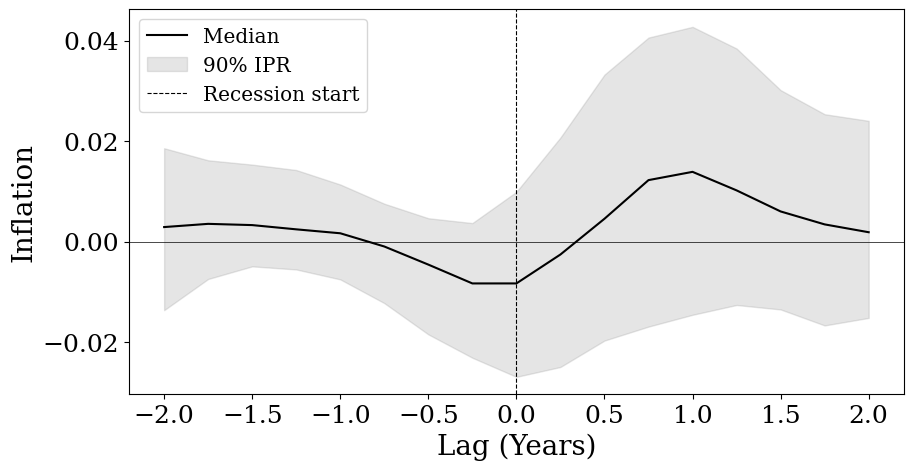}
    \caption{Behaviour of inflation around recessions. Black line is the median, the shaded area is the 90\% inter-percentile range (IPR), and the vertical dotted line marks the recession start.}
    \label{figure_app_3}
\end{figure}

Using a similar methodology to show the Minskyan debt cycles, we demonstrate that the model reproduces Fisher's debt-deflation dynamic \citep{fisher1933debt_deflation}. Given that the debt ratio peaks before a recession, as evidenced in Fig.~\ref{figure_app_2}, we next show in Fig.~\ref{figure_app_3} that there is, on average, a period of deflation during recessions in the model. Additionally, 28.65\% of recessions have a debt-deflation mechanism, a peak in the debt ratio before the recession and a period of deflation during the recession.

\subsubsection{GDP growth distribution}
\label{app:stylised_fact_gdp_growth}
\begin{figure}[!htb]
    \centering
    \includegraphics[width=0.6\textwidth]{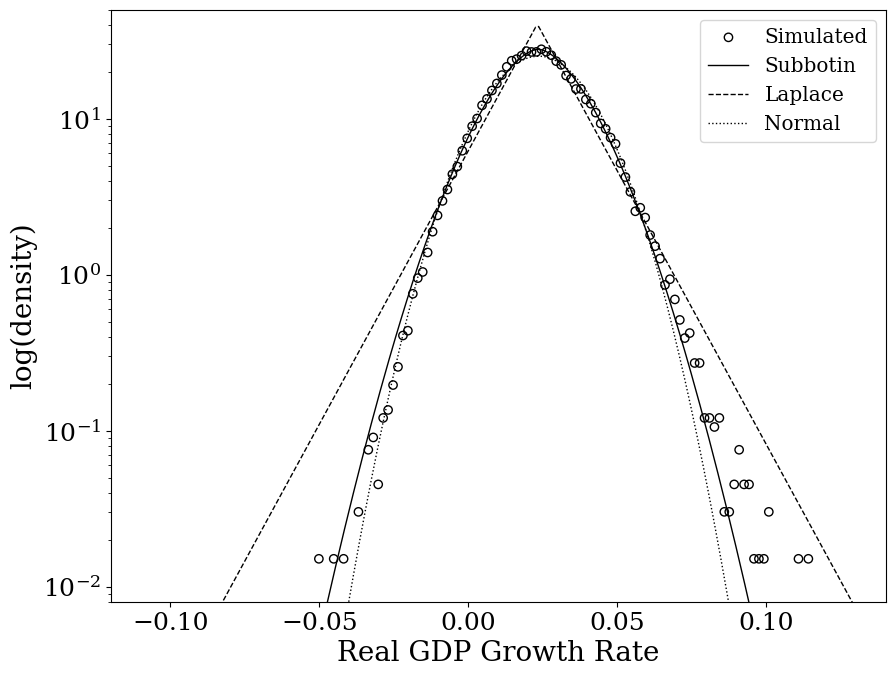}
    \caption{Distribution of real GDP growth over all simulations for the baseline Growth S1 scenario (100 bins with 40,000 observations). Solid line is the Subbotin (exponential power) fit, dashed line is the Laplace distribution fit, dotted line is the normal distribution fit.}
    \label{figure_app_4}
\end{figure}

In line with empirical data on the distribution of real GDP growth rates \citep{fagiolo2008output_distributions,castaldi2009patterns}, Fig.~\ref{figure_app_4} shows in a log-linear plot that real GDP growth rates of the PG-DYNAMIN model has fat-tails larger than a normal distribution and is best characterised by a Subbotin (exponential power) distribution. Additionally, each statistical test of normality is rejected at the 1\% significance level in \ref{app:stat_tests}.

\subsubsection{Recession duration}
\label{app:stylised_fact_recessions}
\begin{figure}[!htb]
    \centering
    \includegraphics[width=0.9\textwidth]{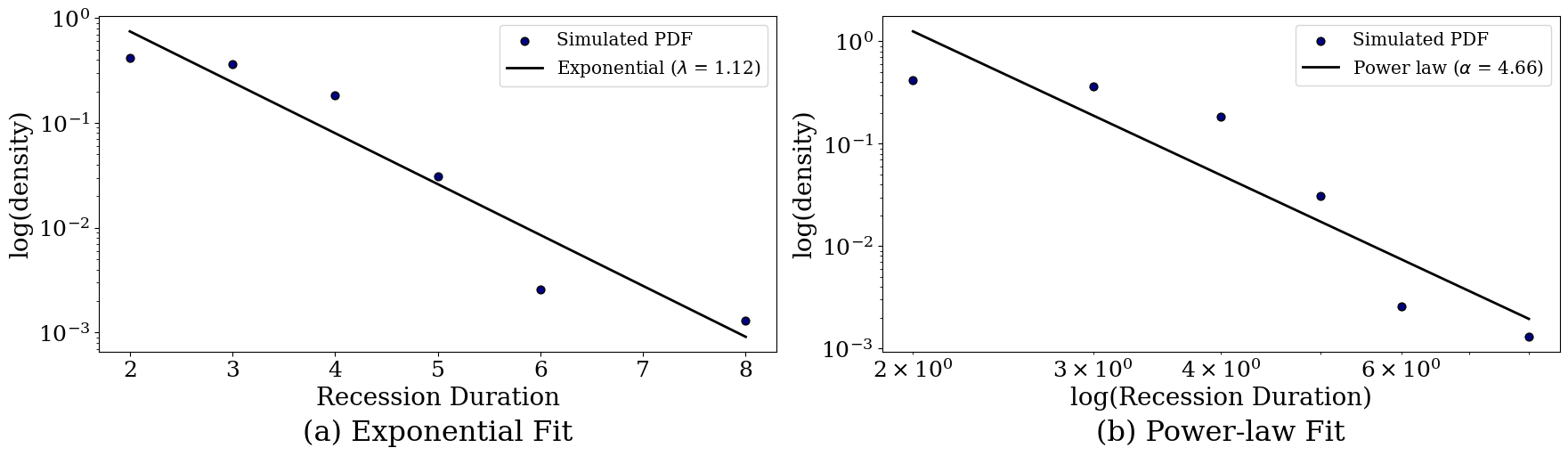}
    \caption{Panel (a) shows the exponential fit to the distribution of recession durations over all simulations for the baseline Growth S1 scenario. Panel (b) shows the power-law fit to the distribution of recession durations over all simulations for the baseline Growth S1 scenario.}
    \label{figure_app_5}
\end{figure}

The duration of recessions has been well documented as having a large tail, and being best fit by an exponential distribution \citep{ausloos2004recession_durations,wright2005duration}. We compared fitting exponential and power-law distributions and found, as seen in Fig.~\ref{figure_app_5}, that according to a comparison of the $R^2$ and root mean squared error (RMSE), the exponential distribution was the better fit. The exponential distribution produced $R^2=0.913$ and $\text{RMSE}=0.681$, whereas for the power-law distribution, $R^2=0.843$ and $\text{RMSE}=0.913$.

\subsubsection{Autocorrelation of macroeconomic variables}
\label{app:stylised_fact_acf}
\begin{figure}[!htb]
    \centering
    \includegraphics[width=0.9\textwidth]{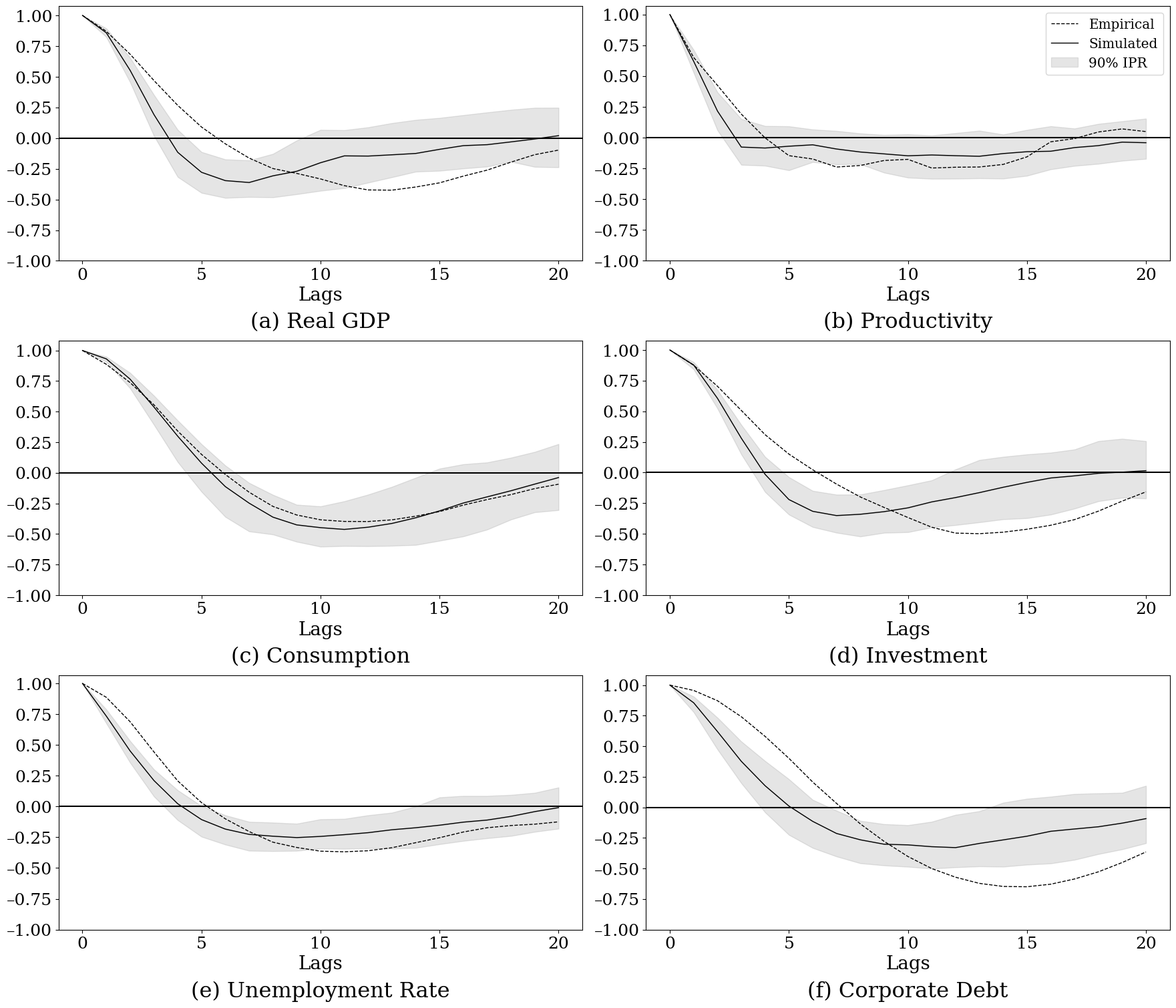}
    \caption{Autocorrelation of the baseline Growth S1 scenario and empirical time-series. The dashed line is the empirical autocorrelation, the solid line is the median autocorrelation across all simulations, and the shaded area is the 90\% inter-percentile range (IPR) across all simulations.}
    \label{figure_app_6}
\end{figure}

In line with the stylised facts of business cycles from \cite{fiorito1994stylized}, Fig.~\ref{figure_app_6} shows that the autocorrelation of the cyclical component of the Hodrick-Prescott filtered time-series for real GDP, aggregate productivity, real consumption, real investment, the unemployment rate, and corporate debt display similarly levels to those found in empirical US time-series data. 

\subsubsection{Cross-correlation of macroeconomic variables}
\label{app:stylised_fact_ccf}
\begin{figure}[!htb]
    \centering
    \includegraphics[width=0.9\textwidth]{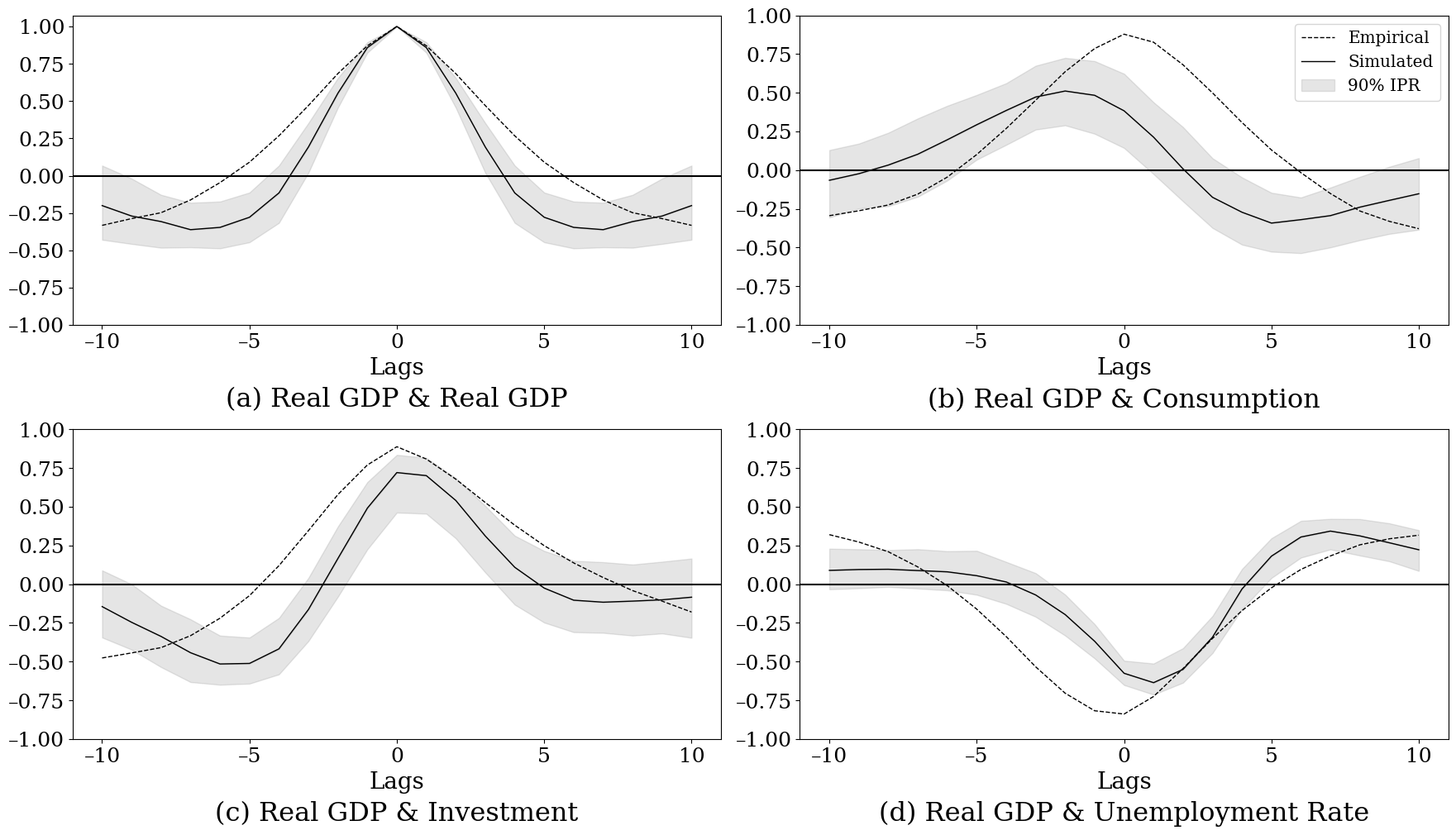}
    \caption{Cross correlation of the baseline Growth S1 scenario and empirical time-series. The dashed line is the empirical cross-correlation, the solid line is the median cross-correlation across all simulation, and the shaded area is the 90\% inter-percentile range (IPR) across all simulations.}
    \label{figure_app_7}
\end{figure}

In line with business cycle evidence from \cite{stock1999business}, in Fig.~\ref{figure_app_7} we show that the cross-correlation of the cyclical component of the Hodrick-Prescott filtered time-series for real GDP and real GDP, real GDP and consumption, real GDP and real investment, and real GDP and the unemployment rate have similar levels to those found in empirical US time-series data. 

\subsubsection{Volatility hierarchy of macro growth rates}
\label{app:stylised_fact_volatility_hierarchy}
Evidence from \cite{stock1999business} suggest that the growth rate of real investment is more volatile than the growth rate of real GDP, which is in turn more volatile than the growth rate of real consumption. In the model, we demonstrate that this hierarchy of growth rate volatilities is replicated. We find that the average standard deviation (std. dev.) of real investment growth across simulations is 5.51\% (0.46\% std. dev. across simulations), the average standard deviation of real GDP growth rates is 1.58\% (0.15\% std. dev. across simulations), and the average standard deviation of real consumption is 1.51\% (0.18\% std. dev. across simulations). Hence, on average the volatility hierarchy is replicated, however, there is not a significant difference between real GDP and real consumption growth volatilities. 

\subsubsection{Relationships between macroeconomic variables}
\label{app:stylised_fact_relationships}
\begin{figure}[!htb]
    \centering
    \includegraphics[width=0.9\textwidth]{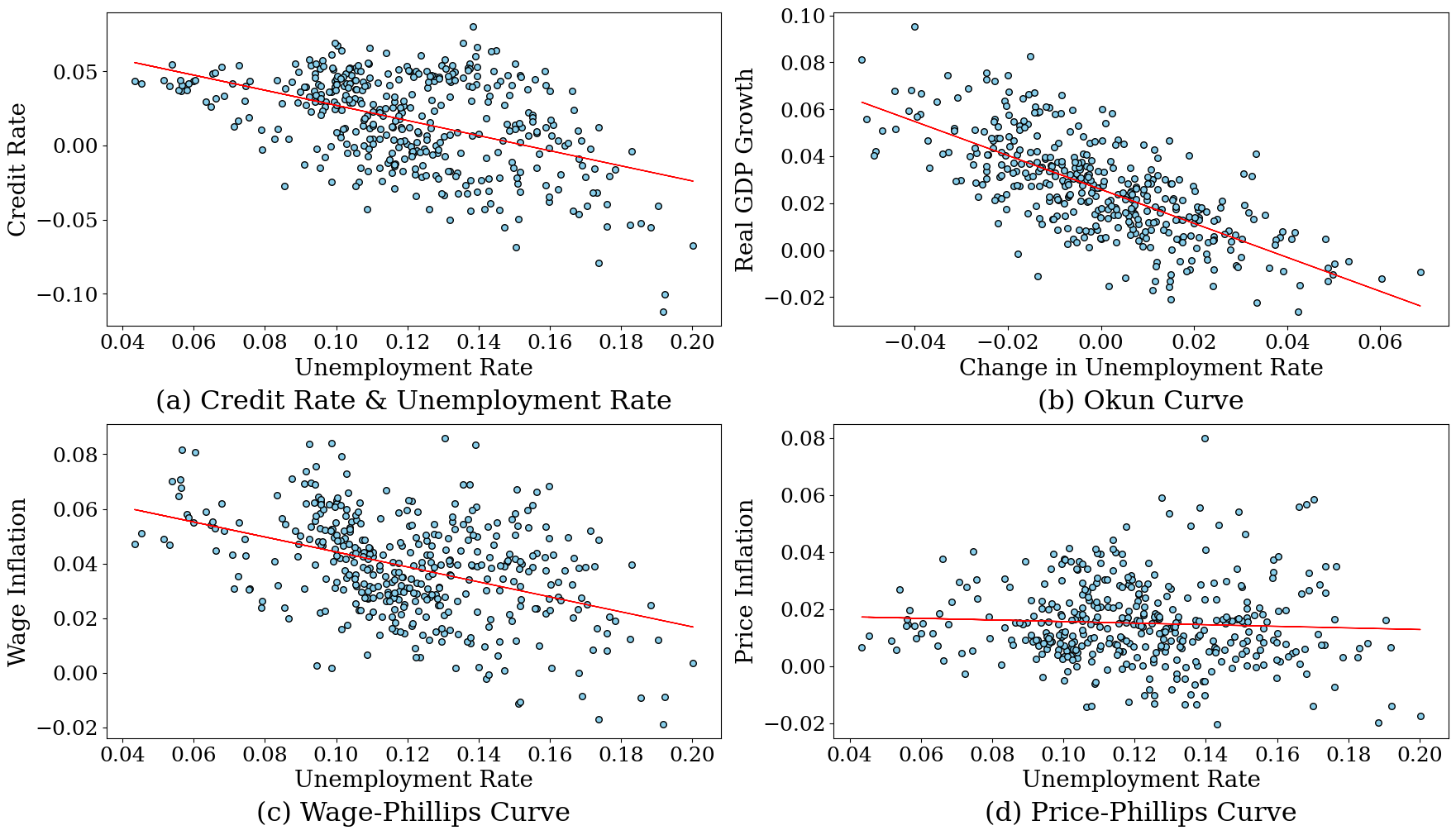}
    \caption{Stylised facts of the relationship between key economic variables for a typical Growth S1 run ($s=5$). Panel (a) shows the relationship between the credit rate and the unemployment rate. Panel (b) shows the Okun curve. Panel (c) shows the wage-related Phillips curve. Finally, Panel (d) shows the price-related Phillips curve.}
    \label{figure_app_8}
\end{figure}

We demonstrate in Fig.~\ref{figure_app_8} that a randomly chosen simulation ($s=5$) replicates some of the relationships between macroeconomic variables seen in real world data; similar relationships were also found for other randomly chosen simulations. Panel (a) shows the relationship between the credit rate (change in debt over real GDP) and the unemployment rate as posited by \cite{keen2014endogenous}, with intercept $=0.0782$, slope $=0.5101$, and $R^2=0.2165$. Panel (b) shows a strong Okun curve \citep{okun1962potential_gnp}, with intercept $=0.0259$, slope $=-0.7236$, and $R^2=0.4863$. Panel (c) shows the wage-Phillips curve being stronger, as originally analysed by \cite{phillips1958relation}, with intercept = $=0.0717$, slope = $-0.2742$, and $R^2=0.1805$. Finally, Panel (d) shows a weak inflation-Phillips curve echoing empirical results in \cite{stock2020slack}, with intercept $=0.0184$, slope $=-0.0280$, and $R^2=0.0032$.

\subsection{Microeconomic stylised facts}

We also show that along with the macroeconomic stylised facts, the PG-DYNAMIN model is able to replicate some well known microeconomic empirical regularities. 

\subsubsection{Distribution of firm growth rates}
\label{app:stylised_fact_firm_growth}
\begin{figure}[!htb]
    \centering
    \includegraphics[width=0.9\textwidth]{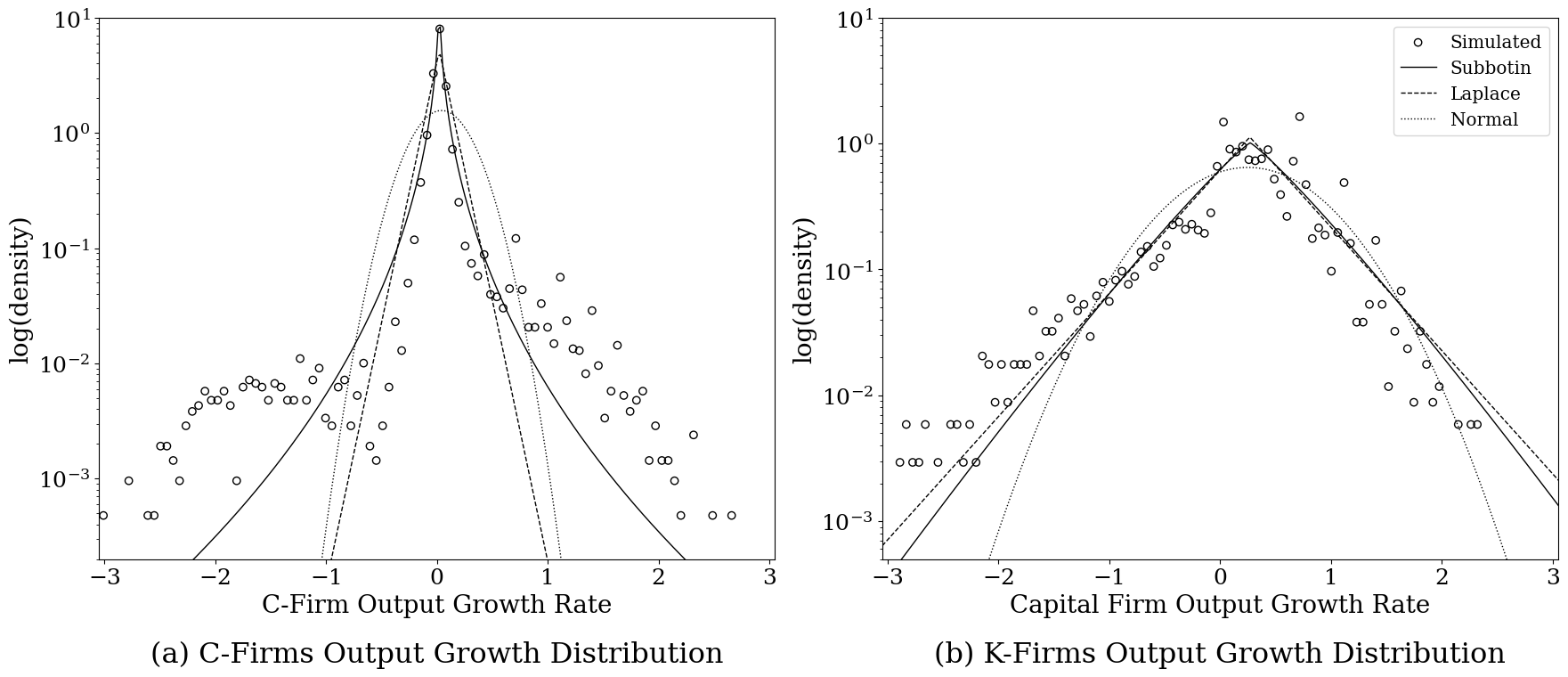}
    \caption{Distribution of firm output growth rates in log-linear plots. Solid line is the Subbotin (exponential power) fit, dashed line is the Laplace distribution fit, dotted line is the normal distribution fit. Panel (a) shows the distribution of C-firm output growth rates over all simulations for the baseline Growth S1 scenario, taken at a midpoint snapshot in quarter 1 of year 50, $t=600$ and $s\in\{1,2,...100\}$ (100 bins with 13,984 observations). Panel (b) shows the distribution of K-firm output growth rates over all simulations of the baseline Growth S1 scenario, taken at a midpoint snapshot in quarter 1 of year 50, $t=600$ and $s\in\{1,2,...100\}$ (100 bins with 2,161 observations).}
    \label{figure_app_9}
\end{figure}

It has been well studied that the output growth rates of firms tends to follow tent-shaped distributions with large tails \citep{bottazzi2003common,bottazzi2006firm_distributions}. Fig.~\ref{figure_app_9} shows the distribution of C-firm and K-firm output growth rates. It can be noted from both distributions that they replicate the tent shape, and this is particularly evident for K-firms. The distribution for C-firms has a larger tail and a hump in the left tail, likely due to the inclusion of small C-firms, which are usually excluded in empirical studies. Statistical tests of normality for both C-firm and K-firm output growth rate distributions are rejected at the 1\% significance level, see \ref{app:stat_tests}. 

\subsubsection{Distribution of firm size}
\label{app:stylised_fact_firm_size}
\begin{figure}[!htb]
    \centering
    \includegraphics[width=0.9\textwidth]{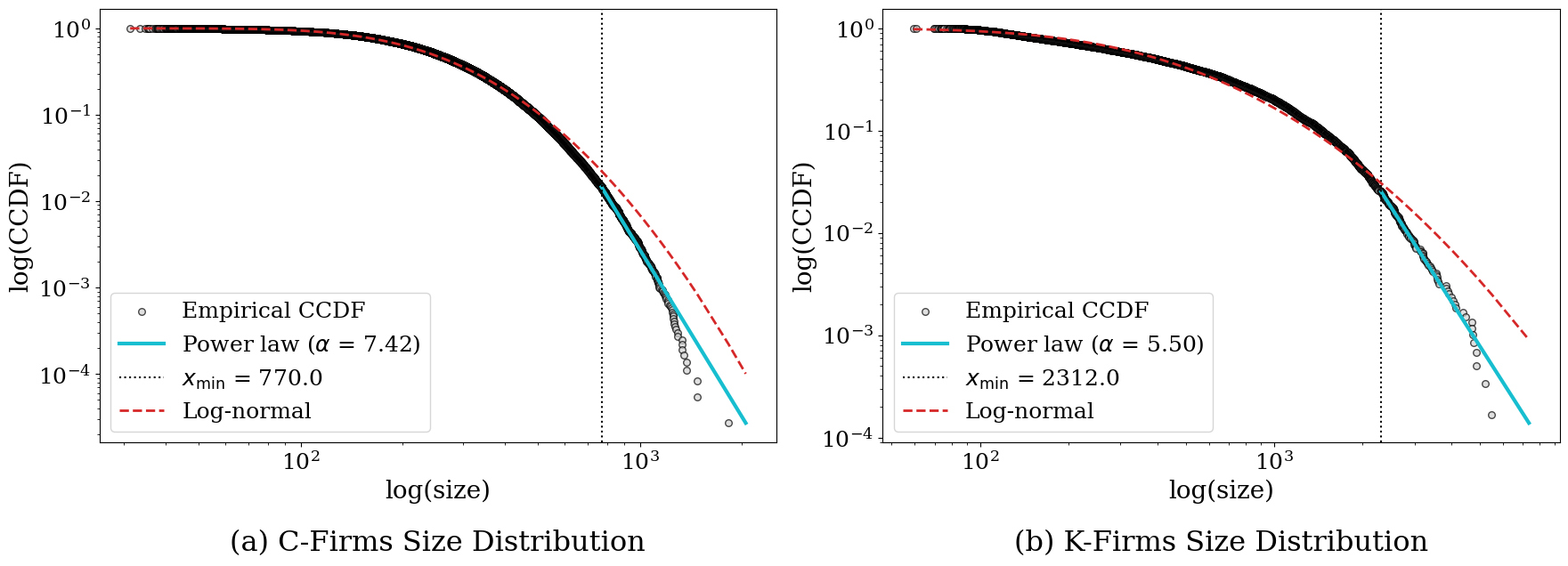}
    \caption{Distribution of firm size (real output) in log-log plots. Panel (a) shows the distribution of C-firm size and Panel (b) shows the distribution of K-firm size over all simulations for the baseline Growth S1 scenario, taken at a midpoint snapshot in quarter 1 of year 50, $t=600$ and $s\in\{1,2,...100\}$ (100 bins with 13,984 and 2,161 observations, respectively). Dots are the binned simulated distribution, solid blue line is the power-law fit, and dashed red line is the log-normal fit.}
    \label{figure_app_10}
\end{figure}

We show the distribution of firm size (measured as real output) for both C-firms and K-firms in Fig.~\ref{figure_app_10}. There has been some debate since \cite{gibrat1931inegalites} on the distribution of firm size, whether the distribution follows a log-normal or power-law/Zipf distribution \citep{axtell2001zipf}, see \cite{wit2005firm_size} for a discussion of firm size distributions and models that generate them. The main feature of the firm size distribution in the empirical literature is a right-skewed distribution with a large tail, which we show to be replicated in Fig.~\ref{figure_app_10}, where both the log-normal and power-law distributions fit the data well. Statistical tests of normality for both C-firm and K-firm size distributions are rejected at the 1\% significance level, see \ref{app:stat_tests}.

\subsubsection{Distribution of bank size}
\label{app:stylised_fact_bank_size}
\begin{figure}[!htb]
    \centering
    \includegraphics[width=0.6\textwidth]{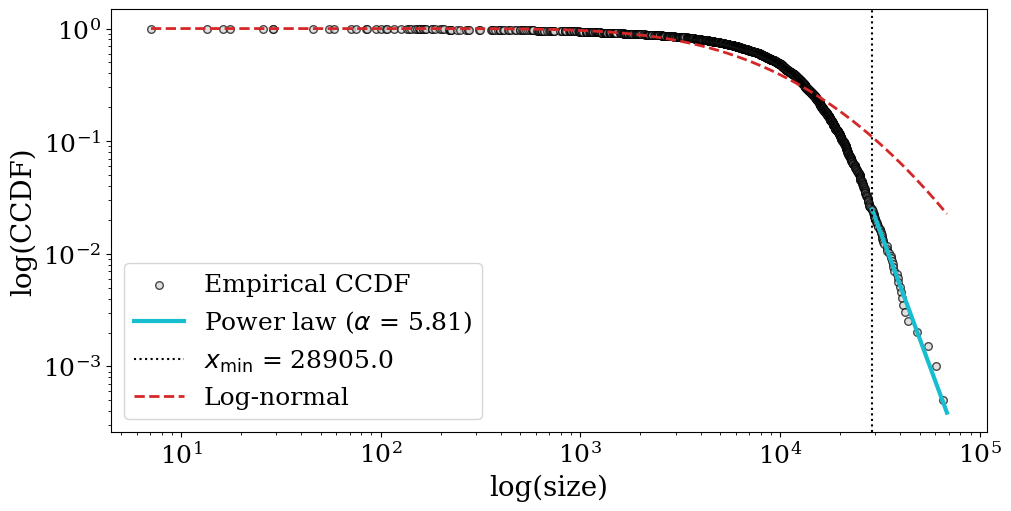}
    \caption{Distribution of bank size (real loans) in a log-log plot over all simulations for the baseline Growth S1 scenario, taken at a midpoint snapshot in quarter 1 of year 50, $t=600$ and $s\in\{1,2,...100\}$ (100 bins with 1,937 observations). Dots are the binned simulated distribution, the solid blue line is the power-law fit, and the dashed red line is the log-normal fit.}
    \label{figure_app_11}
\end{figure}

There is less empirical literature on the distribution of bank sizes compared to firm sizes. However, empirical studies such as \cite{ennis2001bank_size} show that the size distribution of banks is also right-skewed and has a large tail. It is shown in Fig.~\ref{figure_app_11} that the distribution of bank sizes in the model replicates these empirical findings. Statistical tests of normality for the size distribution of banks are rejected at the 1\% significance level, see \ref{app:stat_tests}.

\subsubsection{Investment lumpiness}
\label{app:stylised_fact_investment_lumpiness}
To check that the model reproduces lumpy investment rates, as observed by \cite{doms1998capital}, we tested for skew and kurtosis in their distribution. We found that the distribution of investment rates is indeed skewed at the 1\% significance level, and that there is excess kurtosis at the 1\% significance level.

\subsubsection{Statistical Tests}
\label{app:stat_tests}

Statistical tests of normality for each distributional variable discussed in Section \ref{sec:stylised_facts} - small p-values here imply the distributions are not normal. 

\begin{table}[!htb]
    \centering
    \begin{tabular}{lcccccc}
    \toprule
     \textbf{Variable} & \multicolumn{2}{c}{\textbf{KS Test}} &  \multicolumn{2}{c}{\textbf{SW Test}} & \multicolumn{2}{c}{\textbf{AD Test}}\\
     \cmidrule(l{10pt}r{5pt}){2-3}
     \cmidrule(l{10pt}r{5pt}){4-5}
     \cmidrule(l{10pt}r{5pt}){6-7}
     & stat. & $p$-value & stat. & $p$-value & stat. & $p$-value \\
    \midrule
    Real GDP Growth
        & 0.489 & 0.000 & 0.996 & 0.000 & 31.769 & 0.000 \\
    C-firm Output Growth
        & 0.411 & 0.000 & 0.490 & 0.000 & 5983.680 & 0.000 \\
    K-firm Output Growth
        & 0.289 & 0.000 & 0.944 & 0.000 & 83.878 & 0.000 \\
    C-firm Size 
        & 1.000 & 0.000 & 0.906 & 0.000 & 700.038 & 0.000 \\
    K-firm Size 
        & 1.000 & 0.000 & 0.771 & 0.000 & 369.609 & 0.000 \\
    Bank Size
        & 1.000 & 0.000 & 0.913 & 0.000 & 25.093 & 0.000 \\
    \bottomrule
    \end{tabular} 
\caption{Statistical tests of normality for baseline scenario (Growth S1). Column 2 shows the test statistic (stat.) and $p$-value for the Kolmogorov–Smirnov (KS) test, column 3 shows these for the Shapiro-Wilk (SW) test, and column 4 shows these for the Anderson-Darling (AD) test.}
\label{tab:norm_tests}
\end{table}

\section{Measures}
\subsection{CPI Inflation}
\label{app:cpi_inflation}
The consumer price index (CPI) is defined using a Paasche price index of C-firm prices weighted by output:
\begin{equation}
    CPI(t) = \frac{\sum^{\mathbf{N}_C}_{i=1}P_i(t)Y_i(t)}{\sum^{\mathbf{N}_C}_{i=1}P_i(0)Y_i(t)},
\end{equation}
where $\mathbf{N}_C$ is the total number of C-firms, $Y_i(t)$ is the output of C-firm $i$ in period $t$, $P_i(t)$ is the price of C-firm $i$ in period $t$, and $P_i(0)$ is the initial price of C-firm $i$. Then, CPI inflation, $g_P(t)$ is defined as the log difference of the CPI:
\begin{equation}
    g_P(t)=\ln(CPI(t)) - \ln(CPI(t-1)).
\end{equation}

\subsection{Gini Coefficient}
\label{app:gini}
The statistic for the Gini Coefficient, which measures household wealth inequality, is calculated using the method in \cite{dixon1987bootstrapping}, where, given an ordered list of household wealth, the Gini Coefficient at time $t$ can be defined as: 
\begin{equation}
    G(t) = \frac{1}{\mathbf{N}_H}\Bigg(\mathbf{N}_H+1 - 2\Bigg(\frac{\sum^{\mathbf{N}_H}_{h=1}(\mathbf{N}_H+1-h)M_{h}(t)}{\sum^{\mathbf{N}_H}_{h=1}M_{h}(t)}\Bigg)\Bigg),
\end{equation}
where $\mathbf{N}_H$ is the number of households and $M_{h}(t)$ is the wealth of household $h$ at time $t$.

\subsection{Crisis Probability \& Severity}
\label{app:crisis_measures}
To calculate the probability of a crisis during a given year, we must first compute a crisis indicator function per time period $t$ for a given simulation $s$:
\begin{equation}
    CR_s(t)=
    \begin{cases}
        1 & \text{if $g^Y_s(t)<-0.03$}\\
        0 & \text{otherwise},
    \end{cases}
\end{equation}
where $s\in\{1,...,S\}$ is the simulation, $g^Y_s(t)=\ln(Y_s(t))-\ln(Y_s(t-n))$ is real GDP growth in simulation $s$ in time period $t$, where $n=1/\Delta t$ is the number of periods in a year ($n=4$ for our parameter specification). Given that each time period is a quarter of a year, the yearly crisis indicator function is defined as:
\begin{equation}
    CR^y_s(t)=\max_{k\in\{0,...,n-1\}} CR_s(t+k).
\end{equation}
Then, the probability of a crisis for a given simulation $s$ is given by the average across the total number of time periods $T$:
\begin{equation}
    \Pr(CR^y_s(t)=1)=\frac{1}{T} \sum^T_{\tau=1} CR^y_s(\tau).
\end{equation}

To measure the severity of crises, we first define contiguous crisis spells. Let $\mathcal{T}_{s,k}$ denote the set of time periods belonging to crisis spell $k$ in simulation $s$, spells being defined as consecutive periods for which $CR^y_s(t)=1$. The severity of crisis spell $k$ in simulation $s$ is then given by the cumulative growth shortfall relative to the crisis threshold in crisis spell $k$:
\begin{equation}
    CS_{s,k} = \sum_{t \in \mathcal{T}_{s,k}} \left(-0.03 - g_s(t)\right).
\end{equation}
The average crisis severity in simulation $s$ is defined as the mean severity across all crisis spells:
\begin{equation}
    \overline{CS}_s = \frac{1}{K_s} \sum_{k=1}^{K_s} CS_{s,k},
\end{equation}
where $K_s$ is the number of crisis spells in simulation $s$.

\subsection{Hymer-Pashigian Instability Index (HPI)}
\label{app:HPI}
The Hymer-Pashigian Instability Index (HPI) is a measure of the instability of market shares by \citep{hymer1962turnover}: 
\begin{equation}
    HPI(t) = \sum^{\mathbf{N}_A}_{a=1} |ms_{a}(t) - ms_{a}(t-1)|,
\end{equation}
where $\mathbf{N}_A$ is the total number of agents in market $A\in\{C,K,B\}$ ($\mathbf{N}_C$ for the number of C-firms, $\mathbf{N}_K$ for the number of K-firms, and $\mathbf{N}_B$ for the number of banks) and $ms_{a}(t)$ is the market share of agent $a\in\{i,j,b\}$ (C-firms $a=i$, K-firms $a=j$, or banks $a=b$). HPI is defined on the interval $HPI(t)\in[0,1]$. When the HPI is close to 0, this indicates a relatively stable market, and when the HPI is close to 1, this indicates a relatively unstable market.

\subsection{Normalised Herfindahl-Hirschman Index (HHI)}
\label{app:HHI}
The Herfindahl-Hirschman Index (HHI) is defined in \citep{hirschman1945national} and independently developed by Orris C. Herfindahl in his unpublished doctoral thesis ``Concentration in the U.S. Steel Industry" (Columbia University, 1950): 
\begin{equation}
    HHI(t) = \sum^{\mathbf{N}_A}_{a=1} ms^2_{a}(t),
\end{equation}
where $\mathbf{N}_A$ is the total number of agents in market $A\in\{C,K,B\}$ and $ms_{a}(t)$ is the market share of agent $a\in\{i,j,b\}$. Furthermore, because $HHI(t)\in[1/\mathbf{N}_A,1]$, we normalise $HHI(t)$ such that $HHI^*(t)\in[0,1]$, given by:
\begin{equation}
    HHI^*(t) = \frac{HHI(t) - \frac{1}{\mathbf{N}_A}}{1-\frac{1}{\mathbf{N}_A}}.
\end{equation}
Thus, for $HHI^*(t)$ close to 0, this indicates a highly competitive market, and for $HHI^*(t)$ close to 1, this indicates a highly concentrated market.

\subsection{Expected Systemic Loss (ESL)}
\label{app:expected_systemic_loss}
We follow \cite{poledna2015systemic_risk} to construct a measure of expected systemic loss (ESL) in the bank-firm credit network. To construct such a measure we must first calculate the DebtRank on a bipartite graph between banks and firms. We follow the methodology in \cite{aoyama2013debtrank}. DebtRank gives a numeric value for the amount of economic value that could potentially be affected by a node in a network. DebtRank is a measure that is applied to the bank layer in a network, therefore we only investigate the effect that banks have on the network. 

An edge connecting bank $b$ and firm $\iota$ ($\iota=i$ for C-firms and $\iota=j$ for K-firms) in period $t$ is associated with the total lending (credit) of bank $b$ to firm $\iota$, defined as:
\begin{equation}
    C_{b\iota}(t) = \sum_{\ell\in\mathcal{L}_{\iota b}(t)}L_{\iota b,\ell}(t),
\end{equation}
where $L_{\iota b,\ell}(t)$ is the value of the $\ell$th loan that firm $\iota$ has outstanding with bank $b$, and $\mathcal{L}_{\iota b,\ell}(t)$ is the set of loans that firm $\iota$ has outstanding with bank $b$ in period $t$. Let the matrix $\mathbf{C}(t)$ represent the credit network between banks and firms in time $t$, where the elements are given by $(\mathbf{C}(t))_{b\iota}=C_{b\iota}(t)$, hence for $\mathbf{N}_B$ banks and $\mathbf{N}_F$ firms in the credit market, the matrix $\mathbf{C}(t)$ is of size $(\mathbf{N}_B \times \mathbf{N}_F)$.

Similarly to \cite{aoyama2013debtrank}, we define two network propagation matrices in period $t$, the bank propagation matrix $\mathbf{W}_B(t)$ and the firm propagation matrix $\mathbf{W}_F(t)$, with elements:
\begin{equation}
    (\mathbf{W}_B(t))_{b\iota} = \frac{C_{b\iota}(t)}{C_{b}(t)},
\end{equation}
\begin{equation}
    (\mathbf{W}_F(t))_{\iota b} = \frac{C_{b\iota}(t)}{C_{\iota}(t)}.
\end{equation}
$C_b$ is bank $b$'s loans (total loans to firms), defined as the row sum of matrix $\mathbf{C}(t)$: 
\begin{equation}
    C_b(t) = \sum^{\mathbf{N}_F}_{\iota=1}C_{b\iota}(t).
\end{equation}
$C_{\iota}$ is firm $\iota$'s debt (total loans from banks), defined as the column sum of matrix $\mathbf{C}$:
\begin{equation}
    C_\iota(t) = \sum^{\mathbf{N}_B}_{b=1}C_{b\iota}(t).
\end{equation}
Hence, the bank propagation matrix $\mathbf{W}_B(t)$ is of size $(\mathbf{N}_B \times \mathbf{N}_F)$ and is row stochastic such that $\sum^{\mathbf{N}_F}_{\iota=1}(\mathbf{W}_B(t))_{b\iota}=1$. Similarly, the firm propagation matrix $\mathbf{W}_F(t)$ is of size $(\mathbf{N}_F \times \mathbf{N}_B)$ and is also row stochastic, such that $\sum^{\mathbf{N}_B}_{b=1}(\mathbf{W}_B(t))_{\iota b}=1$.

The DebtRank is calculated at each period $t$ of the simulation, where it is initialised at time period $\tau=t$ and runs until the algorithm stops at some time period $\tau=t+T$. To calculate the DebtRank of the network for time period $t$ of the simulation, each node is associated with two variables, the distress of the node $h_k(\tau)\in[0,1]$ ($k=\iota$ for firms and $k=b$ for banks) and the state of the node $s_k(\tau)\in\{U,D,I\}$, with three discrete states, undistressed ($U$), distressed ($D$), and inactive ($I$). At the initial time period $\tau=t$, the set of distressed nodes is given by $\mathcal{D}_t$. Hence, the initial conditions are given by:
\begin{equation}
    \begin{aligned}
        h_k(t)&=\psi \quad \quad \forall k\in \mathcal{D}_t,\\
        h_k(t)&=0 \quad \quad \forall k\notin \mathcal{D}_t,\\
        s_k(t)&=D \quad \quad \forall k\in \mathcal{D}_t,\\
        s_k(t)&=U \quad \quad \forall k\notin \mathcal{D}_t.
    \end{aligned}
\end{equation}
For simplicity, we set $\psi=1$, such that the initial set of distressed nodes are maximally distressed (bankrupt). The distress and state variables are then updated as:

\begin{equation}
    \begin{aligned}
        h_k(\tau)&=\min\left\{ 1, h_k(\tau-1) + \sum_{\ell|s_\ell(\tau-1)=D} W_{\ell k}h_\ell(\tau-1) \right\},\\
        s_k(\tau)&=
        \begin{cases}
            D & \text{if $h_k(\tau)>0$ and $s_k(\tau-1) \neq I$} \\
            I & \text{if $s_k(\tau-1) = D$} \\
            s_k(\tau-1) & \text{else}.
        \end{cases}
    \end{aligned}
\end{equation}
Only banks are initially distressed, therefore, firm $h_\iota(\tau)$ variables are updated in parallel, then all $s_\iota(\tau)$, followed by all bank $h_b(\tau)$ and $s_b(\tau)$ variables. Furthermore, after a finite number of steps $T$, the dynamics stop as each nodes state is either $s_k(t+T)=U$ or $s_k(t+T)=I$. As in \cite{aoyama2013debtrank}, the bank DebtRank is calculated as a weighted sum of the distress, with weights equal to the bank's asset size:
\begin{equation}
\label{eq:bank_debtrank}
    DR^B_{\mathcal{D}_t}(t)=\sum^{\mathbf{N}_B}_{b=1,\notin \mathcal{D}_t}h_b(t+T) \frac{L_{b}(t)+R_{b}(t)}{\sum^{\mathbf{N}_B}_{b'=1,\notin \mathcal{D}_t}L_{b'}(t)+R_{b'}(t)},
\end{equation}
where bank $b$'s assets are given by $L_{b}(t)+R_{b}(t)$ from Eq.~\ref{eq:bank_balance_sheet}, and $L_{b}(t)$ are bank $b$'s loans and $R_{b}(t)$ are bank $b$'s reserves in time $t$.

Similarly, the firm DebtRank is calculated as a weighted sum of the distress, with weights equal to firms' asset size, which is broken into the sum of C-firms DebtRank and K-firms DebtRank. Additionally, because no firms are included in the set of initially distressed nodes, the sums simplify to:
\begin{equation}
\label{eq:firm_debtrank}
    DR^F_{\mathcal{D}_t}(t)=\sum^{\mathbf{N}_C}_{i=1}h_i(t+T) \frac{M_{i}(t)+KE_{i}(t)}{\sum^{\mathbf{N}_C}_{i'=1}M_{i'}(t)+KE_{i'}(t)} + \sum^{\mathbf{N}_K}_{j=1}h_j(t+T) \frac{M_{j}(t)}{\sum^{\mathbf{N}_K}_{j'=1}M_{j'}(t)},
\end{equation}
where C-firm $i$'s assets are given by $M_{i}(t)+KE_{i}(t)$ from Eq.~\ref{eq:cfirm_balance_sheet}, $M_{i}(t)$ are C-firm $i$'s deposits and $KE_{i}(t)$ is C-firm $i$'s capital expenditure in time $t$, and K-firm $j$'s assets are only given by their deposits, $M_{j}(t)$ from Eq.~\ref{eq:kfirm_balance_sheet}. 

The ESL in period $t$ gives a measure of the total economic value, $V(t)$, that could be lost multiplied by the probability of that loss occurring. Following, \cite{poledna2015systemic_risk} the ESL is defined as:
\begin{equation}
    ESL(t)=V(t)\sum_{\mathcal{D}\in\mathcal{P}(\mathcal{B})}\prod_{b\in\mathcal{D}}p_b(t)\prod_{b'\in\mathcal{B}\backslash\mathcal{D}}(1-p_{b'}(t))R_{\mathcal{D}},
\end{equation}
where $p_b(t)$ is the probability of bank $b$, defined as the historical default probability of bank $b$ in period $t$, $\mathcal{D}$ is the initial set of distressed banks, $R_{\mathcal{D}}$ is the total DebtRank of the initial set of distressed banks, and $\mathcal{P}(\mathcal{B})$ is the power set of the total set of banks $b\in\mathcal{B}$. The above equation for the ESL is computationally expensive. Hence, given that we find default probabilities to be low (see Fig.~\ref{figure_5}), we can use the ESL approximation derived by \cite{poledna2015systemic_risk}:
\begin{equation}
    ESL(t)\approx \sum_{b=1}^{\mathbf{N}_B}p_b(t)(DR^B_b(t)V^B(t) + DR^F_b(t)V^F(t)),
\end{equation}
where $DR^B_b(t)$ is the DebtRank associated with the bank layer and $DR^F_b(t)$ is the DebtRank associated with the firm layer if bank $b$ were to default. $V^B(t)$ is the total value of the bank layer, and $V^F(t)$ is the total value of the firms layer:
\begin{equation}
    \begin{aligned}
        V^B(t)&=\sum_{b=1}^{\mathbf{N}_B}L_{b}(t)+R_{b}(t)\\ V^F(t)&=\sum_{i=1}^{\mathbf{N}_C}M_i(t)+KE_i(t)+\sum_{j=1}^{\mathbf{N}_K}M_j(t),
    \end{aligned}
\end{equation}
where $L_b(t)$ and $R_b(t)$ are bank loans to firms and bank reserves, respectively, $M_i(t)$ and $KE_i(t)$ are C-firm deposits and capital expenditure, respectively, and $M_j(t)$ are K-firm deposits. Additionally, $p_b(t)$ is bank $b$'s probability of default in time $t$, which is calculated as the average number of defaults of bank $b$ in period $t$ across the $S$ simulations. 

\section{Discussion Results}
\label{app:discussion_results}
Table \ref{tab:inflation_stats} shows simulation averages for aggregate productivity growth ($g_a$), nominal wage inflation ($g_w$), CPI inflation ($g_P$), and the difference between wage inflation and productivity growth ($g_w-g_a$) for each scenario. It can be noted that CPI inflation is well approximated by the difference between wage inflation and productivity growth.

\begin{table}[!htbp]
    \centering
        \begin{tabular}{lcccc}
        \toprule
        \textbf{Variable} & \textbf{Growth S1} & \textbf{Growth S2} & \textbf{Zero-Growth S1} & \textbf{Zero-Growth S2}\\
        \midrule
        $g_a$       & 0.0236 & 0.0224 & 0.0048 & 0.0050 \\
        $g_w$       & 0.0380 & 0.0373 & 0.0316 & 0.0316 \\
        $g_P$       & 0.0156 & 0.0158 & 0.0271 & 0.0269 \\
        $g_w-g_a$   & 0.0144 & 0.0149 & 0.0267 & 0.0266 \\
        \bottomrule
        \end{tabular} 
    \caption{Average values across time periods and simulations for aggregate productivity growth ($g_a$), nominal wage inflation ($g_w$), CPI inflation ($g_P$), and the difference between wage inflation and productivity growth ($g_w-g_a$) for each scenario.}
    \label{tab:inflation_stats}
\end{table}

%% file: references.bib
@article{barrett2018stability,
   title = {{Stability of Zero-growth Economics Analysed with a Minskyan Model}},
   volume = {146},
   issn = {09218009},
   language = {en},
   journal = {Ecological Economics},
   author = {Barrett, Adam B.},
   month = apr,
   year = {2018},
   pages = {pp. 228--239}
}

@article{dosi2010schumpeter,
  title = {{Schumpeter meeting Keynes: A Policy-Friendly Model of Endogenous Growth and Business Cycles}},
  volume = {34},
  issn = {01651889},
  shorttitle = {Schumpeter Meeting {Keynes}},
  language = {en},
  number = {9},
  journal = {Journal of Economic Dynamics and Control},
  author = {Dosi, Giovanni and Fagiolo, Giorgio and Roventini, Andrea},
  month = sep,
  year = {2010},
  pages = {pp. 1748-1767},
}

@article{botte2021transition,
  title = {{Modelling Transition Risk Towards an Agent-Based, Stock-Flow Consistent Framework}},
  language = {en},
  author = {Botte, Florian and Ciarli, Tommaso and Foxon, Tim and Jackson, Andrew and Jackson, Tim and Valente, Marco},
  pages = {pp. 1-65},
  year = {2021},
  journal = {Rebuilding Macroeconomics Working Paper Series},
  volume = {40},
}

@article{gaffeo2008emergent_macro,
  title={{Adaptive Microfoundations for Emergent Macroeconomics}},
  author={Gaffeo, Edoardo and Delli Gatti, Domenico and Desiderio, Saul and Gallegati, Mauro},
  journal={Eastern Economic Journal},
  volume={34},
  number={4},
  pages={441--463},
  year={2008},
  publisher={Springer}
}

@article{assenza2015macro_abm,
  title = {{Emergent Dynamics of a Macroeconomic Agent Based Model with Capital and Credit}},
  volume = {50},
  issn = {01651889},
  language = {en},
  journal = {Journal of Economic Dynamics and Control},
  author = {Assenza, Tiziana and Delli Gatti, Domenico and Grazzini, Jakob},
  month = jan,
  year = {2015},
  pages = {pp. 5-28},
}

@article{jackson2015growth_imperative,
  title={{Does credit create a ‘growth imperative’? A quasi-stationary economy with interest-bearing debt}},
  author={Jackson, Tim and Victor, Peter A},
  journal={Ecological Economics},
  volume={120},
  pages={32--48},
  year={2015},
  publisher={Elsevier}
}

@article{malmaeus2017potential,
  title = {{Potential Consequences on the Economy of Low or No Growth - Short and Long Term Perspectives}},
  volume = {134},
  issn = {09218009},
  language = {en},
  journal = {Ecological Economics},
  author = {Malmaeus, J. Mikael and Alfredsson, Eva C.},
  year = {2017},
  pages = {pp. 57-64},
}

@article{dixon1987bootstrapping,
  title = {{Bootstrapping the Gini Coefficient of Inequality}},
  volume = {68},
  issn = {0012-9658},
  number = {5},
  journal = {Ecology},
  author = {Dixon, Philip M. and Weiner, Jacob and Mitchell-Olds, Thomas and Woodley, Robert},
  year = {1987},
  pages = {pp. 1548--1551},
}

@book{piketty2014capital,
  title={{Capital in the Twenty-First Century}},
  author={Piketty, Thomas},
  year={2014},
  address = {Cambridge, Massachusetts},
  publisher={Harvard University Press},
}

@incollection{keynes1930granchildren,
  title={{Economic Possibilities for our Grandchildren}},
  author={Keynes, John Maynard},
  booktitle={Essays in Persuasion},
  pages={321-332},
  year={1930},
  publisher={Springer}
}

@article{steffen2015planetary,
  title={{Planetary boundaries: Guiding human development on a changing planet}},
  author={Steffen, Will and Richardson, Katherine and Rockstr{\"o}m, Johan and Cornell, Sarah E and Fetzer, Ingo and Bennett, Elena M and Biggs, Reinette and Carpenter, Stephen R and De Vries, Wim and De Wit, Cynthia A and others},
  journal={Science},
  volume={347},
  number={6223},
  year={2015},
  publisher={American Association for the Advancement of Science},
}

@book{minsky1986stabilizing,
  title={{Stabilizing an Unstable Economy}},
  author={Minsky, Hyman P},
  volume={1},
  year={1986},
  publisher={Yale University Press},
  address={New Haven},
}

@article{keen1995finance,
  title={{Finance and Economic Breakdown: Modeling Minsky’s “Financial Instability Hypothesis”}},
  author={Keen, Steve},
  journal={Journal of Post Keynesian Economics},
  volume={17},
  number={4},
  pages={pp. 607--635},
  year={1995},
  publisher={Taylor \& Francis}
}

@article{russo2007industrial,
  title={{Industrial Dynamics, Fiscal Policy and R\&D: Evidence from a Computational Experiment}},
  author={Russo, Alberto and Catalano, Michele and Gaffeo, Edoardo and Gallegati, Mauro and Napoletano, Mauro},
  journal={Journal of Economic Behavior \& Organization},
  volume={64},
  number={3-4},
  pages={pp. 426--447},
  year={2007},
  publisher={Elsevier}
}

@article{dosi2009microfoundations,
  title={{The microfoundations of business cycles: an evolutionary, multi-agent model}},
  author={Dosi, Giovanni and Fagiolo, Giorgio and Roventini, Andrea},
  journal={Journal of Evolutionary Economics},
  volume={18},
  number={1},
  pages={pp. 413--432},
  year={2008},
  publisher={Springer}
}

@article{dosi2006evolutionary,
  title={{An Evolutionary Model of Endogenous Business Cycles}},
  author={Dosi, Giovanni and Fagiolo, Giorgio and Roventini, Andrea},
  journal={Computational Economics},
  volume={27},
  number={1},
  pages={pp. 3--34},
  year={2006},
  publisher={Springer}
}

@article{dosi2013income,
  title={{Income Distribution, Credit and Fiscal Policies in an Agent-Based Keynesian Model}},
  author={Dosi, Giovanni and Fagiolo, Giorgio and Napoletano, Mauro and Roventini, Andrea},
  journal={Journal of Economic Dynamics and Control},
  volume={37},
  number={8},
  pages={pp. 1598--1625},
  year={2013},
  publisher={Elsevier}
}

@article{gatti2005new,
  title={{A New Approach to Business Fluctuations: Heterogeneous Interacting Agents, Scaling laws and Financial Fragility}},
  author={Delli Gatti, Domenico and Di Guilmi, Corrado and Gaffeo, Edoardo and Giulioni, Gianfranco and Gallegati, Mauro and Palestrini, Antonio},
  journal={Journal of Economic behavior \& organization},
  volume={56},
  number={4},
  pages={pp. 489--512},
  year={2005},
  publisher={Elsevier}
}

@article{delli_gatti2020rising,
  title={{Rising to the Challenge: Bayesian Estimation and Forecasting Techniques for Macroeconomic Agent Based Models}},
  author={Delli Gatti, Domenico and Grazzini, Jakob},
  journal={Journal of Economic Behavior \& Organization},
  volume={178},
  pages={pp. 875--902},
  year={2020},
  publisher={Elsevier}
}

@article{gatti2003financial,
  title={{Financial Fragility, Patterns of Firms’ Entry and Exit and Aggregate Dynamics}},
  author={Delli Gatti, Domenico and Gallegati, Mauro and Giulioni, Gianfranco and Palestrini, Antonio},
  journal={Journal of economic behavior \& organization},
  volume={51},
  number={1},
  pages={pp. 79--97},
  year={2003},
  publisher={Elsevier}
}

@article{gatti2010financial,
  title={{The Financial Accelerator in an Evolving Credit Network}},
  author={Delli Gatti, Domenico and Gallegati, Mauro and Greenwald, Bruce and Russo, Alberto and Stiglitz, Joseph E},
  journal={Journal of Economic Dynamics and Control},
  volume={34},
  number={9},
  pages={pp. 1627--1650},
  year={2010},
  publisher={Elsevier}
}

@article{hymer1962turnover,
  title={{Turnover of Firms as a Measure of Market Behavior}},
  author={Hymer, Stephen and Pashigian, Peter},
  journal={The Review of Economics and Statistics},
  pages={pp. 82--87},
  volume = {44},
  number ={1},
  year={1962},
  publisher={The MIT Press},
}

@book{hirschman1945national,
  title={{National Power and the Structure of Foreign Trade}},
  author={Hirschman, Albert O},
  year={1945},
  publisher={University of California Press},
  address={Los Angeles}
}

@article{dosi2019endogenous,
  title={{Endogenous Growth and Global Divergence in a Multi-Country Agent-Based Model}},
  author={Dosi, Giovanni and Roventini, Andrea and Russo, Emanuele},
  journal={Journal of Economic Dynamics and Control},
  volume={101},
  pages={pp. 101--129},
  year={2019},
  publisher={Elsevier}
}

@article{wiedmann2020scientists,
  title={{Scientists’ warning on affluence}},
  author={Wiedmann, Thomas and Lenzen, Manfred and Key{\ss}er, Lorenz T and Steinberger, Julia K},
  journal={Nature Communications},
  volume={11},
  number={1},
  pages={1--10},
  year={2020},
  publisher={Nature Publishing Group}
}

@article{haberl2020systematic,
  title={{A systematic review of the evidence on decoupling of GDP, resource use and GHG emissions, part II: synthesizing the insights}},
  author={Haberl, Helmut and Wiedenhofer, Dominik and Vir{\'a}g, Doris and Kalt, Gerald and Plank, Barbara and Brockway, Paul and Fishman, Tomer and Hausknost, Daniel and Krausmann, Fridolin and Leon-Gruchalski, Bartholom{\"a}us and others},
  journal={Environmental Research Letters},
  volume={15},
  number={6},
  pages={065003},
  year={2020},
  publisher={IOP Publishing}
}

@article{cahen2016ecological,
  Author = {Cahen-Fourot, Louison and Lavoie, Marc},
  Title = {{Ecological monetary economics: A post-Keynesian critique}},
  Journal = {{Ecological Economics}},
  Year = {{2016}},
  Volume = {{126}},
  Pages = {{163-168}},
  Month = {{JUN}},
  ISSN = {{0921-8009}},
  EISSN = {{1873-6106}},
  ORCID-Numbers = {{Cahen-Fourot, Louison/0000-0003-2580-1982}},
  Unique-ID = {{WOS:000376215400016}},
}

@article{elder2019design,
  title={{The Design of Environmental Priorities in the SDG's}},
  author={Elder, Mark and Olsen, Simon H{\o}iberg},
  journal={Global Policy},
  volume={10},
  pages={70--82},
  year={2019},
  publisher={Wiley Online Library}
}

@article{jackson2020transition,
  title={{The Transition to a Sustainable Prosperity - A Stock-Flow-Consistent Ecological Macroeconomic Model for Canada}},
  author={Jackson, Tim and Victor, Peter A},
  journal={Ecological Economics},
  volume={177},
  pages={106787},
  year={2020},
  publisher={Elsevier}
}

@article{rosenbaum2015zerogrowth,
  Author = {Rosenbaum, Eckehard},
  Title = {{Zero Growth and Structural Change in a Post Keynesian Growth Model}},
  Journal = {Journal of Post Keynesian Economics},
  Year = {2015},
  Volume = {37},
  Number = {4},
  Pages = {623-647},
  Month = {SUM},
  ISSN = {0160-3477},
  EISSN = {1557-7821},
  Unique-ID = {WOS:000358660800006},
}

@article{binswanger2009growth,
  Author = {Binswanger, Mathias},
  Title = {{Is There a Growth Imperative in Capitalist Economies? A Circular Flow Perspective}},
  Journal = {Journal of Post Keynesian Economics},
  Year = {2009},
  Volume = {31},
  Number = {4},
  Pages = {707-727},
  Month = {SUM},
  ISSN = {0160-3477},
  Unique-ID = {WOS:000268599100010},
}

@book{georgescuroegen1972entropy,
  title={{The Entropy Law and the Economic Process}},
  author={Nicholas Georgescu-Roegen},
  year={1972},
  address={Cambridge},
  publisher={Cambridge University Press},
}

@book{daly1996beyond,
  title={{Beyond Growth: The Economics of Sustainable Development}},
  author={Daly, Herman E},
  year={1996},
  publisher={Beacon Press}
}

@article{battiston2012debtrank,
  title={{Debtrank: Too Central to Fail? Financial Networks, the Fed and Systemic Risk}},
  author={Battiston, Stefano and Puliga, Michelangelo and Kaushik, Rahul and Tasca, Paolo and Caldarelli, Guido},
  journal={Scientific reports},
  volume={2},
  number={1},
  pages={1--6},
  year={2012},
  publisher={Nature Publishing Group}
}

@TechReport{aoyama2013debtrank,
  author={Aoyama, Hideaki and Battiston, Stefano and Fujiwara, Yoshi},
  title={{DebtRank Analysis of the Japanese Credit Network}},
  year={2013},
  month={Oct},
  institution={Research Institute of Economy, Trade and Industry (RIETI)},
  type={Discussion Paper Series},
  number={No. 13-E-087}
}

@article{richardson2023earth,
  title={{Earth Beyond Six of Nine Planetary Boundaries}},
  author={Richardson, Katherine and Steffen, Will and Lucht, Wolfgang and Bendtsen, J{\o}rgen and Cornell, Sarah E and Donges, Jonathan F and Dr{\"u}ke, Markus and Fetzer, Ingo and Bala, Govindasamy and Von Bloh, Werner and others},
  journal={Science advances},
  volume={9},
  number={37},
  pages={eadh2458},
  year={2023},
  publisher={American Association for the Advancement of Science}
}

@article{rockstrom2009safe,
  title={{A Safe Operating Space for Humanity}},
  author={Rockstr{\"o}m, Johan and Steffen, Will and Noone, Kevin and Persson, {\AA}sa and Chapin, F Stuart and Lambin, Eric F and Lenton, Timothy M and Scheffer, Marten and Folke, Carl and Schellnhuber, Hans Joachim and others},
  journal={nature},
  volume={461},
  number={7263},
  pages={472--475},
  year={2009},
  publisher={Nature Publishing Group}
}

@article{caiani2016benchmark_model,
  title={{Agent Based-Stock Flow Consistent Macroeconomics: Towards a Benchmark Model}},
  author={Caiani, Alessandro and Godin, Antoine and Caverzasi, Eugenio and Gallegati, Mauro and Kinsella, Stephen and Stiglitz, Joseph E},
  journal={Journal of Economic Dynamics and Control},
  volume={69},
  pages={375--408},
  year={2016},
  publisher={Elsevier}
}

@book{godley2006monetary_economics,
  title={{Monetary Economics: An Integrated Approach to Credit, Money, Income, Production and Wealth}},
  author={Godley, Wynne and Lavoie, Marc},
  year={2007},
  address={New York},
  publisher={Palgrave MacMillan},
}

@book{soddy1926wealth,
  title={{Wealth, Virtual Wealth and Debt: The Solution of the Economic Paradox}},
  author={Soddy, Frederick},
  year={1926},
  address={London},
  publisher={Allen \& Unwin}
}

@article{ward2016decoupling,
  title={Is Decoupling GDP Growth from Environmental Impact Possible?},
  author={Ward, James D and Sutton, Paul C and Werner, Adrian D and Costanza, Robert and Mohr, Steve H and Simmons, Craig T},
  journal={PloS one},
  volume={11},
  number={10},
  pages={e0164733},
  year={2016},
  publisher={Public Library of Science}
}

@article{alcott2005jevons,
  title={Jevons' paradox},
  author={Alcott, Blake},
  journal={Ecological Economics},
  volume={54},
  number={1},
  pages={9--21},
  year={2005},
  publisher={Elsevier}
}

@article{slamervsak2024post_growth,
  title={{Post-growth: A viable path to limiting global warming to 1.5°C}},
  author={Slamer{\v{s}}ak, Aljo{\v{s}}a and Kallis, Giorgos and O’Neill, Daniel W and Hickel, Jason},
  journal={One Earth},
  volume={7},
  number={1},
  pages={44--58},
  year={2024},
  publisher={Elsevier}
}

@article{alessandro2020green_growth,
  title={{Feasible Alternatives to Green Growth}},
  author={D’Alessandro, Simone and Cieplinski, Andr{\'e} and Distefano, Tiziano and Dittmer, Kristofer},
  journal={Nature Sustainability},
  volume={3},
  number={4},
  pages={329--335},
  year={2020},
  publisher={Nature Publishing Group UK London}
}

@article{axtell2025abm,
  title={Agent-based modeling in economics and finance: Past, present, and future},
  author={Axtell, Robert L and Farmer, J Doyne},
  journal={Journal of Economic Literature},
  volume={63},
  number={1},
  pages={197--287},
  year={2025},
  publisher={American Economic Association 2014 Broadway, Suite 305, Nashville, TN 37203-2425}
}

@article{myers1984capital_structure,
  author = {Myers, Stewart C.},
  title = {The Capital Structure Puzzle},
  journal = {The Journal of Finance},
  volume = {39},
  number = {3},
  pages = {574-592},
  year = {1984}
}

@article{stockhammer2023debt_cycles,
  title={{Debt-GDP cycles in historical perspective: the case of the USA (1889--2014)}},
  author={Stockhammer, Engelbert and Gouzoulis, Giorgos},
  journal={Industrial and Corporate Change},
  volume={32},
  number={2},
  pages={317--335},
  year={2023},
  publisher={Oxford University Press UK}
}

@article{fagiolo2008output_distributions,
  title={{Are output growth-rate distributions fat-tailed? Some evidence from OECD countries}},
  author={Fagiolo, Giorgio and Napoletano, Mauro and Roventini, Andrea},
  journal={Journal of Applied Econometrics},
  volume={23},
  number={5},
  pages={639--669},
  year={2008},
  publisher={Wiley Online Library}
}

@article{bottazzi2003common,
  title={Common properties and sectoral specificities in the dynamics of US manufacturing companies},
  author={Bottazzi, Giulio and Secchi, Angelo},
  journal={Review of Industrial Organization},
  volume={23},
  number={3},
  pages={217--232},
  year={2003},
  publisher={Springer}
}

@article{bottazzi2006firm_distributions,
  title={Explaining the distribution of firm growth rates},
  author={Bottazzi, Giulio and Secchi, Angelo},
  journal={The RAND Journal of Economics},
  volume={37},
  number={2},
  pages={235--256},
  year={2006},
  publisher={Wiley Online Library}
}

@article{axtell2001zipf,
  title={{Zipf Distribution of U.S. Firm Sizes}},
  author={Axtell, Robert L},
  journal={science},
  volume={293},
  number={5536},
  pages={1818--1820},
  year={2001},
  publisher={American Association for the Advancement of Science}
}

@article{wit2005firm_size,
  title={{Firm size distributions: An overview of steady-state distributions resulting from firm dynamics models}},
  author={de Wit, Gerrit},
  journal={International Journal of Industrial Organization},
  volume={23},
  number={5-6},
  pages={423--450},
  year={2005},
  publisher={Elsevier}
}

@article{berg2015stock,
  title={{A stock-flow consistent input-output model with applications to energy price shocks, interest rates, and heat emissions}},
  author={Berg, Matthew and Hartley, Brian and Richters, Oliver},
  journal={New journal of physics},
  volume={17},
  number={1},
  pages={015011},
  year={2015},
  publisher={IOP Publishing}
}

@article{parrique2019decoupling,
  title={{Decoupling Debunked: Evidence and arguments against green growth as a sole strategy for sustainability.}},
  author={Parrique, Timoth{\'e}e and Barth, Jonathan and Briens, Fran{\c{c}}ois and Kerschner, Christian and Kraus-Polk, Alejo and Kuokkanen, Anna and Spangenberg, Joachim H},
  journal={European Environment Bureau},
  year={2019}
}

@article{keen2014endogenous,
  title={Endogenous money and effective demand},
  author={Keen, Steve},
  journal={Review of Keynesian Economics},
  volume={2},
  number={3},
  pages={271--291},
  year={2014},
  publisher={Edward Elgar Publishing Ltd}
}

@article{lengnick2013baseline,
  title={{Agent-based macroeconomics: A baseline model}},
  author={Lengnick, Matthias},
  journal={Journal of Economic Behavior \& Organization},
  volume={86},
  pages={102--120},
  year={2013},
  publisher={Elsevier}
}

@article{klimek2015bail,
  title={{To bail-out or to bail-in? Answers from an agent-based model}},
  author={Klimek, Peter and Poledna, Sebastian and Farmer, J Doyne and Thurner, Stefan},
  journal={Journal of Economic Dynamics and Control},
  volume={50},
  pages={144--154},
  year={2015},
  publisher={Elsevier}
}

@article{lamperti2018faraway,
  title={{Faraway, So Close: Coupled Climate and Economic Dynamics in an Agent-Based Integrated Assessment Model}},
  author={Lamperti, Francesco and Dosi, Giovanni and Napoletano, Mauro and Roventini, Andrea and Sapio, Alessandro},
  journal={Ecological Economics},
  volume={150},
  pages={315--339},
  year={2018},
  publisher={Elsevier}
}

@article{lamperti2021three,
  title={Three green financial policies to address climate risks},
  author={Lamperti, Francesco and Bosetti, Valentina and Roventini, Andrea and Tavoni, Massimo and Treibich, Tania},
  journal={Journal of Financial Stability},
  volume={54},
  pages={100875},
  year={2021},
  publisher={Elsevier}
}

@article{dosi2017evolutionary,
  title={{Micro and macro policies in the Keynes + Schumpeter evolutionary models}},
  author={Dosi, Giovanni and Napoletano, Mauro and Roventini, Andrea and Treibich, Tania},
  journal={Journal of Evolutionary Economics},
  volume={27},
  number={1},
  pages={63--90},
  year={2017},
  publisher={Springer}
}

@article{reissl2025dsk,
  title={{The DSK stock-flow consistent agent-based integrated assessment model}},
  author={Reissl, Severin and Fierro, Luca E and Lamperti, Francesco and Roventini, Andrea},
  journal={Ecological Economics},
  volume={236},
  pages={108641},
  year={2025},
  publisher={Elsevier}
}

@article{lamperti2020climate,
  title={{Climate change and green transitions in an agent-based integrated assessment model}},
  author={Lamperti, Francesco and Dosi, Giovanni and Napoletano, Mauro and Roventini, Andrea and Sapio, ALESSANDRO},
  journal={Technological Forecasting and Social Change},
  volume={153},
  pages={119806},
  year={2020},
  publisher={Elsevier}
}

@article{mcleay2014money,
  title={Money creation in the modern economy},
  author={McLeay, Michael and Radia, Amar and Thomas, Ryland},
  journal={Bank of England quarterly bulletin},
  pages={Q1},
  year={2014}
}

@article{werner2014money,
  title={Can banks individually create money out of nothing?—The theories and the empirical evidence},
  author={Werner, Richard A},
  journal={International review of financial analysis},
  volume={36},
  pages={1--19},
  year={2014},
  publisher={Elsevier}
}

@techreport{minsky1992financial,
  title={{The financial instability hypothesis}},
  author={Minsky, Hyman P},
  year={1992},
  type={Working Paper},
  number={No. 74},
  institution={Levy Economics Institute of Bard College, Annandale-on-Hudson, NY}
}

@article{minsky1977financial,
  title={{The Financial Instability Hypothesis: An Interpretation of Keynes and an Alternative to “Standard” Theory}},
  author={Minsky, Hyman P},
  journal={Challenge},
  volume={20},
  number={1},
  pages={20--27},
  year={1977},
  publisher={Taylor \& Francis}
}

@article{kallis2025postgrowth,
  title={{Post-growth: the science of wellbeing within planetary boundaries}},
  author={Kallis, Giorgos and Hickel, Jason and O’Neill, Daniel W and Jackson, Tim and Victor, Peter A and Raworth, Kate and Schor, Juliet B and Steinberger, Julia K and {\"U}rge-Vorsatz, Diana},
  journal={The Lancet Planetary Health},
  volume={9},
  number={1},
  pages={E62--E78},
  year={2025},
  publisher={Elsevier}
}

@article{lauer2025comparative,
  title={A comparative review of de-and post-growth modeling studies},
  author={Lauer, Arthur and Capell{\'a}n-P{\'e}rez, I{\~n}igo and Wergles, Nathalie},
  journal={Ecological Economics},
  volume={227},
  pages={108383},
  year={2025},
  publisher={Elsevier}
}

@article{poledna2023economic_forecasting,
  title={Economic forecasting with an agent-based model},
  author={Poledna, Sebastian and Miess, Michael Gregor and Hommes, Cars and Rabitsch, Katrin},
  journal={European Economic Review},
  volume={151},
  pages={104306},
  year={2023},
  publisher={Elsevier}
}

@article{hommes2025canvas,
  title={{CANVAS: A Canadian behavioral agent-based model for monetary policy}},
  author={Hommes, Cars and He, Mario and Poledna, Sebastian and Siqueira, Melissa and Zhang, Yang},
  journal={Journal of Economic Dynamics and Control},
  volume={172},
  pages={104986},
  year={2025},
  publisher={Elsevier}
}

@article{chen2022regression,
  title={A regression-based calibration method for agent-based models},
  author={Chen, Siyan and Desiderio, Saul},
  journal={Computational Economics},
  volume={59},
  number={2},
  pages={687--700},
  year={2022},
  publisher={Springer}
}

@book{chancel2021inequality,
  title={{World Inequality Report 2022}},
  author={Chancel, Lucas and Piketty, Thomas and Saez, Emmanuel and Zucman, Gabriel},
  year={2021},
  publisher={World Inequality Lab}
}

@book{kalecki1954dynamics,
  title={{Theory of Economic Dynamics: An Essay on Cyclical and Long-Run Changes in Capitalist Economy}},
  author={Kalecki, Micha{\l}},
  year={1954},
  publisher={George Allen and Unwin},
  address={London}
}

@article{poledna2015systemic_risk,
  title={The multi-layer network nature of systemic risk and its implications for the costs of financial crises},
  author={Poledna, Sebastian and Molina-Borboa, Jos{\'e} Luis and Mart{\'\i}nez-Jaramillo, Seraf{\'\i}n and Van Der Leij, Marco and Thurner, Stefan},
  journal={Journal of Financial Stability},
  volume={20},
  pages={70--81},
  year={2015},
  publisher={Elsevier}
}

@article{hickel2020green,
  title={{Is Green Growth Possible?}},
  author={Hickel, Jason and Kallis, Giorgos},
  journal={New Political Economy},
  volume={25},
  number={4},
  pages={469--486},
  year={2020},
  publisher={Taylor \& Francis}
}

@incollection{goodwin1967growth_cycle,
  author={Goodwin, Richard M.},
  title={{A Growth Cycle}},
  editor={Feinstein, C. H.},
  booktitle={Socialism, Capitalism and Economic Growth},
  publisher={Cambridge University Press},
  address={Cambridge},
  year={1967},
  pages={54--58},
}

@book{nitzan2009casp,
  title={{Capital as power: A study of Order and Creorder}},
  author={Nitzan, Jonathan and Bichler, Shimshon},
  year={2009},
  address = {New York},
  publisher={Routledge}
}

@article{herr2022currency,
  title={Currency hierarchy and underdevelopment},
  author={Herr, Hansj{\"o}rg and Nettekoven, Zeynep},
  journal={European Journal of Economics and Economic Policies},
  volume={19},
  number={2},
  pages={238--259},
  year={2022},
  publisher={Edward Elgar Publishing Ltd}
}

@book{cohen2018geography,
  title={{The Geography of Money}},
  author={Cohen, Benjamin J},
  year={2018},
  publisher={Cornell University Press}
}

@article{orsi2025currency_hierarchy,
  title={{Currency hierarchy and the nature of peripheral currencies’ internationalization}},
  author={Orsi, Bianca and Kaltenbrunner, Annina and Dymski, Gary},
  journal={Journal of Post Keynesian Economics},
  pages={1--27},
  year={2025},
  publisher={Taylor \& Francis}
}

@article{dosi2021public,
  title={{Public policies and the art of catching up: matching the historical evidence with a multicountry agent-based model}},
  author={Dosi, Giovanni and Roventini, Andrea and Russo, Emanuele},
  journal={Industrial and Corporate Change},
  volume={30},
  number={4},
  pages={1011--1036},
  year={2021},
  publisher={Oxford University Press}
}

@article{ciarli2010effect,
  title={The effect of consumption and production structure on growth and distribution. A micro to macro model},
  author={Ciarli, Tommaso and Lorentz, Andr{\'e} and Savona, Maria and Valente, Marco},
  journal={Metroeconomica},
  volume={61},
  number={1},
  pages={180--218},
  year={2010},
  publisher={Wiley Online Library}
}

@article{simon1957compensation,
  title={{The Compensation of Executives}},
  author={Simon, Herbert A},
  journal={Sociometry},
  volume={20},
  number={1},
  pages={32--35},
  year={1957},
  publisher={JSTOR}
}

@article{lydall1959distribution,
  title={{The Distribution of Employment Incomes}},
  author={Lydall, Harold F},
  journal={Econometrica: Journal of the Econometric Society},
  pages={110--115},
  year={1959},
  publisher={JSTOR}
}

@article{fix2018hierarchy,
  title={Hierarchy and the power-law income distribution tail},
  author={Fix, Blair},
  journal={Journal of Computational Social Science},
  volume={1},
  number={2},
  pages={471--491},
  year={2018},
  publisher={Springer}
}

@article{edwards2025towards,
  title={Towards modelling post-growth climate futures: a review of current modelling practices and next steps},
  author={Edwards, Alex and Brockway, Paul and Bickerstaff, Karen and Nijsse, Femke JMM},
  journal={Environmental Research Letters},
  year={2025},
  publisher={IOP Publishing}
}

@article{weber2023sellers_inflation,
  title={Sellers’ inflation, profits and conflict: why can large firms hike prices in an emergency?},
  author={Weber, Isabella M and Wasner, Evan},
  journal={Review of Keynesian Economics},
  volume={11},
  number={2},
  pages={183--213},
  year={2023},
  publisher={Edward Elgar Publishing Ltd}
}

@techreport{chatterjee2015credit_risk,
  author={Chatterjee, Somnath},
  title={Modelling credit risk},
  institution={Bank of England},
  year={2015},
  type={{Centre for Central Banking Studies}}
}

@article{stock1999business,
  title={{Business cycle fluctuations in US macroeconomic time series}},
  author={Stock, James H and Watson, Mark W},
  journal={Handbook of macroeconomics},
  volume={1},
  pages={3--64},
  year={1999},
  publisher={Elsevier}
}

@article{stock2020slack,
  title={{Slack and cyclically sensitive inflation}},
  author={Stock, James H and Watson, Mark W},
  journal={Journal of Money, Credit and Banking},
  volume={52},
  number={S2},
  pages={393--428},
  year={2020},
  publisher={Wiley Online Library}
}

@article{fiorito1994stylized,
  title={Stylized facts of business cycles in the G7 from a real business cycles perspective},
  author={Fiorito, Riccardo and Kollintzas, Tryphon},
  journal={European economic review},
  volume={38},
  number={2},
  pages={235--269},
  year={1994},
  publisher={Elsevier}
}

@inproceedings{okun1962potential_gnp,
  author={Okun, Arthur M.},
  title={{Potential GNP: Its Measurement and Significance}},
  booktitle={{Proceedings of the Business and Economics Statistics Section}},
  year={1962},
  publisher={American Statistical Association},
  pages={98--104}
}

@article{phillips1958relation,
  title={{The relation between unemployment and the rate of change of money wage rates in the United Kingdom, 1861-1957}},
  author={Phillips, Alban W},
  journal={Economica},
  volume={25},
  number={100},
  pages={283--299},
  year={1958}
}

@article{doms1998capital,
  title={{Capital Adjustment Patterns in Manufacturing Plants}},
  author={Doms, Mark and Dunne, Timothy},
  journal={Review of Economic Dynamics},
  volume={1},
  number={2},
  pages={409--429},
  year={1998},
  publisher={Elsevier}
}

@article{bartelsman2000understanding,
  title={Understanding productivity: Lessons from longitudinal microdata},
  author={Bartelsman, Eric J and Doms, Mark},
  journal={Journal of Economic literature},
  volume={38},
  number={3},
  pages={569--594},
  year={2000},
  publisher={American Economic Association}
}

@article{ausloos2004recession_durations,
  title={The durations of recession and prosperity: does their distribution follow a power or an exponential law?},
  author={Ausloos, Marcel and Mi{\'s}kiewicz, Janusz and Sanglier, Mich{\`e}le},
  journal={Physica A: Statistical Mechanics and its Applications},
  volume={339},
  number={3-4},
  pages={548--558},
  year={2004},
  publisher={Elsevier}
}

@article{fisher1933debt_deflation,
  title={The debt-deflation theory of great depressions},
  author={Fisher, Irving},
  journal={Econometrica: Journal of the Econometric Society},
  pages={337--357},
  year={1933},
  publisher={JSTOR}
}

@article{castaldi2009patterns,
  title={The patterns of output growth of firms and countries: Scale invariances and scale specificities},
  author={Castaldi, Carolina and Dosi, Giovanni},
  journal={Empirical Economics},
  volume={37},
  number={3},
  pages={475--495},
  year={2009},
  publisher={Springer}
}

@article{wright2005duration,
  title={The duration of recessions follows an exponential not a power law},
  author={Wright, Ian},
  journal={Physica A: Statistical Mechanics and its Applications},
  volume={345},
  number={3-4},
  pages={608--610},
  year={2005},
  publisher={Elsevier}
}

@techreport{ennis2001bank_size,
  title={{On the Size Distribution of Banks}},
  author={Ennis, Huberto M},
  institution={Federal Reserve Bank of Richmond},
  type={Economic Quarterly},
  year={2001},
  pages={1--25},
  number={85(4)}
}

@book{gibrat1931inegalites,
  author    = {Gibrat, Robert},
  title     = {Les in{\'e}galit{\'e}s {\'e}conomiques},
  year      = {1931},
  publisher = {Librairie du Recueil Sirey},
  address   = {Paris}
}

@article{hodrick1997postwar,
  title={Postwar US business cycles: an empirical investigation},
  author={Hodrick, Robert J and Prescott, Edward C},
  journal={Journal of Money, credit, and Banking},
  pages={1--16},
  year={1997},
  publisher={JSTOR}
}

@article{deissenberg2008eurace,
  title={{EURACE: A massively parallel agent-based model of the European economy}},
  author={Deissenberg, Christophe and Van Der Hoog, Sander and Dawid, Herbert},
  journal={Applied mathematics and computation},
  volume={204},
  number={2},
  pages={541--552},
  year={2008},
  publisher={Elsevier}
}
